\documentclass[aps,                % American Physical Society (APS) styling
               prc,                % customize styling for APS journals
               showpacs,           % Control display of PACS
               notitlepage,
               onecolumn,          % One-column formatting
               10pt,               % 10 pt text size
               twoside,            % Two side paper
               floatfix,           % Avoid float problems when twocolumn is used
               superscriptaddress] % Associate authors with applications via superscript numbers
               {revtex4-1}

\usepackage{mathptmx}
\usepackage[T1]{fontenc}
\usepackage{geometry}         % To modify page geometry (margin, text width, ...)
\usepackage{fancyhdr}         % To modify headlines and footlines
\usepackage{graphicx}         % For pictures and cartoons
\usepackage[svgnames]{xcolor} % To color text
\usepackage{hyperref}         % To add URL and link
\usepackage{amsmath}          % Various features to facilitate writing math formula
\usepackage{amssymb}          % Various mathematical symbols
\usepackage{amsfonts}         % Various mathematical font (\math(bb)(bf)...)
\usepackage{amstext}          % To ute normal text in formula (\text)
\usepackage{textcomp}         % For Copyright
\usepackage{slashed}
\hypersetup{colorlinks=true,linkcolor=Blue,citecolor=Blue,urlcolor=Blue}

\DeclareMathAlphabet{\mathcal}{OMS}{cmsy}{m}{n}

\newcommand{\bs}[1]{p\!\!\!\!\! p \!\!\!\!\! p\!\!\!\!\! p}

\numberwithin{equation}{section}                                % number equations to the section

\fancypagestyle{firststyle}
{%
  \fancyhf{}
  \fancyhead[C]{PHYSICAL REVIEW C {\bf 87}, 034912 (2013)} % Left Odd
  \fancyfoot[L]{\small 0556-2813/2013/87(1)/034912(25)} % Left
  \fancyfoot[C]{\small 034912-\thepage} % Centered number of the page in footline
  \fancyfoot[R]{\small \textcopyright 2013 American Physical Society} % Right
}

\begin{document}

\title{\texorpdfstring{\begin{minipage}[c]{\textwidth}\centering Collisional processes of on-shell and off-shell heavy quarks in vacuum and in the Quark-Gluon-Plasma \end{minipage}}{Collisional processes of on-shell and off-shell heavy quarks in vacuum and in the Quark-Gluon-Plasma}}

\author{H.~Berrehrah}
\email{berrehrah@fias.uni-frankfurt.de}
\affiliation{\begin{minipage}[c]{\textwidth}Frankfurt Institute for Advanced Studies, Johann Wolfgang Goethe Universit\"at, Ruth-Moufang-Strasse 1,\end{minipage}\\
60438 Frankfurt am Main, Germany \\ \vspace{0.5mm}}

\author{E.~Bratkovskaya}
\email{brat@th.physik.uni-frankfurt.de}
\affiliation{\begin{minipage}[c]{\textwidth}Frankfurt Institute for Advanced Studies, Johann Wolfgang Goethe Universit\"at, Ruth-Moufang-Strasse 1,\end{minipage}\\
60438 Frankfurt am Main, Germany \\ \vspace{0.5mm}}
\affiliation{\begin{minipage}[c]{\textwidth}Institut for Theoretical Physics, Johann Wolfgang Goethe Universit\"at, Max-von-Laue-Str. 1, 60438 Frankfurt am Main, Germany \end{minipage} \\ \vspace{0.5mm}}

\author{W.~Cassing}
\email{wolfgang.cassing@theo.physik.uni-giessen.de}
\affiliation{\begin{minipage}[c]{\textwidth}Institut f\"ur Theoretische Physik, Universit\"at Giessen, 35392 Giessen, Germany\end{minipage} \\\vspace{0.5mm}}

\author{P.B.~Gossiaux}
\email{gossiaux@subatech.in2p3.fr}
\affiliation{\begin{minipage}[c]{0.98\textwidth}Subatech, UMR 6457, IN2P3/CNRS, Universit\'e de Nantes, \'Ecole des Mines de Nantes, 4 rue Alfred Kastler, 44307 Nantes cedex 3, France\end{minipage} \\ \vspace{1.5mm} \normalfont{(Received 24 August 2013; revised 3 February 2014; published 1 May 2014)}
\vspace{2.5mm}}

\author{J.~Aichelin}
\email{aichelin@subatech.in2p3.fr}
\affiliation{\begin{minipage}[c]{0.98\textwidth}Subatech, UMR 6457, IN2P3/CNRS, Universit\'e de Nantes, \'Ecole des Mines de Nantes, 4 rue Alfred Kastler, 44307 Nantes cedex 3, France\end{minipage} \\ \vspace{1.5mm} \normalfont{(Received 24 August 2013; revised 3 February 2014; published 1 May 2014)}
\vspace{2.5mm}}

\author{M.~Bleicher}
\email{bleicher@fias.uni-frankfurt.de}
\affiliation{\begin{minipage}[c]{\textwidth}Frankfurt Institute for Advanced Studies, Johann Wolfgang Goethe Universit\"at, Ruth-Moufang-Strasse 1,\end{minipage}\\
60438 Frankfurt am Main, Germany \\ \vspace{0.5mm}}
\affiliation{\begin{minipage}[c]{\textwidth}Institut for Theoretical Physics, Johann Wolfgang Goethe Universit\"at, Max-von-Laue-Str. 1, 60438 Frankfurt am Main, Germany \end{minipage} \\ \vspace{0.5mm}}

% PACS, the Physics and Astronomy Classification Scheme
\pacs{11.10.Wx, 13.85.Dz, 11.80.-m, 12.38.Mh%
\hfill DOI:\href{http://link.aps.org/doi/10.1103/PhysRevC.89.054901}{10.1103/PhysRevC.89.054901}
}

\begin{abstract}
We study the heavy-quark scattering on partons of the quark-gluon
plasma (QGP), being especially interested in the collisional
(elastic) scattering processes of heavy quarks $Q$ on light quarks
$q$ and gluons $g$. We calculate the different cross sections for
perturbative partons (massless on-shell particles in the vacuum) and
for dynamical quasiparticles (off-shell particles in the QGP medium
as described by the dynamical quasi-particles model ``DQPM'') using
the leading-order Born diagrams. Our results show clearly the effect
of a finite parton mass and width on the perturbative elastic $(q(g)
Q \rightarrow q (g) Q)$ cross sections which depend on temperature
$T$, energy density $\epsilon$, the invariant energy $\sqrt{s}$, and
the scattering angle $\theta$. Our detailed comparisons demonstrate
that the finite width of the quasiparticles in the DQPM---which
encodes the multiple partonic scattering---has little influence on
the cross section for $q Q \rightarrow q Q$ as well as $g Q
\rightarrow g Q$ scattering, except close to thresholds.  Thus, when
studying the dynamics of energetic heavy quarks in a QGP medium the
spectral width of the degrees-of-freedom may be discarded. We have,
furthermore, compared the cross sections from the DQPM with
corresponding results from hard-thermal-loop (HTL) approaches. The
HTL-inspired models---essentially fixing the regulators by
elementary vacuum cross sections and decay amplitudes instead of
properties of the QGP at finite temperature---provide quite
different results especially, with respect to the temperature dependence of
the $qQ$ and $gQ$ cross sections (in all settings). Accordingly, the
transport properties of heavy quarks will be very different as a
function of temperature when compared to DQPM results.
\end{abstract}


\maketitle

  %---------------------------------------------------------------------------------------------------------------------------------------------
\section{Introduction}
\label{Introduction}
%---------------------------------------------------------------------------------------------------------------------------------------------
\vskip -2mm

  The ultimate aim of strongly interacting physics is to study the properties of  the quark-gluon plasma (QGP) produced in relativistic heavy-ion collisions (HICs). The heavy quarks $Q= b,c $ (and correspondingly charm and beauty mesons) are considered to be one of the best probes for such a study. Owing to their large mass in HICs they are produced  dominantly by hard binary initial collisions between the incoming nucleons during the early stage of the reaction when the QGP is formed. Heavy quarks have initially a transverse momentum spectrum which can be calculated perturbatively and which is very different from a thermal spectrum. Their cross section with partons of the plasma is not strong enough to thermalize the heavy quarks during the expansion of the plasma. Therefore, they provide an important observable that probes some properties of the QGP. To this aim, it is useful to study theoretically the heavy-quark dynamics from their production until hadronization and freeze-out to obtain information on the
properties of the plasma from the finally observed spectra.

  The interaction of heavy quarks with the plasma particles is described by (elastic and inelastic) cross sections. Their knowledge in a finite temperature medium allows for the evaluation of several physical quantities, like the collisional and radiative energy loss of heavy quarks, the interaction rates, diffusion coefficients, viscosity, etc. Ultimately, the scattering of heavy quarks with the QGP particles represents the first step to the explicit microscopic dynamics of heavy flavors in the QGP and the hadronic phase. Such a study of the heavy-quark propagation can be performed within the microscopic parton-hadron-string-dynamics (PHSD) transport approach \cite{Wcassing2009EPJS,Cassing:2009vt}, which incorporates explicit partonic degrees of freedom in terms of strongly interacting quasiparticles (quarks and gluons) in line with an equation of state
from lattice QCD, as well as dynamical hadronization and hadronic collision dynamics in the final reaction phase.

 For the QGP phase the natural starting point to calculate the interactions of heavy quarks with the constituents of the plasma, the light quarks and gluons, is perturbative QCD (pQCD). A comparison with data on $R_{AA}$ and the elliptic flow $v_2$ shows that the interaction has to be much stronger than obtained in standard pQCD calculations using a constant coupling constant, zero-mass plasma constituents, and the Debye mass as an infrared regulator of the cross section. This motivates the study of heavy-quark scattering in a finite temperature QCD medium.

 In this study---by dressing the quark and gluon lines with the effective propagators---we derive the off-shell cross sections for the reactions $q Q \rightarrow q Q$ and $g Q \rightarrow g Q$, taking into account the quasiparticle nature of the quarks and gluons based on the dynamical quasiparticle model (DQPM) \cite{PhysRevD.70.034016,0954-3899-31-4-046,Wcassing2009EPJS}. These cross sections are compared with standard pQCD cross sections. We find that the finite mass of the effective quasiparticles does not only screen the singularities typical for the perturbative cross sections with massless quarks, but also modifies the shape of the scattering cross sections, especially at low momentum transfer $Q$ and at the edges of the phase space. \

 This paper aims at an effective-theory approach for the derivation of the off-shell cross sections for the interaction of massive dynamical quasiparticles as constituents of the finite-temperature strongly interacting medium (sQGP). Therefore, and to fix the notion of off-shell particles in our calculation, we start out in Sec. \ref{pPartons-DQP} with a presentation of the description of massive and off-shell particles in the DQPM. We describe the parton spectral
function by a finite mass and width in the hot QCD medium. Sections \ref{qQProcess} and \ref{gQProcess} present the differential cross sections for heavy-quark $Q$ elastic scattering on light dressed quarks $q$ (and gluons $g$) in the on-shell and off-shell limits. In these sections we compare the $qQ$ and $gQ$ elastic scatterings in vacuum with those in the QGP medium, considering both the on-shell and off-shell limits, as well as light and heavy quarks and gluons as collision partners. For completeness the case of on-shell partons will include also  massless light quarks and gluons. We start in these sections by demonstrating that we reproduce the well-known pQCD $qQ$ and $gQ$ elastic cross sections as well as those determined by Gossiaux and Aichelin following a hard-thermal-loop-inspired approach (HTL-GA) \cite{Gossiaux:2004qw,PhysRevC.78.014904,Gossiaux2009203c,Gossiaux:2009hr,Gossiaux:2009mk,Gossiaux:2010yx,Gossiaux:2006yu} for massless light quarks and gluons. Then, these particles are dressed by effective masses and the elastic $qQ$ and $gQ$ cross sections are evaluated (this approach is called \emph{``DpQCD''} for dressed pQCD). We end by addressing the
case of off-shell $qQ$ and $gQ$ elastic scattering, where the DQPM quark, antiquark and gluon masses, nonperturbative spectral functions and self-energies for different temperatures of the medium are employed. For this purpose, we use parametrizations of the quark and gluon propagators provided by the DQPM matched to reproduce lattice quantum chromodynamics (lQCD) data. The corresponding cross sections are labeled by \emph{``IEHTL''} for ``infrared enhanced hard thermal loop''.

\begin{table}[h!]
\centerline{
\begin{tabular}{|c|c|c|c|c|} \hline
                         &  $q Q$ & $q Q$  & $g Q$  & $g Q$
 \\ %\hline
                         &(On-shell)&(Off-shell) &(On-shell)  &(Off-shell)
 \\ \hline
 Naive pQCD              &  Sec. \ref{qQProcess.B}            &            & Sec. \ref{gQProcess.B}    &
 \\ \hline
 HTL-GA                  &  Sec. \ref{qQProcess.B}, \ref{qQProcess.C}      &            & Sec. \ref{gQProcess.B}, \ref{gQProcess.C}    &
 \\ \hline
 DpQCD                   &  Sec. \ref{qQProcess.C}, \ref{qQProcess.D}          &            & Sec. \ref{gQProcess.C}, \ref{gQProcess.D}    &
 \\ \hline
 IEHTL                   &               &  Sec. \ref{qQProcess.D}          &     &   Sec. \ref{gQProcess.D}
 \\ \hline
\end{tabular}}
\caption{Overview of the on-shell and off-shell heavy-quark elastic cross sections following the
different approaches studied in this article.}
\label{tab:SummaryqQgQ}
\end{table}

  The off-shell cross sections are compared to the perturbative ones throughout Sects. \ref{qQProcess} and \ref{gQProcess}, where in the limit of a high hard scale $Q^2$ the off-shell cross sections are shown to approach the perturbative ones. Finally, in Sec. \ref{Summary}, we present our conclusions, summarize the main results, and point out future applications. In Table \ref{tab:SummaryqQgQ} we present the outline of our systematic study in compact form.
  %---------------------------------------------------------------------------------------------------------------------------------------------
\section{Perturbative partons \emph{versus} dynamical quasi-particles}
\label{pPartons-DQP}
%---------------------------------------------------------------------------------------------------------------------------------------------
\vskip -2mm

%---------------------------------------------------------------------------------------------------------------------------------------------
\subsection[Perturbative partons]{Perturbative partons} %---------------------------------------------------------------------------------------------------------------------------------------------

  The scattering of heavy quarks in vacuum and in a QGP medium in lowest-order QCD perturbation theory (pQCD) has  extensively been studied in the literature \cite{PhysRevD.17.196,Combridge1979429}. The application of the lowest-order QCD perturbation theory (pQCD) to these collisions \cite{Combridge1979429} has been motivated by the low value of the effective coupling (which has been considered as constant or running for moderate values of $Q^2$) at a time when the QGP has been considered as a system of weakly interacting partons.

  In the last decade, experiments at BNL Relativistic Heavy Ion Collider (RHIC) and CERN Large Hadron Collider (LHC) have shown that the QGP, produced in ultrarelativistic HICs, is a strongly interacting system and recent theoretical and experimental studies have improved our understanding of its properties. On the theoretical side, the estimates of the temperatures $T$, which are expected to be currently achieved in HICs at RHIC and LHC, are not large compared to the QCD scale $\Lambda_{QCD}$ \cite{PhysRevD.31.545}. This has the peculiar consequence that the coupling  cannot be considered as small anymore. The experimental studies at RHIC have indicated that the new medium created in ultrarelativistic Au+Au collisions is a strongly interacting many-body system, interacting even stronger than hadronic matter. In addition, lattice QCD results \cite{Karsch2002199} have also shown that the high-temperature plasma phase is a medium of interacting partons which are strongly screened and influenced by non perturbative effects even at temperatures as high as 10 $T_c$. From these observations, the concept of perturbatively interacting massless quarks and gluons as constituents of the QGP, which scatter according to the leading (Born) diagrams, had to be reconsidered.

  A first development---aiming to treat non perturbative effects in heavy-quark scattering---was given by Braaten \emph{et al}. \cite{PhysRevD.44.R2625,PhysRevD.44.1298} using a powerful resummation technique based on reordering perturbation theory by expanding correlation functions in terms of effective propagators and vertices instead of bare ones \cite{PhysRevD.44.2774}. In this approach, the singularities in the cross-section calculation are regularized by the thermal masses of quarks and gluons, which are, in turn, determined by the one-loop leading order result in thermal perturbation theory, denoted as the hard-thermal-loop (HTL) approach. Later, Gossiaux \emph{et al}. extended this approach
\cite{Gossiaux:2004qw,PhysRevC.78.014904,Gossiaux2009203c,Gossiaux:2009hr,Gossiaux:2009mk,Gossiaux:2010yx,Gossiaux:2006yu}.

  Following the latter direction and to consider all the effects of the non-perturbative nature of the sQGP constituents, i.e., the large coupling, the multiple scattering etc., we refrain from a fixed-order thermal loop calculation relying on perturbative self-energies (calculated in the limit of infinite temperature) to fix the in-medium masses of the quarks and gluons and pursue instead a more phenomenological approach. The multiple strong interactions of quarks and gluons in the sQGP are encoded in their effective propagators with broad spectral functions. The effective propagators, which can be interpreted as resummed propagators in a hot QCD environment, have been extracted from lattice data in the scope of the DQPM \cite{Wcassing2009EPJS,Cassing:2009vt,Cassing2007365}.

  We note that the majority of previous studies of heavy-quark scattering have considered the QGP partons as massless \cite{PhysRevD.17.196}. Only heavy quarks have been massive
\cite{Combridge1979429} in the scattering processes $q Q \rightarrow q Q$, $g Q \rightarrow g Q$. Therefore, divergence problems are encountered especially in the $t$ channel (cf. Sec. \ref{qQProcess} and \ref{gQProcess}). Several attempts have been advanced to remedy this problem, either by introducing an infrared  cut-off in the integration over the momentum transfer squared $t$ \cite{Combridge1979429} or by generating  a finite mass for the exchanged gluon \cite{PhysRevC.78.014904,PhysRevD.44.R2625,PhysRevD.44.1298}, as in
the HTL approach. The value of the introduced cut off or of the gluon mass has been fixed according to phenomenological considerations. Therefore, the uncertainty in these scales is large, which also leads to corresponding uncertainties in the cross sections.

%---------------------------------------------------------------------------------------------------------------------------------------------
\subsection[Partons in the Dynamical Quasi-Particle Model (DQPM)]{Partons in the Dynamical Quasi-Particle Model (DQPM)} %---------------------------------------------------------------------------------------------------------------------------------------------

  The DQPM describes QCD properties in terms of the single-particle Green's functions [in the sense of a two-particle irreducible (2PI) approach] and leads to the notion of the constituents of the sQGP being strongly interacting massive effective quasi-particles with broad spectral functions (owing to the interaction rates). The strategy for the determination of parton masses and widths within the DQPM approach is to fit the analytical expression of the dynamical quasiparticle entropy density $s^{DQP}$  to the lQCD entropy density \emph{``$s^{lQCD}$''} determined numerically. Based on the studies of Peshier \cite{PhysRevD.70.034016,0954-3899-31-4-046} and Cassing \cite{Cassing2007365,Wcassing2009EPJS}, the first step of the DQPM in calculating $s^{DQP}$ is well determined for the case of a scalar theory with (retarded) scalar gluon and fermion propagators. The dynamical quasiparticle entropy density $s^{DQP}$ is then fitted to lQCD data, which makes it possible to fix the few parameters present in $s^{DQP}$ which is given by
{\setlength\arraycolsep{0pt}
\begin{eqnarray}
\label{equ:Sec1.1.1}
s^{DQP}  = \ - \sum_{i=g,q,\bar{q}} \!\int\!\frac{d \omega}{2 \pi} \frac{d^3p}{(2 \pi)^3} & & \frac{\partial n_{B/F,\bar{F}}}{\partial T} \times \left( \Im\ln(- \Delta_i^{-1}) + \Im \Pi_i \,\Re \Delta_i \right),
%\nonumber
\end{eqnarray}}
  where $n_B$, $n_F$ ($n_{\bar{F}}$) denote the Bose and Fermi distribution functions for gluons and quarks (antiquarks), respectively, while $\Delta_i$ ($\Pi_i$) is the retarded propagator (self-energy) of the particle $i$. The first part in (\ref{equ:Sec1.1.1}) gives the quasiparticle contribution to the entropy density, whereas the second part $\sim \Im \Pi_i$ is the interaction contribution. With the restriction to scalar degrees of freedom we explicitly discard the longitudinal gluon. This might be legitimate because at finite temperature the contribution of the longitudinal gluon to the general thermodynamic potential is sub dominant compared to the contribution from the transverse gluons. In the future, however, the full Lorentz structure of the gluon propagator will have to be considered.

  The variation of parton masses as a function of the medium properties is described by the spectral functions, which are (except of a factor) identical to the imaginary part of the retarded propagator. These are no longer $\delta$ functions in the invariant mass squared (as in the case for bare masses) \cite{Cassing:2009vt} but (in the more general case) related to the imaginary part of the trace of the effective propagator $D_{\mu}^{\nu}$ and to the partonic self-energies $\Sigma_{\nu}^{\mu}$ as:
{\setlength\arraycolsep{0pt}
\begin{eqnarray}
\label{equ:Sec2.1}
\hspace{-0.7cm} A (p) \ \propto \Im D_{\mu}^{\mu} (p) \propto \frac{\Im \Sigma_{\mu}^{\mu} (p)}{[p^2 - \Re \Sigma_{\mu}^{\mu} (p)]^2 + [\Im \Sigma_{\mu}^{\mu} (p)]^2},
%\nonumber\\
%& & {}
\end{eqnarray}}
  where $\Re \sum$ and $\Im \sum$ denote the real and imaginary parts of the (4$\times$4) self-energy. For the current analysis,  we use the approximation of momentum-independent real and imaginary parts of the self-energy, which are---for a given temperature $T$---proportional to the parton mass and width, respectively (cf. \cite{Wcassing2009EPJS}). The propagator $\Delta$ is expressed in the Lehmann representation in terms of the spectral function $\Delta (p) = \int \frac{d \omega}{2 \pi} \frac{A (\omega, \bs{p})}{p_0 -\omega}$, which allows finally the evaluation of the entropy functional $s^{DQP}$ (\ref{equ:Sec1.1.1}) for a given form of the spectral function.

  An often-used ansatz to model a non zero width is obtained by replacing the free spectral function $A_0 (p) = 2 \pi [\delta (\omega - p^2)^2 - \delta (\omega + p^2)^2]$ by a Lorentzian form. This Lorentzian parametrization of the partonic spectral functions $A_{i} (\omega_{i})$, where $i$ is the parton species and $\omega_i^2 = m_i^2 + \bs{p}_i^2$, is given by
{\setlength\arraycolsep{-10pt}
\begin{eqnarray}
\label{equ:Sec2.2}
A_i^L (\omega_i) & & \ = \frac{\gamma_i}{\tilde{E}_i} \biggl(\frac{1}{(\omega_i - \tilde{E}_i)^2 + \gamma_i^2} -
\frac{1}{(\omega_i + \tilde{E}_i)^2 + \gamma_i^2} \biggr)  \equiv \frac{4 \omega_i \gamma_i}{(\omega_i^2 - \bs{p}_i^2 - M_i^2)^2 +
4 \gamma_i^2 \omega_i^2},
\end{eqnarray}}
  with $\tilde{E}_i^2 (\bs{p}_i) = \bs{p}_i^2 + M_i^2 - \gamma_i^2$ and $i \in [g, q, \bar{q}, Q, \bar{Q}]$. The spectral functions $A_i^L (\omega_i)$ are antisymmetric in $\omega_i$ and normalized as:
{\setlength\arraycolsep{0pt}
\begin{eqnarray}
\label{equ:Sec2.3}
& & \hspace*{-1cm} \int_{- \infty}^{+ \infty} \frac{d \omega_i}{2 \pi} \ \omega_i \ A_i^L (\omega_i, \bs{p}) = \!\!\! \int_0^{+ \infty} \frac{d \omega_i}{2 \pi} \ 2 \omega_i \ A_i^L (\omega_i, \bs{p}_i) =
1 ,
%\nonumber\\
%& & {}
\end{eqnarray}}
  where $M_i$, $\gamma_i$ are the dynamical quasiparticle mass (i.e., pole mass) and width of the spectral function for particle $i$, respectively. They are directly related to the real and imaginary parts of the related self-energy, e.g., $\Pi_i = M_i^2 - 2 i \gamma_i \omega_i$, \cite{Cassing:2009vt}. In the off-shell approach, $\omega_i$ is an independent variable and related to the \emph{``running mass''} $m_i$ by: $\omega_i^2 = m_i^2 + \bs{p}_i^2$. Therefore, one has
{\setlength\arraycolsep{0pt}
\begin{eqnarray}
& & \hspace*{-1cm} \sum_i \int \frac{d^4 p_i}{(2 \pi)^4} \ A_i (p_i) = \sum_i \int_0^{+ \infty} \frac{d \omega_i}{2 \pi} \int \frac{d^3 p_i}{(2 \pi)^3} \ A_i^L (\omega_i, \bs{p}_i).
%\nonumber\\
%& & {}
\label{equ:Sec2.4}
\end{eqnarray}}
  The mass (for gluons and quarks) is assumed to be given by the thermal mass in the asymptotic high-momentum regime. By considering the effect on the entropy, which is known from lattice calculations, the width of the partons, which in the perturbative limit is given by $\gamma \approx g^2 \ln (g^{-1} T)$, should be sizable at intermediate temperatures \cite{PhysRevD.44.R2625,0954-3899-31-4-046,PhysRevD.70.034016,Cassing:2009vt}. Hence, the functional forms of $M_g$, $M_q$, $\gamma_g$, and $\gamma_q$ are given by ($\gamma_g$ and $\gamma_q$ are given here for zero quark potential $\mu_q = 0$)
{\setlength\arraycolsep{-6pt}
\begin{eqnarray}
\label{equ:Sec2.5}
& & M_g^2 (T) = \frac{g^2 (T/T_c)}{6} \Biggl((N_c + \frac{1}{2} N_f) T^2 + \frac{N_c}{2} \sum_q \frac{\mu_q^2}{\pi^2} \Biggr),
\nonumber\\
& & {} M_q^2 (T) = \frac{N_c^2 - 1}{8 N_c} g^2 (T/T_c) \Biggl( T^2 + \frac{\mu_q^2}{\pi^2} \Biggr),
\\
& & {}  \gamma_g (T) = \frac{1}{3} N_c \frac{g^2 (T/T_c) T}{8 \pi} \ln \left(\frac{2 c}{g^2 (T/T_c)} + 1 \right),
\nonumber\\
& & {} \gamma_q (T) = \frac{1}{3} \frac{N_c^2 - 1}{2 N_c} \frac{g^2 (T/T_c) T}{8 \pi} \ln \left(\frac{2 c}{g^2 (T/T_c)} + 1 \right).
\nonumber
\end{eqnarray}}
  The physical processes contributing to the width $\gamma_g$ are both $g g \rightarrow g g$, $g q \rightarrow g q$ scattering, as well as splitting and fusion reactions $g g \rightarrow g$, $g g \rightarrow g g g$, $g g g \rightarrow g g g g$, or $g \rightarrow q \bar{q}$, etc. On the fermion side elastic fermion-fermion scattering $f f \rightarrow f f$, where $f$ stands for a quark $q$ or antiquark $\bar{q}$, fermion-gluon scattering $f g \rightarrow f g$, gluon bremsstrahlung $f f \rightarrow f f g$, or quark-antiquark fusion $q \bar{q} \rightarrow g$, etc. Note, however, that the explicit form of Eq. (\ref{equ:Sec2.5}) is derived for hard two-body scatterings only. It is worth pointing out that the ratio of the masses to their widths $\gamma/M \sim g \ln (2 c/g^2 + 1)$ approaches zero only asymptotically for $T \rightarrow \infty$ such that the width of the quasiparticles is smaller but comparable to the pole mass slightly above $T_c$ up to all HIC energy scales.

  The coupling constant $g^2 (T/T_c)$ in (\ref{equ:Sec2.5}) is considered here as depending on the medium temperature and for $T > T_s$ is given by
{\setlength\arraycolsep{0pt}
\begin{eqnarray}
\label{equ:Sec2.6}
& & \hspace{-0.1cm} \displaystyle{g^2 (T/T_c) = \frac{48 \pi^2}{(11 N_c - 2 N_f) \ln \left(\lambda^2 (\frac{T}{T_c} - \frac{T_s}{T_c})^2 \right)}  \hspace{0.3cm}  T > T^{\star} = 1.19 \ T_c},
\nonumber\\
& & {} \hspace{-0.1cm} \displaystyle{g^2 (T/T_c) \rightarrow g^2 (T^{\star}/T_c)
\left(\frac{T^{\star}}{T} \right)^{3.1}  \hspace{0.7cm}  T < T^{\star} = 1.19 \ T_c.}
\end{eqnarray}}
  We mention that the form of the running coupling specified in Eq. (\ref{equ:Sec2.6}) for low temperatures ($T < 1.19 T_c$) is fully introduced by hand to fit the equation of state of lattice QCD for 2+1 flavors from Refs. \cite{Borsanyi:2010bp,Borsanyi:2010cj} also down to temperatures of 120 MeV where partonic degrees of freedom are no longer expected to persist. For our actual studies we only address temperatures above $T_c$.  Once the three free parameters in Eq. (\ref{equ:Sec2.6}) are fixed, the resulting coupling constant may tentatively be employed for a model study of heavy-quark scattering.

  The DQPM quark mass and width in (\ref{equ:Sec2.5}) are fixed for the $u$ and $d$ light quarks. For the other flavors (especially $s$ and $c$ quarks), one adopts \cite{Cassing:2009vt}
{\setlength\arraycolsep{0pt}
\begin{eqnarray}
\label{equ:Sec2.7}
& & M_s (T) = M_{u,d} (T) + 0.045 \ \textrm{GeV}, \  \  M_c (T) = M_{u,d} (T) + 1.3 \ \textrm{GeV},
\nonumber\\
& & {} \gamma_s (T) = \gamma_{u, d} (T) = \gamma_c (T),
\end{eqnarray}}
  where the assumption $\gamma_{u, d} (T) = \gamma_c (T)$ is not extracted from lattice QCD and is expected to represent an upper limit for the width of the $c$ quark. Because $\gamma_c \ll M_c$ the charm quark should be considered as a ``good quasiparticle.'' The actual value for the $s$-quark has been fixed by the kaon-to-pion ratio at top CERN Super Proton Synchrotron (SPS) energy using PHSD calculations in comparison to experimental data \cite{Cassing:2009vt}. The transverse momentum spectra are compatible with using the same width for $u,d,s$ quarks. This assumption then has been used also for all heavy-ion studies from low SPS to LHC energies in the PHSD transport model and lead to a good description of the strange hadron multiplicities and spectra.

  Using the expressions (\ref{equ:Sec2.2}) and (\ref{equ:Sec2.5})--(\ref{equ:Sec2.7}), the analytical expression of the dynamical quasiparticle entropy density $s^{DQP}$---to be
fitted to the lQCD entropy density \emph{``$s^{lQCD}$''}---is given explicitly as \cite{Wcassing2009EPJS,0954-3899-31-4-046}
{\setlength\arraycolsep{0pt}
\begin{eqnarray}
\label{equ:Sec1.1.8}
& & s^{DQP}  = \ s^{(0),DQP} + \Delta s^{DQP},
%\nonumber\\
%& & {}
\end{eqnarray}}
\vspace*{-0.5cm}
 with:
\begin{widetext}
{\setlength\arraycolsep{0pt}
\begin{eqnarray}
\label{equ:Sec1.1.9}
s^{(0),DQP} & & = \ d_g \ \frac{1}{T} \int \frac{d^3 p}{(2 \pi)^3} \biggl(-T\ln(1-e^{-\omega_{m_g}/T}) + \omega_{m_g} \, n_B \left(\omega_{m_g}/T \right) \biggr) 
\nonumber\\
& & {} \ +  d_q \ \frac{1}{T} \int \frac{d^3 p}{(2 \pi)^3} \biggl(T\ln(1-e^{-\omega_{m_q}/T}) + \omega_{m_q} \, n_F \left(\omega_{m_q}/T \right) \biggr)
\nonumber\\
& & {} \ + d_{\bar{q}} \ \frac{1}{T} \int \frac{d^3 p}{(2 \pi)^3} \biggl(T\ln(1-e^{-\omega_{m_{\bar{q}}}/T}) + \omega_{m_{\bar{q}}} \, n_{\bar{F}} \left(\omega_{m_{\bar{q}}}/T \right) \biggr),
%\nonumber\\
%& & {}
\end{eqnarray}}
\vspace*{-0.5cm}
and
{\setlength\arraycolsep{0pt}
\begin{eqnarray}
\label{equ:Sec1.1.10}
\Delta s^{DQP} & & = \ d_g \ \int \frac{d \omega}{(2 \pi)} \frac{d^3 p}{(2 \pi)^3} \frac{\partial n_B}{\partial T}\, \left(\arctan\biggl(\frac{2\gamma\omega}{\omega_{m_g}^2-\omega^2} \biggr) - \frac{2\gamma\omega(\omega_{m_g}^2-\omega^2)}{(\omega^2-\omega_{m_g}^2)^2+(2\gamma\omega)^2} \right)
\nonumber\\
& & {} \ + d_{q} \int \frac{d \omega}{(2 \pi)} \frac{d^3 p}{(2 \pi)^3} \frac{\partial n_F}{\partial T}\, \left(\arctan\biggl(\frac{2\gamma\omega}{\omega_{m_q}^2-\omega^2} \biggr) - \frac{2\gamma\omega(\omega_{m_q}^2-\omega^2)}{(\omega^2-\omega_{m_q}^2)^2+(2\gamma\omega)^2} \right)
\nonumber\\
& & {} \ + d_{\bar{q}} \int \frac{d \omega}{(2 \pi)} \frac{d^3 p}{(2 \pi)^3} \frac{\partial n_{\bar{F}}}{\partial T}\,
\left(\arctan\biggl(\frac{2\gamma\omega}{\omega_{m_{\bar{q}}}^2-\omega^2} \biggr) - \frac{2\gamma\omega(\omega_{m_{\bar{q}}}^2-\omega^2)}{(\omega^2-\omega_{m_{\bar{q}}}^2)^2
+(2\gamma\omega)^2} \right)\! ,
%\nonumber\\
%& & {}
\end{eqnarray}}
\end{widetext}
  where $d_i$ is the degeneracy factor of the particle and $\omega_i^2 = m_i^2 + \bs{p}_i^2$. The contribution $s^{(0),DQP}$ has the form of the entropy for a massive quasiparticle while the (subleading) contribution $\Delta s^{DQP}$ is attributable to a nontrivial imaginary part of the self-energy.

\begin{figure}[h!]
   \begin{center}
     \begin{minipage}[h!]{0.6\linewidth}
\includegraphics[width=8.3cm, height=7.3cm]{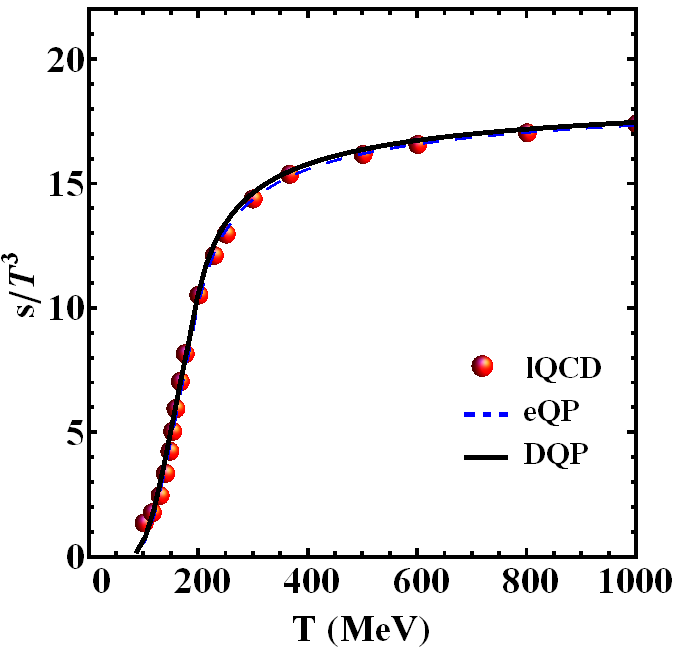}
     \end{minipage}\hfill
     \begin{minipage}[h!]{0.4\linewidth}
%\vspace{-3mm}
\caption{\emph{(Color online) The fit of the DQPM entropy density
$s^{DQP}$ [Eqs. (\ref{equ:Sec1.1.8})--(\ref{equ:Sec1.1.10})] to lattice data of Refs. \cite{Borsanyi:2010bp,Borsanyi:2010cj} for the entropy density $s^{lQCD}$. Here eQP stays for the effective quasiparticle contribution where $s^{DQP} = s^{(0),DQP}$ and DQP for the total dynamical quasiparticle entropy density, including the (subleading) contribution from the finite width.} \label{fig:slQCD-DQPM}}
   \end{minipage}
   \end{center}
\end{figure}

  The fit of the DQPM entropy density $s^{DQP}$ (\ref{equ:Sec1.1.8})--(\ref{equ:Sec1.1.10}) to lattice data for the entropy density $s^{lQCD}$ is shown in Fig.  \ref{fig:slQCD-DQPM} and suggests for $N_c = N_f = 3$ the following values of the parameters contained in the expressions (\ref{equ:Sec2.5}) and (\ref{equ:Sec2.6}) \cite{Cassing:2009vt}:
{\setlength\arraycolsep{0pt}
\begin{eqnarray}
\label{equ:Sec2.8}
& & \hspace*{-1cm} T_c = 158 \ \textrm{MeV}; \ T_s = 0.56 \ T_c; \ c = 14.4; \ \lambda = 2.42.
%\nonumber\\
%& & {}
\end{eqnarray}}
  The parametrization of the light quark and gluon DQPM widths given by Eq. (\ref{equ:Sec2.5}) with the parameters (\ref{equ:Sec2.8}) is obtained by fitting the lattice data of Refs.
\cite{Borsanyi:2010bp,Borsanyi:2010cj}. This parametrization has been used in the PHSD calculations in  Ref. \cite{PhysRevC.87.024901}. However, one has to keep in mind that
earlier fits to the lattice data from Refs. \cite{Kaczmarek:2004gv,Okamoto:1999hi,Karsch2000447} have led to different parametrizations of the DQPM widths which had been used in
Refs. \cite{Wcassing2009EPJS,Cassing:2009vt}.

\begin{figure}[h!] %tbh!
\begin{center}
    \begin{minipage}[h!]{0.6\linewidth}
     \includegraphics[height=7cm, width=7.5cm]{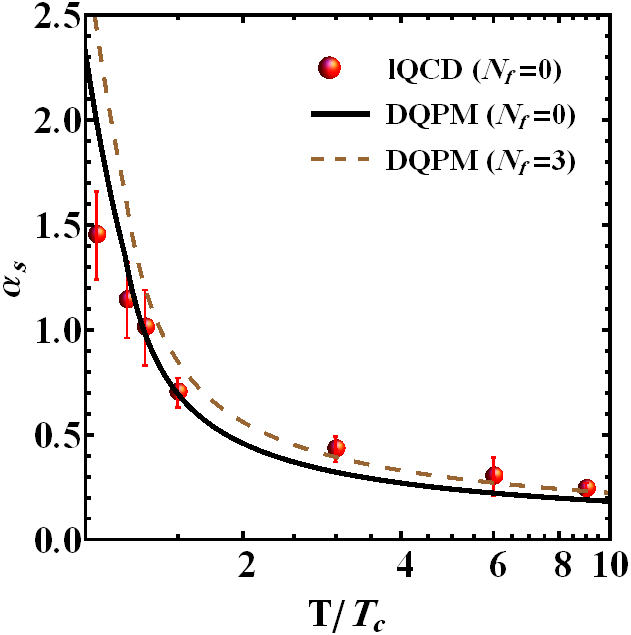} \hspace{5mm}
     \end{minipage}\hfill
     \begin{minipage}[h!]{0.4\linewidth}
%\vspace{-4mm}
\caption{\emph{(Color online) The running coupling constant $\alpha_s = g^2/(4 \pi)$ (\ref{equ:Sec2.6}) as a function of $T/T_c$ in the lQCD for $N_f = 0$ (red spheres) \cite{Kaczmarek:2004gv} and in the DQPM for $N_f = 0$ (black line) and $N_f = 3$ (dashed brown line).}\label{fig:DQPM-alpha}}
   \end{minipage}
\end{center}
\end{figure}

  The running coupling constant $\alpha= \alpha_s$ is presented in Fig. \ref{fig:DQPM-alpha} as a function of $T/T_c$. One sees that $\alpha_s$ is much larger than 1 near $T_c$ and non perturbative effects are most pronounced at these temperatures. The DQPM provides a good parametrization of the QCD running coupling as a function of temperature in the non perturbative regime for temperatures at least up to $10 T_c$. Note that close to $T_c$ the full coupling calculated on the lattice increases with decreasing temperature much
faster than the pQCD regime.

\begin{figure}[h!] %tbh!
\begin{center}
    \begin{minipage}[h!]{0.6\linewidth}
     \includegraphics[height=7cm, width=7.5cm]{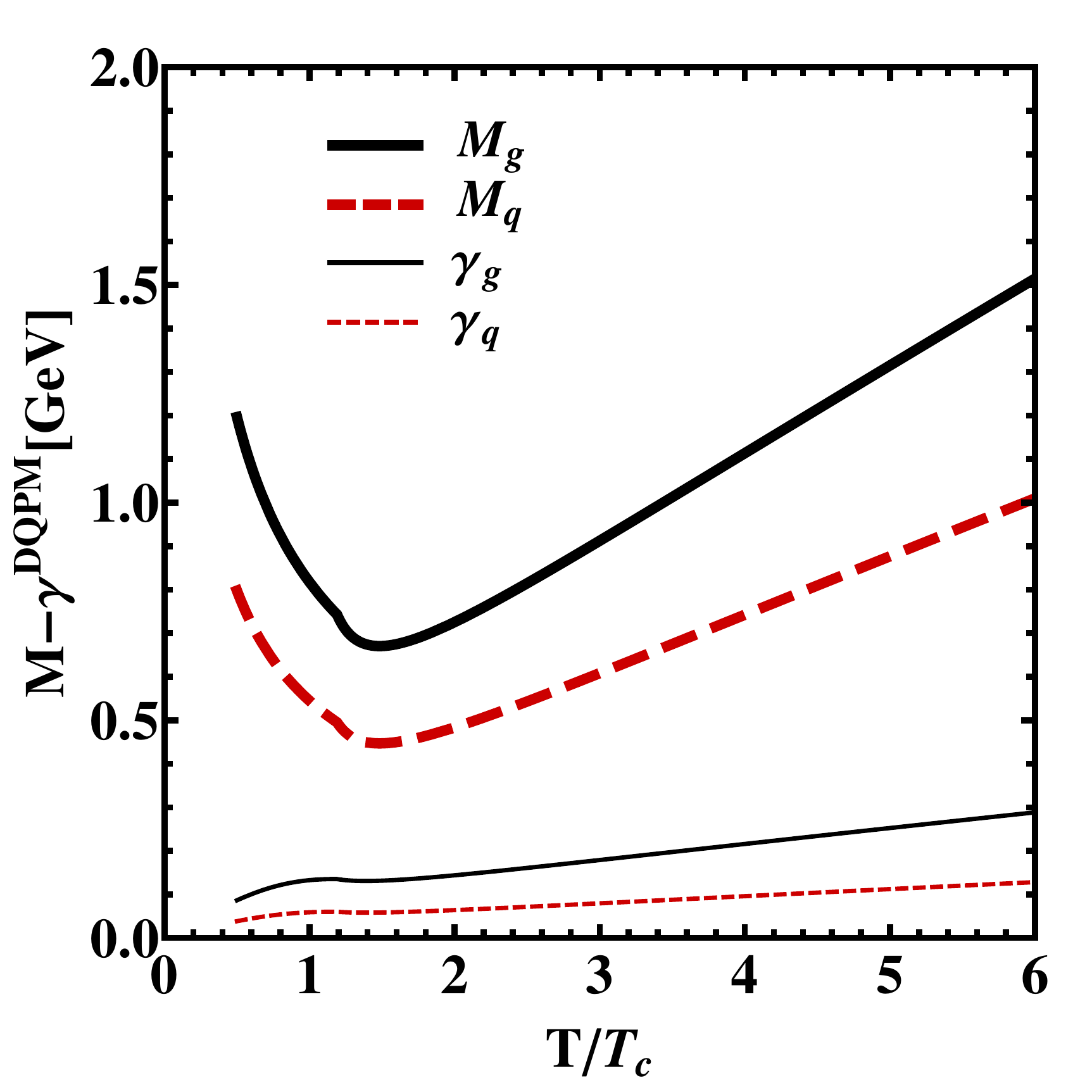}
     \end{minipage}\hfill
     \begin{minipage}[h!]{0.4\linewidth}
\vspace{-4mm}
\caption{\emph{(Color online) The DQPM pole masses and widths for the light quarks ($M_q$, $\gamma_q$) and gluons ($M_g$, $\gamma_g$) given by Eq. (\ref{equ:Sec2.5}).}}
\label{fig:DQPM-MGamma}
   \end{minipage}
\end{center}
\end{figure}

  The DQPM masses and widths for the light quarks and gluon, given by Eq. (\ref{equ:Sec2.5}), are presented in Fig. \ref{fig:DQPM-MGamma}. Because the width is found to be much smaller as the pole mass, the excitations can well be considered as quasiparticles. This could be expected from the general parametric behavior $\gamma \sim g^2 \ln (g^{-1} + 1)$ when extrapolating to larger coupling ``$g$''.

  For larger $T$, after a shallow minimum at $T \approx 1.2 T_c$, the width $\gamma$ increases very slowly with $T$ even for very large $T$ it is to a good accuracy proportional to the temperature (and also to the mass). This underlines the fact that in this range of temperatures quasiparticle models can provide an effective description. However, up to rather large temperatures the coupling is large: Terms of higher order in $g$ contribute significantly in the resummed entropy. We note additionally that a couple of
correlators have been computed in the last years within the DQPM to find further constraints in comparison to lQCD, i.e., the shear and bulk viscosities, the heat conductivity, and the electric conductivity \cite{Marty:2013ita}, as well as the electromagnetic correlator in comparison to lQCD. All studies indicate that the effective model assumptions inherent in the DQPM (with only three parameters) appear to match rather well a variety of QCD properties from lQCD.

\begin{figure}[h!]
\begin{center}
% \hskip -1em
  \includegraphics[height=5.5cm, width=5.7cm]{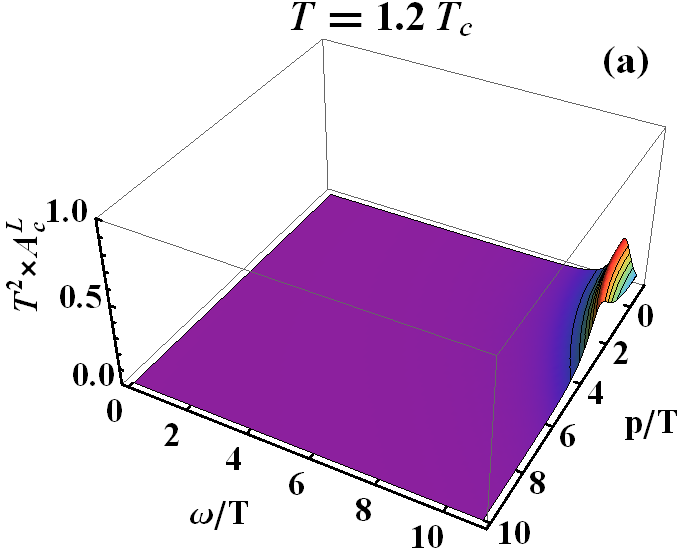}
% \vskip 1.2em
 \hfill
  \includegraphics[height=5.5cm, width=5.7cm]{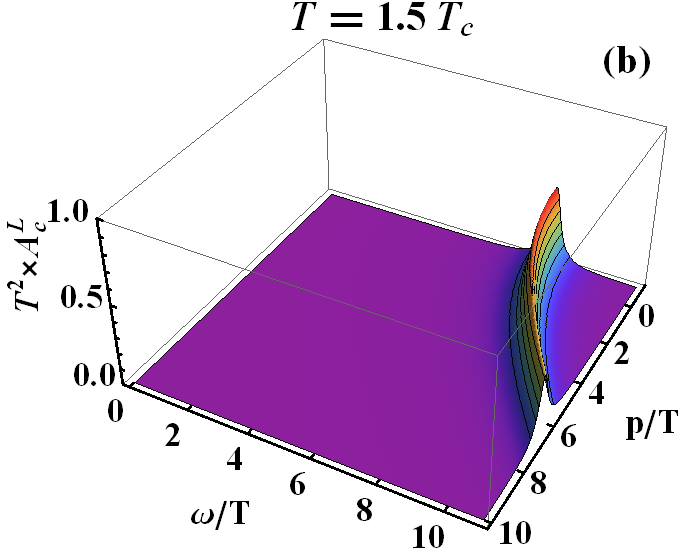}
% \vskip 1.2em
 \hfill
  \includegraphics[height=5.6cm, width=5.8cm]{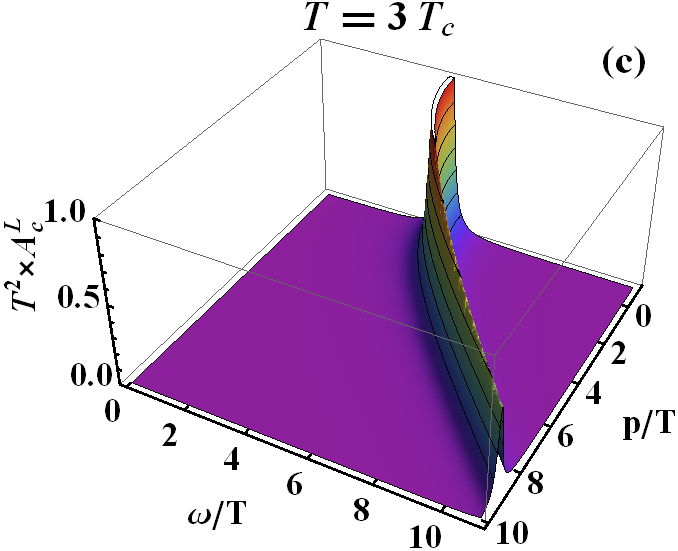}
 \caption{\emph{(Color online) The Lorentzian spectral functions for charm quarks using the
 parameters (\ref{equ:Sec2.7}) for $T/T_c = 1.2, 1.5, 3$ [$\omega$ and $p = \parallel \!\! \bs{p} \!\! \parallel$
 are in units of T and $A_c^L (\omega, p)$ in units of $T^{-2}$].
 Shown here is the full phase space although the present approach can make statements only
 for momenta of the order of a few times $T$.}
\label{fig:3DSpectralFunctionDQPM}}
\end{center}
\end{figure}

  Using the pole masses and widths (\ref{equ:Sec2.5}) and the running coupling  (\ref{equ:Sec2.6}) with the parameters (\ref{equ:Sec2.7}), the Lorentzian spectral function for the different QGP species is completely determined.  Figure \ref{fig:3DSpectralFunctionDQPM} gives a three-dimensional (3D) visualization of the heavy-quark Lorentzian spectral function as a function of $\omega/T$ and $p/T$ for three different temperatures (1.2 $T_c$, 1.5 $T_c$, and 3 $T_c$). When increasing the temperature, the peaks of the spectral functions are approaching the $\omega= \parallel \!\! \bs{p} \!\! \parallel$ region. Owing to the high mass of the charm quark the different spectral functions are almost entirely in the time like region for the temperatures shown in Fig. \ref{fig:3DSpectralFunctionDQPM}, which again points out the quasiparticle nature of the charm quark.

  From Fig. \ref{fig:3DSpectralFunctionDQPM} we see that the Lorentzian-$\omega$ spectral function peaks at small values of $p/T$ at the pole mass of the charm quark $`M_c (T)`$ [$M_{c} (1.2 T_c)=1.8$ GeV, $M_{c} (1.5 T_c)=1.75$ GeV, $M_{c} (3 T_c)=1.9$ GeV]. Therefore, a non relativistic (NR) approximation, where $\omega = \sqrt{m^2 + p^2} \approx m$ for $m \gg p$, is a good approximation. Thus, we can define a Breit-Wigner-$m$ spectral function as a NR approximation to the Lorentzian-$\omega$ form (neglecting $\parallel
\! \bs{p} \! \parallel$ compared to $m$, $m \gg \parallel \! \bs{p} \! \parallel$), where the running parameter is the mass $m$ and not the energy $\omega$ any more. The Breit-Wigner-$m$ spectral function $A_i^{BW} (m_i)$ is defined by
{\setlength\arraycolsep{0pt}
\begin{eqnarray}
& & A_i^{BW} (m_i) = \frac{2}{\pi} \frac{m_i^2 \gamma_i^{\star}}{\left(m_i^2 - M_i^2 \right)^2 + (m_i \gamma_i^{\star})^2},  \ \ \textrm{with} \ \ \int_0^{\infty} d m_i \ A_i (m_i, T) = 1,
\label{equ:Sec2.9}
\end{eqnarray}}
  where $M_i$ and $\gamma_i^{\star}$ are again the dynamical quasiparticle mass (i.e., pole mass) and width, respectively, where $\gamma_i^{\star} = 2 \frac{\sqrt{M_i^2 + p_i^2}}{M_i} \gamma_i \approx 2 \gamma_i$ for $p_i \ll M_i$. Thus, the three DQPM parameters $(T_s, \lambda, c)$ defined for $A^{L} (\omega)$ do not need to be readjusted and can be used for the $A^{BW} (m)$ Breit-Wigner-$m$ spectral function, too. In Eq. (\ref{equ:Sec2.9}), $m \equiv m_i$ is the independent variable and related to $\omega$ by $\omega^2 = m^2 + \bs{p}^2$. The integration over $m$ changes, however, and Eq. (\ref{equ:Sec2.4}) becomes:
{\setlength\arraycolsep{0pt}
\begin{eqnarray}
\hspace*{-1.4cm} \sum_i \!\!\! \int \!\!\!  \frac{d^4 p_i}{(2 \pi)^4} \ A_i (p_i) \! = \!\! \sum_i \!\! \int \!\!\!\! \frac{m_i d m_i}{2 \pi \sqrt{m_i^2 + \bs{p}_i^2}} \!\! \int \!\!\! \frac{d^3 p_i}{(2 \pi)^3} A_i^{BW} \!\!(m_i, \bs{p}_i).
%\nonumber\\
%& & {}
\label{equ:Sec2.10}
\end{eqnarray}}
  The condition $m^2 = \omega^2 - \bs{p}^2 > 0$ used in Eq. (\ref{equ:Sec2.10}) is valid only for the timelike part of the Lorentzian spectral function $A^{L} (\omega)$. In fact, the Lorentzian spectral function $A^{L} (\omega)$, in principle, contains timelike $(t)$ and spacelike $(s)$ information. Generally, one can have $(t)+(s)$ or $(s)+(s)$ interactions going to final $(t)+(s)$ or $(s)+(s)$ states. Purely spacelike processes should contribute to the potential energy of the degrees of freedom and not for real scatterings of $(t)+(t) \rightarrow (t)+(t)$. Because we propagate only timelike particles, we can deal with the spectral function $A^{BW} (m)$ because a small spacelike part of $A^{L} (\omega)$ should contribute to the evaluation of the potential energy (which we discard). Finally, let us point out that by using the Breit-Wigner spectral function $A^{BW} (m)$, we deal only with the scatterings of initial timelike particles to final timelike particles, i.e., $(t)+(t) \rightarrow (t)+(t)$.

  Now, having fixed the parton masses and widths as well as the effective coupling $g^2(T/T_c)$ by fitting the  lQCD results, we can evaluate the $qQ$ and $gQ$ cross sections for finite partons masses at different invariant energy $\sqrt{s}$ and temperature $T$ without encountering divergence problems.

%------------------------------------------------------------------------------------------------------------------------------------------------------------------------------
\subsection[Cross sections for off-shell partons]{Cross sections for off-shell partons} %------------------------------------------------------------------------------------------------------------------------------------------------------------------------------

  In our calculation of the $qQ$ and $gQ$ elastic cross sections for dynamical quasiparticles with a finite width in the spectral function, we also consider the scattering of on-shell massive quasiparticles as well as massless particles for comparison. Depending on what we consider as pQCD or DQPM particles (in and out particles), we have the following situations when studying the process $(1)^{m^{(1)}} + (2)^{m^{(2)}} \rightarrow (3)^{m^{(3)}} + (4)^{m^{(4)}}$ and deduce the corresponding quasielastic infrared enhanced hard thermal loop (IEHTL) cross section $\sigma^{IEHTL}$ by convolution of the modified pQCD cross section $\sigma$, where complex propagators are considered in the transition matrix elements, with the spectral functions, i.e.,
\begin{widetext}
{\setlength\arraycolsep{0pt}
\begin{eqnarray}
& & i) \ \sigma^{IEHTL} (s, m^{(1)}, m^{(2)}) = \int d m^{(3)} \ d m^{(4)} \ \sigma (s, m^{(1)}, m^{(2)}, m^{(3)}, m^{(4)}) \ A_{(3)} (m^{(3)}) \ A_{(4)} (m^{(4)}),
\\
& & {} ii) \ \sigma^{IEHTL} (s, m^{(3)}, m^{(4)}) = \int d m^{(1)} \ d m^{(2)} \ \sigma (s, m^{(1)}, m^{(2)}, m^{(3)}, m^{(4)}) \ A_{(1)} (m^{(1)}) \ A_{(2)} (m^{(2)}),
\nonumber\\
& & {} iii) \ \sigma^{IEHTL} (s) = \int d m^{(1)} \ d m^{(2)} \ d m^{(3)} \ d m^{(4)} \ \sigma (s, m^{(1)}, m^{(2)}, m^{(3)}, m^{(4)}) A_{(1)} (m^{(1)}) \ A_{(2)} (m^{(2)}) \ A_{(3)} (m^{(3)}) \ A_{(4)} (m^{(4)}),
\nonumber
%\nonumber\\
%& & {}
\label{equ:Sec2.11}
\end{eqnarray}}
\end{widetext}
 where in the case $i)$ the initial particles are considered as on-shell particles while the final particles are off-shell particles. In the case $ii)$ we consider the inverse situation of $i)$ i.e., $\sigma^{IEHTL} (s, m^{(1)}, m^{(2)}) \equiv \sigma^{IEHTL} (s, m^{(3)}, m^{(4)})$. The third case $iii)$ considers all interacting particles (initial and final) as DQPM (off-shell) particles. We then convolute the elementary subprocess cross sections (describing heavy-quark--light-quark and gluon interactions) with the spectral functions that characterize the properties of the plasma, the particle virtualities, and their evolution in the finite temperature medium.

 For this purpose, we derive the off-shell cross sections for $qQ$, $gQ$ scattering in the sQGP using the DQPM parametrizations for the quark (gluon) self-energies, spectral functions, and interaction strength. We note that approaches similar in spirit have not yet been performed in the past for the study of heavy-quark scattering in the partonic medium.

 In the context of the hot QGP, the perturbative diagrams for the $qQ$ (or $gQ$) scattering at order $O(\alpha_s)$ are illustrated in Fig.  \ref{fig:qQFeynmanDiagram} (and \ref{fig:gQFeynmanDiagram}). Let us briefly summarize the differences of our approach from the standard pQCD as follows.

\begin{itemize}
\item The full off-shell kinematics---which is different from the on-shell one---is taken into account in particular virtualities (masses and widths) of the partons.
\item Internal and external lines of light and heavy quarks as well as gluons are dressed with non perturbative spectral functions: The cross sections derived for arbitrary masses of all external parton lines are integrated over these virtualities weighted with spectral functions (in line with Refs. \cite{Linnyk:2004mt,Linnyk:2006mv,0954-3899-38-2-025105} for the study of off-shell dilepton production).
\item $qgq$ and $QgQ$ vertices are modified compared to pQCD vertices by replacing the perturbative coupling (either taken as constant or running with respect to the momentum transfer) with the full running coupling that depends on the medium temperature $g^2(T/T_c)$ according to the DQPM parametrization of lattice data (see discussion above).
Concerning the three gluon vertex $ggg$, the reader is referred to Sec. \ref{gQProcess} for more details.
\end{itemize}
  %---------------------------------------------------------------------------------------------------------------------------------------------
\section{$q Q \rightarrow q Q$ \ scattering}
\label{qQProcess}
%---------------------------------------------------------------------------------------------------------------------------------------------
\vskip -2mm

  The processes $q Q \rightarrow q Q$, $g Q \rightarrow g Q$, i.e., the scattering of a heavy quark ($Q$) on light quarks ($q$) and gluons ($g$), of the medium are essential for the study of heavy-quark propagation and thermalization in the QGP. The processes $q Q \rightarrow q Q$, $g Q \rightarrow g Q$ describe as well the eventual knockout of a heavy sea quark of one of the incoming hadrons by a hard collision with a quark ($q$) or a gluon ($g$) from the other.  In this case, these processes (called flavor-excitation
processes; see Combridge \cite{Combridge1979429}) contribute to the production of heavy flavor states (in addition to the usual heavy flavour creation processes $q\bar{q} \rightarrow Q \bar{Q}$ and $g g \rightarrow Q \bar{Q}$). These processes might not be negligible compared to the flavour-creation processes by the fusion
of quarks or gluons ($q \bar{q} \rightarrow Q \bar{Q}$, $g g \rightarrow Q \bar{Q}$) \cite{Combridge1979429}. In this study, however, we do not consider production channels and focus on the heavy-quark scattering.

  The matrix elements for the $q Q \rightarrow q Q$, $g Q \rightarrow g Q$ channels have been calculated for the case of massless partons in Refs. \cite{PhysRevD.17.196,Combridge1979429}. These pQCD cross sections have to be supplemented by two parameters to allow for a quantitative evaluation: the value of the coupling constant $\alpha=\alpha_s$ and the infrared (IR) regulator which renders the cross section infrared finite. In our study we present  values for these two parameters which are based on theoretical considerations. Moreover, we take into account the quasiparticle  nature of the incoming and outgoing particles by incorporating spectral functions. In this way we are able to test to what extent the quasiparticle nature of quarks and gluons will influence the heavy-quark scattering, either attributable to large phase-space corrections, attributable to the width of the spectral function or attributable to a different dependence of the strong coupling on the temperature of the medium.

\begin{figure}[h!]
   \begin{center}
     \begin{minipage}[h!]{0.6\linewidth}
      \includegraphics[height=4.3cm, width=6cm]{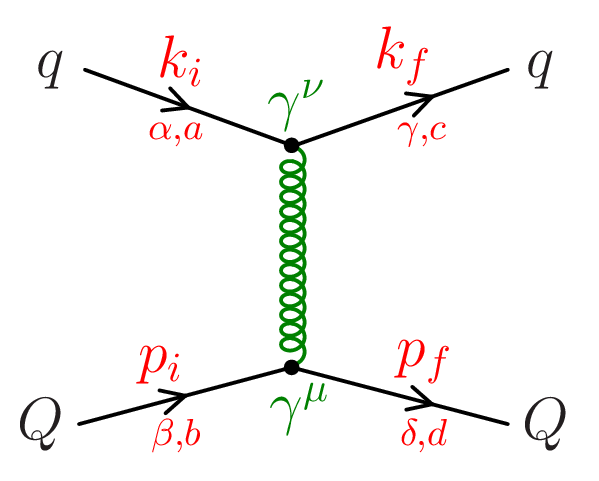}
     \end{minipage}\hfill
     \begin{minipage}[h!]{0.4\linewidth}
       \caption{\emph{(Color online) Feynman diagram for the elastic $q Q \rightarrow q Q$ process.
       Latin (Greek) subscripts denote color (spin) indices. $k_i$ and  $p_i$ ($k_f$ and $p_f$)
       denote the initial (final) 4-momentum of the light quark and heavy quark, respectively.} \label{fig:qQFeynmanDiagram}}
     \end{minipage}
   \end{center}
\end{figure}

  The elastic scattering of a heavy quark with a light quark $q Q \rightarrow q Q$ is described by the $t$-channel Feynman diagram given in Fig. \ref{fig:qQFeynmanDiagram}.
  The process $q Q \rightarrow q Q$ is calculated here to lowest order in the perturbation expansion using the  Feynman rules for massless quarks in Politzer's review \cite{DavidPolitzer1974129}. The color sums are evaluated using the techniques discussed in Ref. \cite{DavidPolitzer1974129};  the spin sums are discussed below. The
\emph{``Feynman gauge''} is used throughout this section for the case of massless gluons; the case of massive gluons is addressed later. The $t$-channel invariant amplitude $\mathcal{M}_t$ is given by
{\setlength\arraycolsep{0pt}
\begin{eqnarray}
\mathcal{M}_t & & \ \equiv \mathcal{M}_t (q_{\alpha}^i Q_{\beta}^k  \rightarrow q_{\alpha}^j Q_{\beta}^l) = \frac{g^2}{t} (T_{i j}^a T_{k l}^a) \bar{u}_{\alpha}^j (k_f) \gamma^{\nu} g_{\mu \nu} u_{\alpha}^i (k_i) \ \bar{u}_{\beta}^l (p_f) \gamma^{\mu} u_{\beta}^k (p_i),
%\nonumber
\label{equ:Sec3.1}
\end{eqnarray}}
  where the color matrices $T_{i j}^a \equiv \frac{1}{2} \lambda_{i j}^a$ are given in  Ref. \cite{DavidPolitzer1974129} while the Latin (Greek) subscripts denote color (spin) indices assigned as in Fig. \ref{fig:qQFeynmanDiagram}. The $\lambda$ matrices are the Gell-Mann
$SU(3)$ matrices.

  We work in the center-of-mass (c.m.) of the $q$-$Q$ system and choose the $(Oz)$ axis to be along the $Q$ axis. The Mandelstam variables are defined as:
{\setlength\arraycolsep{-4pt}
\begin{eqnarray}
& & s = (p_i + k_i)^2 = (p_f + k_f)^2; \hspace{0.3cm} u = (p_i - k_f)^2 = (p_f - k_i)^2; \hspace{0.3cm} t = (k_i - k_f)^2 = (p_i - p_f)^2.
\label{equ:Sec3.2}
\end{eqnarray}}
 We start with the formula for the unpolarized cross section,
{\setlength\arraycolsep{0pt}
\begin{eqnarray}
& & \hspace{-1.5cm} d \sigma \! = \!\! \frac{2 \pi \sum |\bar{\mathcal{M}}_{i \rightarrow f}|^2}{\sqrt{(p_i k_i)^2 - \!(m_q^i M_Q^i)^2}}
\! \frac{d^3 p_f}{(2 \pi)^3} \! \frac{d^3 k_f}{(2 \pi)^3} \delta^4 \!(p_i + \! k_i - \! p_f - \! k_f \!)\!,
%\nonumber\\
%& & {}
\label{equ:Sec3.3}
\end{eqnarray}}
  where $p_i$ and $k_i$ are the incoming light- and heavy-quark 4-momenta with masses $m_q^i$ and $M_Q^i$, respectively; $p_f$, $k_f$ are the 4-momenta of the outgoing particles. The differential cross sections $d \sigma/d t$ and $d \sigma/d \Omega$ then read:
{\setlength\arraycolsep{0pt}
\begin{eqnarray}
& & \frac{d \sigma}{d t} = \frac{1}{64 \pi s} \frac{1}{|\bs{p}_{i, cm}|^2} \sum |\mathcal{M}|^2
%\nonumber\\
%& & {}
\label{equ:Sec3.4}
\end{eqnarray}}
and
{\setlength\arraycolsep{0pt}
\begin{eqnarray}
& & \frac{d \sigma}{d \Omega} = \frac{d \sigma}{d t} \times \frac{d t}{d \Omega} = \frac{1}{64 \pi^2 s} \sum |\mathcal{M}|^2,
%\nonumber\\
%& & {}
\label{equ:Sec3.5}
\end{eqnarray}}
  where $p_{i, cm}$ is the initial momentum of the quarks in the c.m. reference frame. The  total cross section---obtained by integrating Eq. (\ref{equ:Sec3.4}) over $t$ or by integrating Eq. (\ref{equ:Sec3.5}) over $\Omega$---is given by
{\setlength\arraycolsep{0pt}
\begin{eqnarray}
& & \sigma^{qQ} = \int_{t_{min}}^{t_{max}} \frac{1}{64 \pi s} \frac{1}{|\bs{p}_{i, cm}|^2} \sum |\mathcal{M}|^2 \ d t, \hspace{0.5cm} \sigma^{qQ} = \int \frac{1}{64 \pi^2 s} \sum |\mathcal{M}|^2 \ d \Omega.
\label{equ:Sec3.6}
\end{eqnarray}}
  Before computing our new results for off-shell massive degrees of freedom we briefly recall the known cases for comparison (and completeness).

%-------------------------------------------------------------------------------------------------------------------------------------------------------------------------------
\subsection[Massless light quark and massive heavy quark of zero widths]{Massless light quark and massive heavy quark of zero widths}
\label{qQProcess.B}
%-------------------------------------------------------------------------------------------------------------------------------------------------------------------------------

  In this section, we recall the cross section and kinematics of the $qQ$ mechanism in the standard pQCD. For massless light quarks and a massive heavy quark the square of the matrix element (\ref{equ:Sec3.1}) gives---after summing (averaging) over final (initial) colors and spins---the well-known expression \cite{Combridge1979429}:
{\setlength\arraycolsep{0pt}
\begin{eqnarray}
\hspace{-0.1cm} <\!\!|\mathcal{M} (q_{\alpha}^i Q_{\beta}^k  \! \rightarrow \! q_{\alpha}^j Q_{\beta}^l) |^2 \!\!> & & = \frac{1}{9} \sum_{color} \frac{1}{4} \sum_{spin} |\mathcal{M}_t|^2 \! = \! \frac{4 g^4}{9} \! \frac{(M_Q^2 - u)^2 + (s - M_Q^2)^2 + 2 M_Q^2 t}{t^2}\!.
%\nonumber
\label{equ:Sec3.9}
\end{eqnarray}}
  The Mandelstam variables in this case ($m_q = 0$, $M_Q \neq 0$) are given in the c.m. reference frame by:
{\setlength\arraycolsep{0pt}
\begin{eqnarray}
\label{equ:Sec3.10}
& & \hspace{-0.8cm} s = M_Q^2 + 2 \left(p \sqrt{(M_Q^2 + p^2)} + p^2 \right); \ t = - 2 p^2 \left(1 - \cos \theta \right); \hspace{0.1cm} u = M_Q^2 - 2 \! \left(\!p \sqrt{M_Q^2 + p^2} + p^2 \cos \theta \!\right), \ s + t + u = 2 M_Q^2.
\end{eqnarray}}
  For this case, one has $|\bs{p}_{i, cm}|^2 = (s - M_Q^2)^2/4 s$, $t_{min} = - (s - M_Q^2)^2/s $ and $t_{max} = 0$. Again, the total cross section $\sigma^{qQ}$---obtained by integration of $d \sigma/d t$ (\ref{equ:Sec3.4}) over $t$---diverges because of the pole at $t_{max} = 0$ in the expression (\ref{equ:Sec3.6}), which corresponds to ultra soft $q Q$ interactions in the graphs of Fig. \ref{fig:qQFeynmanDiagram}. To render the cross section finite we introduce an infrared regulator $t_{max} = - Q_0^2$ in the upper limit of integration in Eq. (\ref{equ:Sec3.6}) following Refs. \cite{Combridge1979429,Combridge1977234}. For the case of a heavy quark and especially for $q (g) Q \rightarrow q (g) Q$ processes, we assumed $Q_0^2$ to be the value of $Q^2$ for which the heavy-quark sea starts to be generated from zero, i.e., to be proportional to $M_Q^2$. To reproduce the Combridge results \cite{Combridge1979429} for the case ($m_q = 0$, $M_Q \neq 0$), we adopted $Q_0^2 = 1.8 (\frac{M_Q}{1.87})^2 \simeq 0.51 M_Q^2$ \cite{Combridge1979429}. We note in passing that only in this limiting case do we have to introduce the regulator $Q_0$, whereas within the other limits addressed in this section there is no longer a need for such an explicit regulator.

  The kinematical limits for the process $q Q \rightarrow q Q$ are $s = Q^2$. Note that in pQCD, both collinear and intrinsic $k_T$ approaches, the partons are bound by the on-shell condition $p^2 = E^2 - \bs{p}^2 = 0$. In Sec. \ref{qQProcess.D}, we depart from the on-shellness and also consider light and heavy quarks and gluons as dynamical quasiparticles that can assume arbitrary virtualities as distributed according to the DQPM spectral functions.

  Performing the integration (\ref{equ:Sec3.6}) with $t_{max} = - Q_0^2$, we finally obtain
\begin{widetext}
{\setlength\arraycolsep{0pt}
\begin{eqnarray}
\sigma^{q Q} (s) = \frac{4 \pi \alpha^2}{9 (s - M^2)^2} & & \biggl[\left(1 + \frac{2 s}{Q_0^2} \right) \left( \frac{(s - M^2)^2}{s} - Q_0^2\right) - 2 s \ln \frac{(s - M^2)^2}{Q_0^2 s} \biggr].
%\nonumber\\
%& & {}
\label{equ:Sec3.11}
\end{eqnarray}}
\end{widetext}
  One should note that the cut in the integration has the effect of enhancing the threshold from $s \geq M_Q^2$ to $s \geq M_Q^2 + \frac{1}{2} Q_0^2 + \sqrt{M_Q^2 Q_0^2 + \frac{1}{4} Q_0^4}$. Again, to compare to the Combridge results \cite{Combridge1979429} for the case ($m_q = 0$, $M_Q \neq 0$) we consider the case of a fixed and of a running coupling  $\alpha_s (Q^2)$, taken as 
{\setlength\arraycolsep{0pt}
\begin{eqnarray}
\alpha_s (Q^2) = \frac{12 \pi}{27 \ln (Q^2/\Lambda^2)},
%\nonumber\\
%& & {}
\label{equ:Sec3.12}
\end{eqnarray}}
  which is the asymptotic form appropriate for three colors and three flavors. Note that in Ref.\cite{Combridge1979429} the computations have been carried out for $N_f=4$, which only modifies $\displaystyle \alpha_s(Q^2)$ marginally. The parameter $\Lambda$ is taken as $\Lambda = 300$ MeV as in Ref. \cite{Combridge1979429}. To study the influence of the infrared cut off $Q_0^2$  we present our calculations for two different choices: $Q_0^2 = M_Q^2$ or $Q_0^2 = 0.51 M_Q^2$. Because the effective mass of the heavy quark and the choice of the cut-off $Q_0^2$ are uncertain in this regularization scheme, this introduces also some uncertainty in the calculation of
$\sigma^{qQ} (s)$. We see later that it is possible to regularize $\sigma^{qQ} (s)$ also by introducing a finite mass of the exchanged gluon in the $t$-channel (cf. Fig.
\ref{fig:qQFeynmanDiagram}). In principle, such an effective gluon mass has to be considered as an alternative ``free'' parameter; however, this mass may be fixed in some dynamical approximation scheme (as, e.g., in the DQPM).

%-----------------------------------------------------------------------------------------------------------------------------------------------------------------------------
\subsection[Massive light and heavy quarks of zero widths]{Massive light and heavy quarks of zero widths} \label{qQProcess.C} %-----------------------------------------------
%-----------------------------------------------------------------------------------------------------------------------------------------------------------------------------

  Introducing also a mass for the light quark, $m_q^i$, for the initial $q$ and $m_q^f$ for the final $q$ and allowing for different masses of the heavy quark, $M_Q^i$, for the initial $Q$ and $M_Q^f$ for the final $Q$, the squared amplitude---averaged over the initial spin and color degrees of freedom and summed over the final-state spin and color---gives:
\begin{widetext}
{\setlength\arraycolsep{0pt}
\begin{eqnarray}
\label{equ:Sec3.1.14}
\sum |\mathcal{M}|^2 & & \ = \frac{2 g^4}{9 t^2} tr \biggl[\gamma^{\mu} (\slashed{p}_i + M_Q^i) \gamma^{\nu} (\slashed{p}_f + M_Q^f) \biggr] \ tr \biggl[\gamma_{\mu} (\slashed{k}_i + m_q^i) \gamma_{\nu} (\slashed{p}_f + m_q^f) \biggr]
\nonumber\\
& & {} = \frac{2 g^4}{9 t^2} \biggl[ 4 \left(p_f^{\mu} p_i^{\nu} + p_i^{\mu} p_f^{\nu} + g^{\mu \nu} \frac{t}{2} \right) \biggr] \biggl[ 4 \left(k_{f,\mu} k_{i,\nu} + k_{i,\mu} k_{f,\nu} + g_{\mu \nu} \frac{t}{2} \right) \biggr].
\end{eqnarray}}
\end{widetext}
  Because we study here the elastic pQCD $q Q \rightarrow q Q$ cross section, one has $m_q^i \equiv m_q^f = m_q$ and $M_Q^i \equiv M_Q^f = M_Q$. In this case, Eq. (\ref{equ:Sec3.1.14}) becomes:
{\setlength\arraycolsep{0pt}
\begin{eqnarray}
\label{equ:Sec3.11b}
\sum |\mathcal{M}|^2 & & \ = \frac{4 g^4}{9 t^2} \biggl[ (s-M_Q^2-m_q^2)^2 + (u - M_Q^2 - m_q^2)^2 + 2 (M_Q^2 + m_q^2)t \biggr].
\end{eqnarray}}  
   For the cases $m_q^i = m_q^f \neq 0$ and $M_Q^i = M_Q^f \neq 0$, $s$, $u$, and $t$ are defined as:
{\setlength\arraycolsep{0pt}
\begin{eqnarray}
\label{equ:Sec3.1.15}
& & s = M_Q^2 + m_q^2 + 2 \left(\sqrt{(m_q^2 + p^2)(M_Q^2 + p)^2} + p^2 \right),  \hspace{0.5cm} t = - 2 p^2 (1 - \cos \theta)
\nonumber\\
& & {} u = M_Q^2 + m_q^2 - 2 \left(\sqrt{(m_q^2 + p^2)(M_Q^2 + p)^2} + p^2 \cos \theta \right), \hspace{0.5cm} u + s + t = 2 M_Q^2 + 2 m_q^2,
\end{eqnarray}}
  with $|\bs{p}_{i, cm}|^2 = \bigl(m_q^4 - 2 M_Q^2 m_q^2 - 2 s \ m_q^2 + M_Q^4 + s^2 - 2 M_Q^2  s \bigr)/4 s $, $t_{min} = - 4 p^2 = - \bigl( m_q^4-2M_Q^2 m_q^2 - 2 s \ m_q^2 + M_Q^4 + s^2 - 2 M_Q^2 s \bigr)/s$ and $t_{max} = 0$. The divergence of $\sigma^{qQ} (s)$ is regularized again following the discussion in Subsection \ref{qQProcess.B}.

\begin{figure}[h!]
   \begin{center}
     \begin{minipage}[h!]{0.6\linewidth}
\includegraphics[width=8.5cm, height=6.2cm]{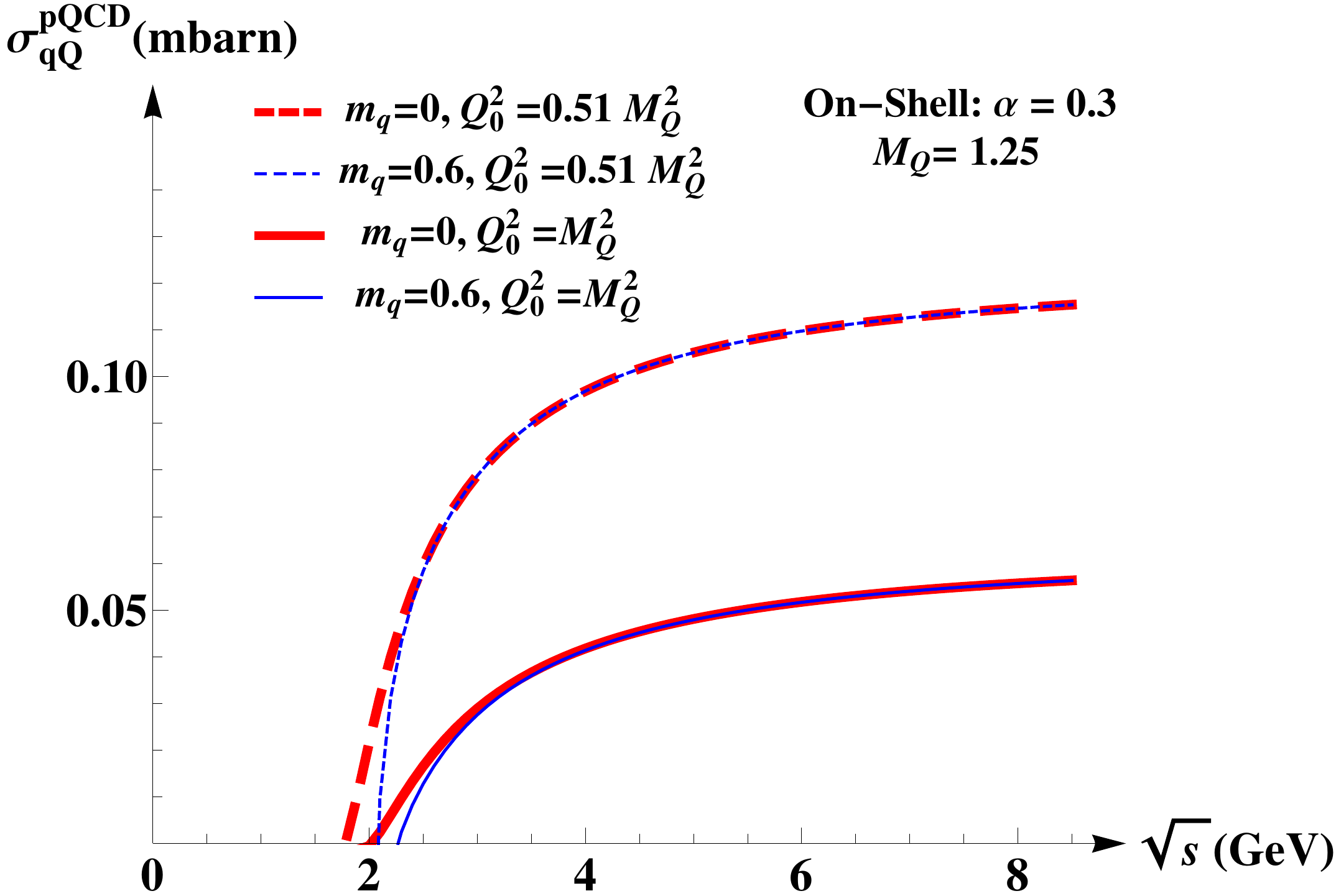}
     \end{minipage}\hfill
     \begin{minipage}[h!]{0.4\linewidth}
       \caption{\emph{(Color online) The cross section  $\sigma_{qQ}^{pQCD}$ for the process $q Q \rightarrow q Q$ with constant coupling
       ($\alpha=\alpha_s = 0.3$) for  light quark masses $m_q =0$ (red) and $m_q=$0.6 GeV (blue) and different regularization cutoffs $Q_0^2 = M_Q^2$ (solid lines) and $Q_0^2 = 0.51 M_Q^2$ (dashed lines). $M_Q = 1.25$ GeV.} \label{fig:SigmapQCD_alphaFixe}}
     \end{minipage}
   \end{center}
\end{figure}

  Figure \ref{fig:SigmapQCD_alphaFixe} displays the pQCD elastic scattering cross sections for the process $q Q \rightarrow q Q$ ($\sigma_{qQ}^{pQCD}$) for a constant coupling ($\alpha_s = 0.3$) and different light and heavy quarks masses. For the heavy quark we have adopted $M_c = 1.25 GeV$ (as in the PDG \cite{Beringer:1900zz}), whereas for the light quark we have used $m_q=0$ and $m_q=0.6$ GeV. The solid line shows the result for the infrared cutoff $Q_0^2= M_Q^2$ and the dashed line the cross section for the lower cutoff $Q_0^2= 0.51 M_Q^2$. We find that the cross sections for the different choices of the infrared regulator differ by at least a factor of two, which demonstrates the importance of
this rather unknown quantity for the perturbative cross section. We also see that at larger c.m. energies $\sqrt{s}$ the influence of the different light quark masses becomes negligible and the results merge while close to threshold the cross sections differ as discussed above.

\begin{figure}[h!]
   \begin{center}
     \begin{minipage}[h!]{0.6\linewidth}
\includegraphics[width=8.5cm, height=6.3cm]{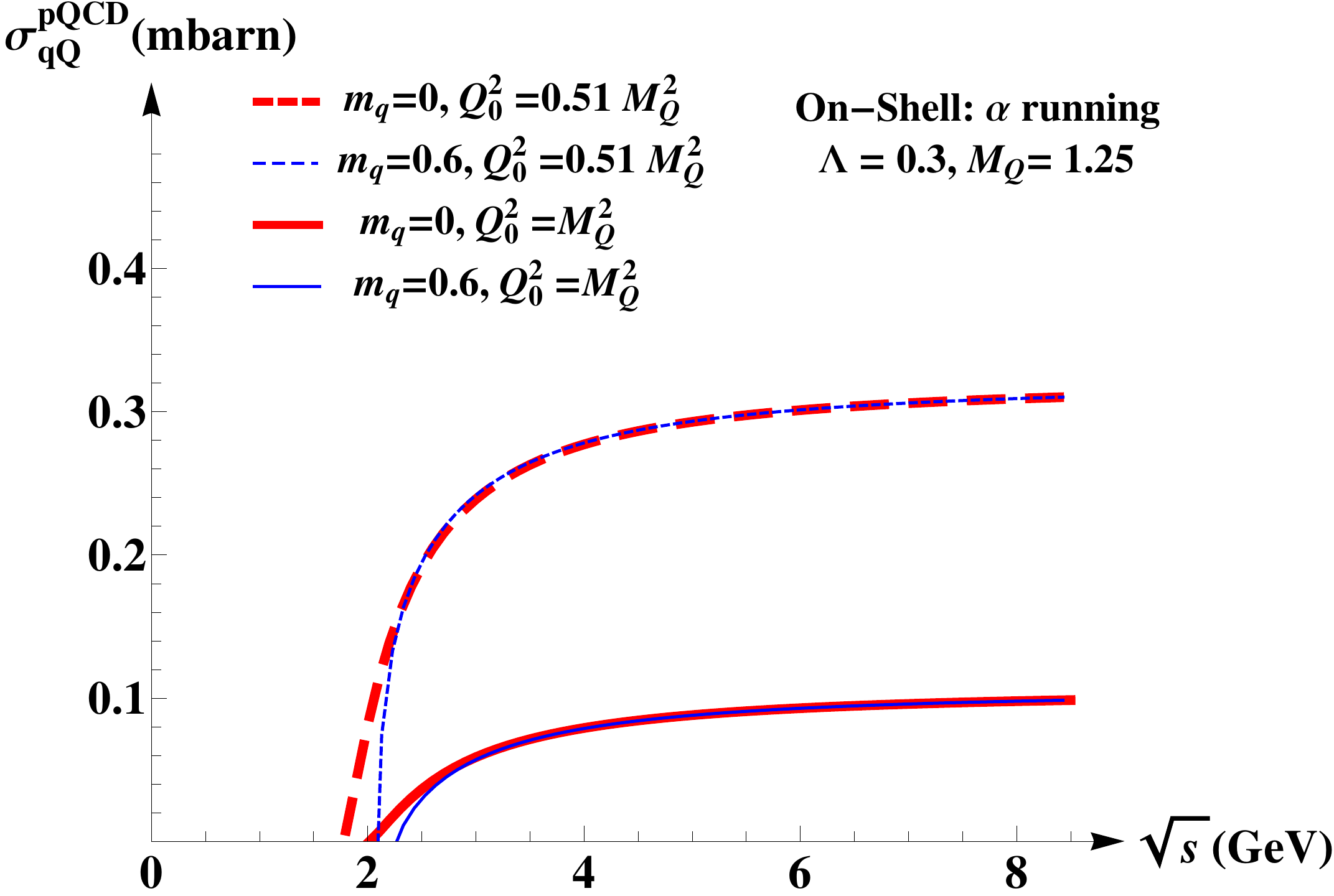}
     \end{minipage}\hfill
     \begin{minipage}[h!]{0.4\linewidth}
       \caption{\emph{(Color online) The cross section $\sigma^{pQCD}$ for the process $q Q \rightarrow q Q$ with running coupling (\ref{equ:Sec3.12}) for different light quarks masses and choices of $Q_0^2$. The color coding is the same as in Fig.\ref{fig:SigmapQCD_alphaFixe}.}
       \label{fig:SigmapQCD_alphaRunning_a_b_c}}
     \end{minipage}
   \end{center}
\end{figure}

  In Fig. \ref{fig:SigmapQCD_alphaRunning_a_b_c}, furthermore, we study the influence of the running coupling (\ref{equ:Sec3.12}) (with $Q^2 = - t$) on the cross section $\sigma_{qQ}^{pQCD}$ as a function of the energy in the c.m.s. $\sqrt{s}$. Compared to the case of a constant coupling we observe an increase of the cross section owing to the different infrared cutoff. Accordingly, the elastic cross section essentially depends on the infrared regulator and to a minor extent on the light quark mass (except for threshold energies).

  We note that the present considerations hold for $T = 0$, which is more an academic limit. In this case we have to introduce an infrared regulator (IR) $\mu^2$ (chosen as $ M_Q^2$ or $0.51 M_Q^2$) to avoid Coulomb singularities in the $t$ channel. In a thermal environment (at finite temperature $T$) the IR regulator $\mu^2$ can be taken as being proportional to the ``thermal gluon mass'' \cite{Weldon}, which implies a screening of the interaction for impact parameters larger than the Debye length $l \sim m_D^{-1}$. In fact, the divergence in the gluon propagator may be interpreted (in the vacuum) as a non zero probability for the heavy quark scattering with light quarks located at spatial infinity. The effective Debye mass has been calculated in thermal quantum field theory by Klimov \cite{VVKlimov} and Weldon \cite{Weldon} and is given by $m_D^2 = (1 + \frac{n_f}{6} g_s T)$, where $n_f$ is the number of flavors. The coefficient of proportionality between the infrared regulator and the thermal gluon mass is not well determined by first principles and one finds several proposals in the literature: In Refs. \cite{Svetitsky:1987gq,Greco:2007sz,vanHees:2004gq,vanHees:2005wb} $\mu^2$ is taken between $g_s^2 T^2$ and $g_s^2 T^2/3$, with $g_s^2 = 4 \pi \alpha_s$, because $\mu^2 = m_D^2/3 = \frac{N_c}{9} (1 + \frac{n_f}{6} g_s T) \approx \frac{(g_s T)^2}{3}$, where $n_f (N_c)$ is the number of flavors (colors).

  We continue with finite temperature regularization schemes and present two approaches in which we can relate these regulators to physical observables. The first approach is based on the work of Peigne and Peshier \cite{Peigne:2008nd} and Gossiaux and Aichelin \cite{PhysRevC.78.014904,Gossiaux2009203c,Gossiaux:2009hr,Gossiaux:2009mk,Gossiaux:2010yx,Gossiaux:2006yu} and the second approach uses the DQPM coupling constant and pole masses for the gluon and light and heavy quarks in the Feynman diagram (Fig. \ref{fig:qQFeynmanDiagram}). We refer to the first as HTL-GA approach and to the second as dressed pQCD (DpQCD). For completeness and transparency, we briefly recall the essential ingredients.

\vspace{0.2cm}
\textbf{I.} \ The Gossiaux-Aichelin approach (HTL-GA) is based on HTL-type calculations and uses the following.

\begin{itemize}
  \itemsep0em
  \item[(i)] A running nonperturbative effective coupling  $\alpha_{eff} (Q^2)$  which  remains finite if the Mandelstam variable $t$  approaches 0 for the timelike sector and by truncating the 1-loop renormalized coupling in the spacelike sector to satisfy the so-called \emph{``universality constraint''} \cite{Gossiaux2009203c},
{\setlength\arraycolsep{0pt}
\begin{eqnarray}
  \alpha_s \rightarrow \alpha_{eff} (Q^2) = \frac{4 \pi}{\beta_0}
  \begin{cases}
   & \hspace*{-0.3cm} \frac{1}{2} - \pi^{-1} \arctan (L_{+}/\pi)  \hspace*{0.4cm} \text{for} \  \  Q^2 > 0  \\
   & \hspace*{-0.3cm} \alpha_{sat}                                \hspace*{1.25cm} \text{for} \  \  - |Q^2|_{sat} < Q^2 < 0 \\
   & \hspace*{-0.3cm} L_{-}^{-1}                                  \hspace*{2cm} \text{for} \  \  Q^2 < - |Q^2|_{sat},
  \end{cases}
\nonumber\\
& & {}
\label{equ:Sec3.16}
\end{eqnarray}}
  with $\beta_0 = 11 - \frac{2}{3} N_f$, $N_f = 3$, $\alpha_{sat} = 1.12$, $|Q^2|_{sat} = 0.14$ GeV$^2$  and $L_{\pm} = \ln (\pm Q^2/\Lambda^2)$. In the timelike sector, the explicit form of the effective running coupling (\ref{equ:Sec3.16}) is determined from electron-positron annihilation and the hadronic decay of $\tau$ leptons \cite{Brodsky:2002nb} and satisfies the so-called ``universality constraint'' \cite{Dokshitzer:2003qj}. It remains finite at $Q^2=0$. In the spacelike sector, it is defined by truncating the 1-loop renormalized coupling at small $Q^2$ to reproduce the ``freeze'' observed, e.g., by Ref.\cite{Deur:2008rf}. It contains implicitly an all-order resummation of perturbation theory.

% The effective running coupling  $\alpha_{eff}(Q^2)$ (\ref{equ:Sec3.16}) is determined from electron-positron annihilation and the non-leptonic decay of $\tau$ leptons and, accordingly, contains implicitly an all-order resummation of perturbation theory.

\item[(ii)] At low momentum the free gluon propagator has to be replaced by a HTL propagator, whereas at high momenta a free gluon propagator is appropriate. As Braaten and Thoma
\cite{PhysRevD.44.R2625,PhysRevD.44.1298} (BT) have shown in QED (as well as in weak-coupling QCD) the total cross section is independent of  the scale  $t^{\star}$ of the transition between the two regimes. This is, however, not the case outside of the weak coupling regime \cite{Romatschke:2003vc}. Gossiaux and Aichelin have extended these calculations to ``strongly coupled'' QCD by introducing a semihard ``global effective propagator'' $(Q^2 - \lambda m_{\textrm{D,eff}}^2)^{-1}$, with $m_{\textrm{D,eff}}^2 (T, Q^2) = \left(1 + \frac{N_f}{6} \right) 4 \pi \alpha_{eff} (Q^2) T^2 \approx (g_s T)^2$, with $g_s^2 = 4 \pi \alpha_s$ in the $|Q^2| > |t^{\star}|$ sector to guarantee a maximal independence of the quark energy loss $d E/dx$ with respect to the unphysical $t^{\star}$ scale.

\item[(iii)] In practice, however, one uses an effective gluon propagator with a 'global' $\mu^2 = \kappa \tilde{m}_D^2$, where $\tilde{m}_D^2 (T) = \frac{N_c}{3} \left(1 +    \frac{N_f}{6} \right) 4 \pi$ $\times \alpha_s  \ (- \tilde{m}_D (T^2)) \ T^2$ for all momentum transfer in $q  Q \rightarrow q Q$ and $g  Q \rightarrow g Q$ scattering. The parameter $\kappa$ is determined by requiring that the energy loss $d E/d x$ obtained with this global effective propagator reproduces the numerical value found with the extended HTL calculation described in (ii). The details of this procedure are found in the Appendixes of  Refs. \cite{PhysRevC.78.014904,Gossiaux:2009hr,Gossiaux:2009mk} and the details of the OGE models labeled \emph{``model E''} in the text of the same reference. We remind the reader that model E was obtained by fixing (for simplicity) $\lambda = 0.11$ for all values of heavy-quark momenta $p_Q$ and deducing an effective gluon squared mass of $\kappa \tilde{m}_D^2$ with $\kappa \approx 0.2$. The Debye mass $\tilde{m}_D$ is determined self-consistently according to
{\setlength\arraycolsep{0pt}
\begin{eqnarray}
\hspace*{-0.5cm} \tilde{m}_D^2 (T) = \frac{N_c}{3} \left(1 + \frac{N_f}{6} \right) 4 \pi \alpha_s (- \tilde{m}_D (T^2))\  T^2.
%\nonumber\\
%& & {}
\label{equ:Sec3.17}
\end{eqnarray}}
  The choices for $\alpha_{eff}$ and $\mu$, motivated by HTL-type calculations, are compatible with  lattice data by Kaczmarek and Zantow \cite{Kaczmarek:2005zp}, where they have studied the potential energy and coupling constant on the lattice in two-flavor QCD by investigating the free energy between two heavy quarks.

  We recall that using these ingredients the authors of Refs. \cite{PhysRevC.78.014904,Gossiaux:2009hr,Gossiaux:2009mk} were able to reproduce the centrality dependence of the experimental ratio $R_{AA}$ in Au+Au collisions at RHIC of heavy mesons up to a factor of 2--3, as well as the experimental value of the elliptic flow $v_2$ of heavy mesons with collisional energy loss only. The remaining factor of two might be attributed to the radiative energy loss of heavy quarks in the QGP medium.

  In the following we use the HTL-GA approach with  $\mu^2=\kappa \tilde{m}_D^2$ and $\kappa \approx 0.2$ . For comparison we also show the cross sections using a fixed coupling constant ($\alpha_s = 0.3$) and the Debye mass ($m_D \approx \zeta g_s T$) with $\zeta = 1$ or $\zeta = 0.5$ as infrared regulators. To emphasize the influence of the running coupling we present also some results for \emph{``model C''} of Refs. \cite{PhysRevC.78.014904,Gossiaux:2009hr,Gossiaux:2009mk} in which the coupling is taken as $\alpha_s(2 \pi T)$ and $\mu^2 = 0.15 m_D^2$.

\end{itemize}

\textbf{II.} \ For the DpQCD approach the $q Q \rightarrow q Q$ cross section is determined  by

\begin{itemize}
  \itemsep0em
  \item[(i)]   the running coupling constant $\alpha_s (T)$ (\ref{equ:Sec2.6}),
  \item[(ii)]  the DQPM pole masses for the incoming and outgoing quarks and gluons. The DQPM pole mass serves also as an infrared regulator in the gluon propagator.
\end{itemize}

\begin{figure}[h!]
       \begin{center}
       \includegraphics[width=8.5cm,height=6.6cm]{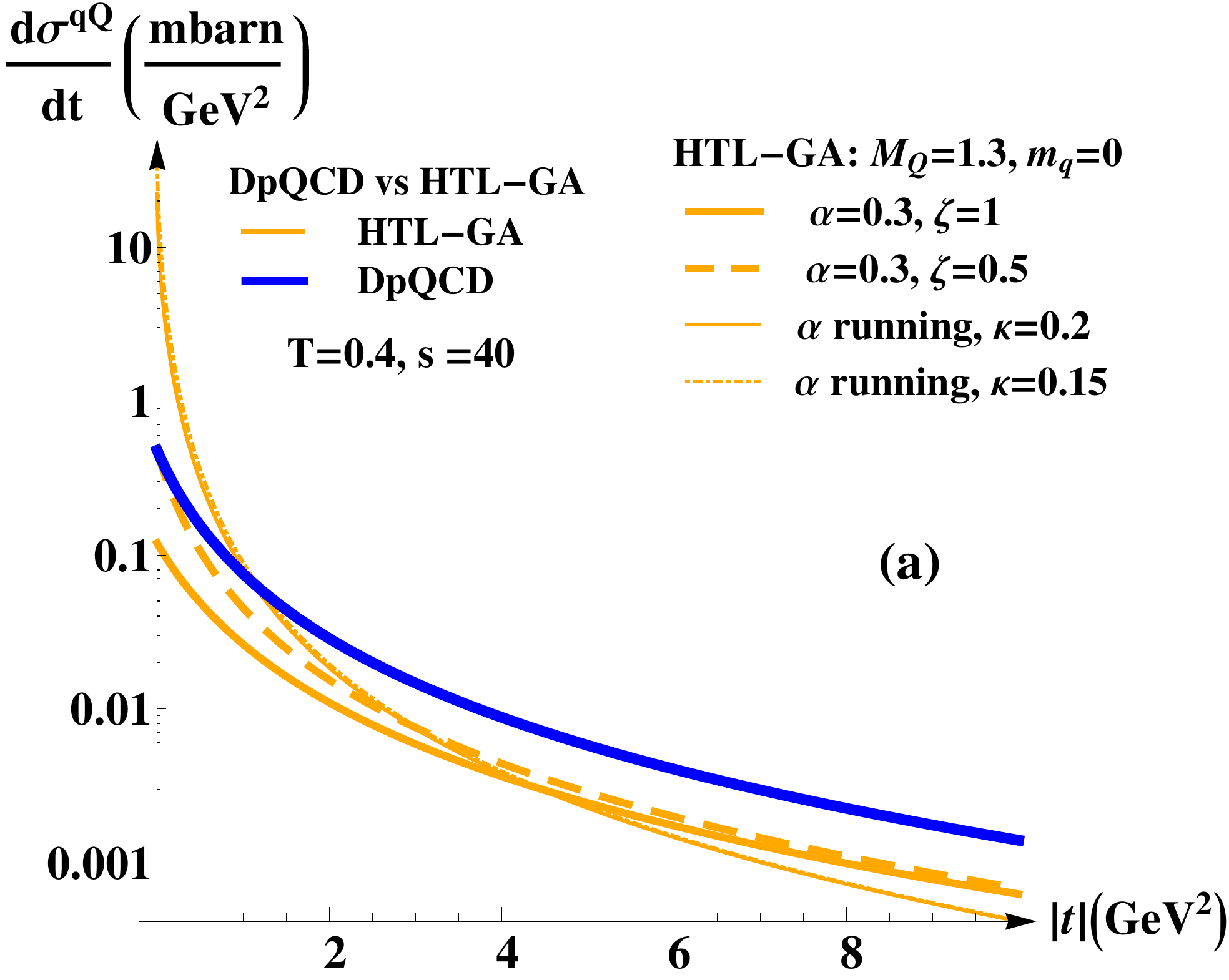}\hfill
%       \vspace*{0.4cm}\\
       \includegraphics[width=8.5cm,height=6.8cm]{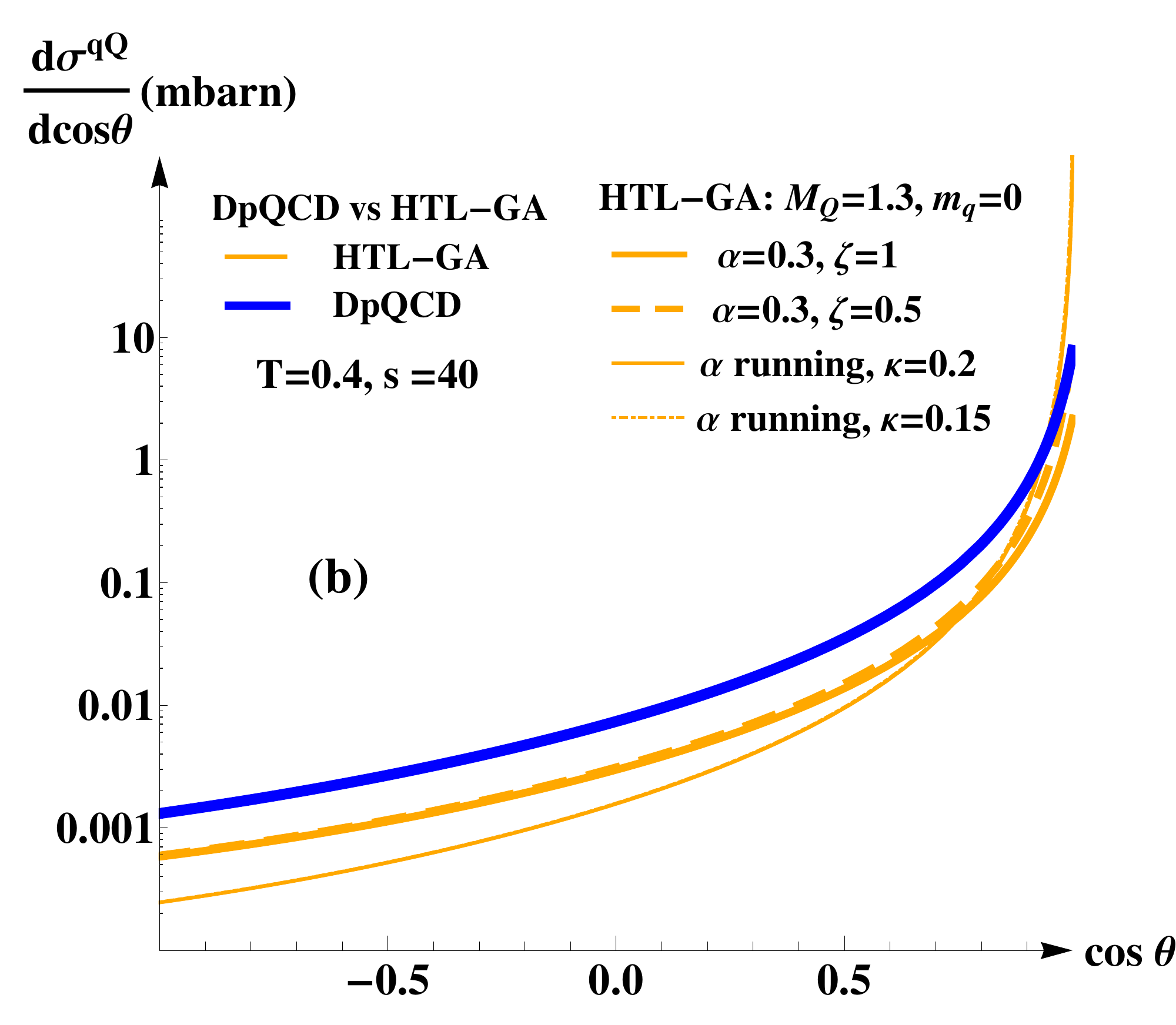}
       \end{center}
       \vspace{-0.4cm}
\caption{\emph{(Color online) Differential elastic cross sections $d\sigma\!/\!dt$ (a) and $d \sigma\!/\!d cos \theta$ (b) for $c q \rightarrow c q$ scattering employing different choices for the strong coupling and the infrared regulator for a QGP of $T = 0.4$ GeV and invariant energy squared $s=$ 40 GeV$^2$ (see legend). $m_q$ and $M_Q$ are given in GeV.} \label{fig:dSigmadtqQDpQCD-HTLapproach}}
\end{figure}

  For our numerical calculations we consider the scattering of a (high-momentum) heavy quark with a light quark in a QGP at temperature $T$ = 0.4  GeV with invariant energy $\sqrt{s} = \sqrt{40}$ GeV. The differential cross sections $d\sigma\!/\!dt$  and $d \sigma\!/\!d cos \theta$) for the two approaches with different parametrizations are displayed in Figs. \ref{fig:dSigmadtqQDpQCD-HTLapproach}-(a) and \ref{fig:dSigmadtqQDpQCD-HTLapproach}-(b). It is evident that both, a running coupling and a lower IR regulator in the HTL-GA, increase the cross section at small $t$ or forward angles, whereas the increase at high $t$ is rather moderate but also the differential cross section is very low. This increase, nevertheless, will be visible in the case of $g Q$ reactions owing to the $u$ channel (cf. Sec. \ref{gQProcess.C}). The DpQCD gives a flatter $t$ (a) and angular distribution (b) as the HTL-GA model for the lower regularization scale ($\zeta=0.5$). Note that for $\alpha = 0.3$ and the infrared regulator $\zeta$=1 both approaches give a similar angular distribution, but differ in magnitude by about a factor of three in this case.

\begin{figure}[h!]
\begin{center}
\includegraphics[width=8.7cm,height=6.7cm]{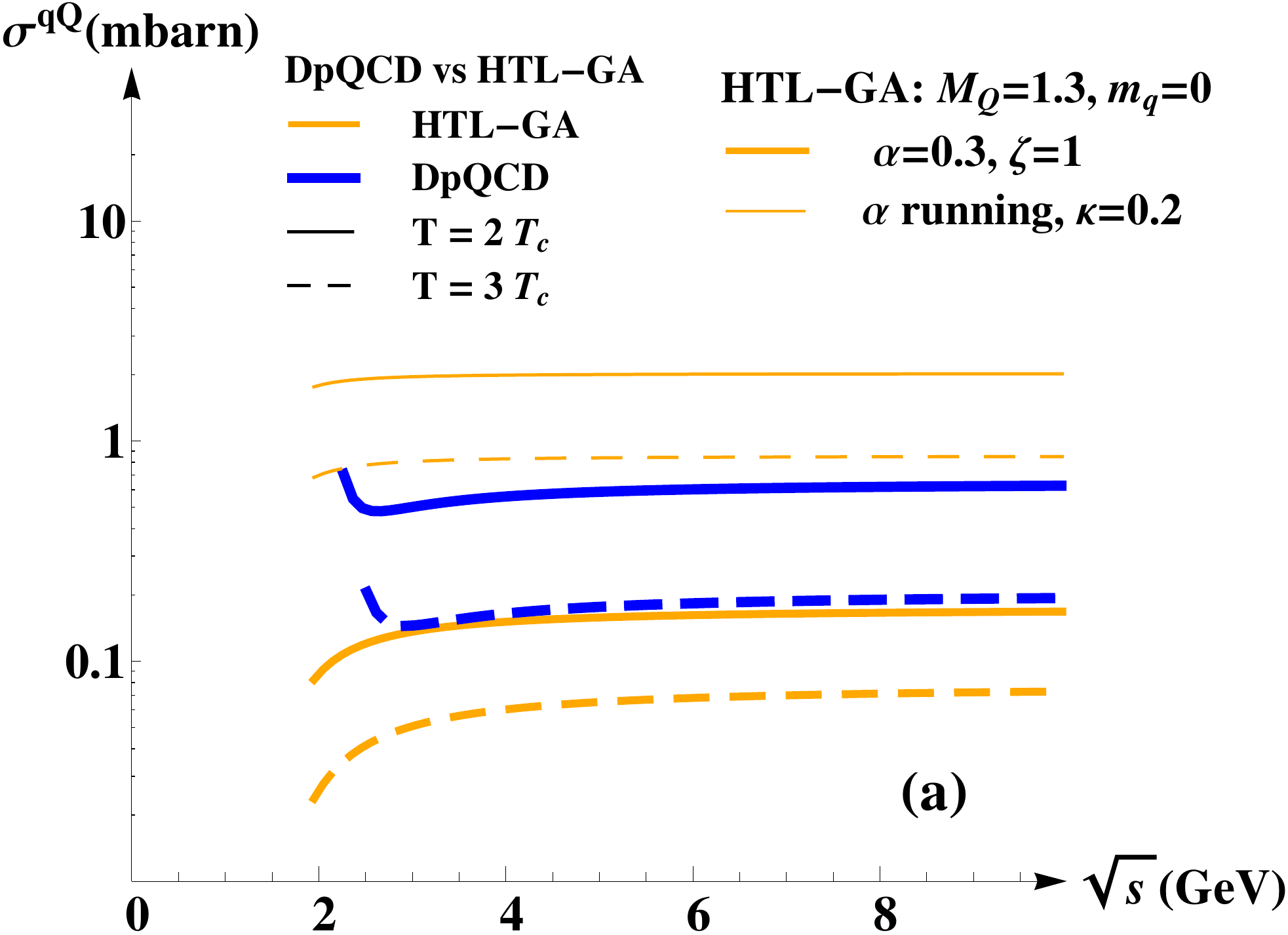} \hspace*{0.3cm}
\includegraphics[width=8.5cm,height=6.8cm]{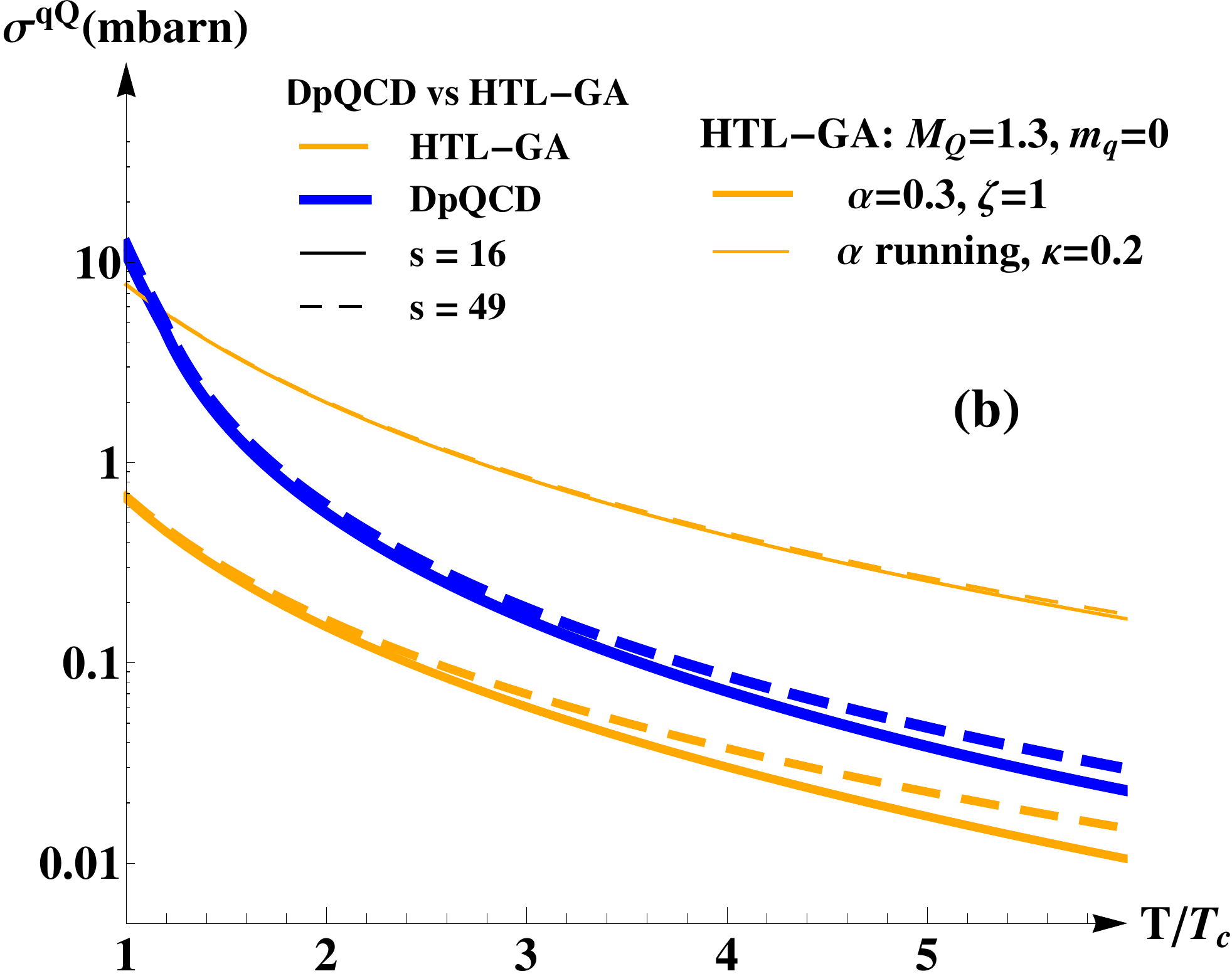}
%\end{minipage}
    \vspace{-0.1cm}
\caption{\emph{(Color online) Comparison of $\sigma^{qQ}$ calculated within the HTL-GA and DpQCD approaches for different settings (see legend). $m_q$ and $M_Q$ are given in GeV. (a) As a function of $\sqrt{s}$ for different temperatures [$T = 2 T_c$ (solid lines), $T= 3 T_c$ (dashed lines)]. (b) As a function of the scaled temperature $T/T_c$, with $T_c = 0.158$ GeV, for different energies in the c.m. [$\sqrt{s}$ = 4 GeV (solid lines) and  7 GeV (dashed lines)].}}
\label{fig:SigmaqQDpQCD-HTLapproach}
\end{center}
\end{figure}

\begin{figure}[h!]
   \begin{center}
     \begin{minipage}[h!]{0.6\linewidth}
\includegraphics[width=8.5cm, height=6.9cm]{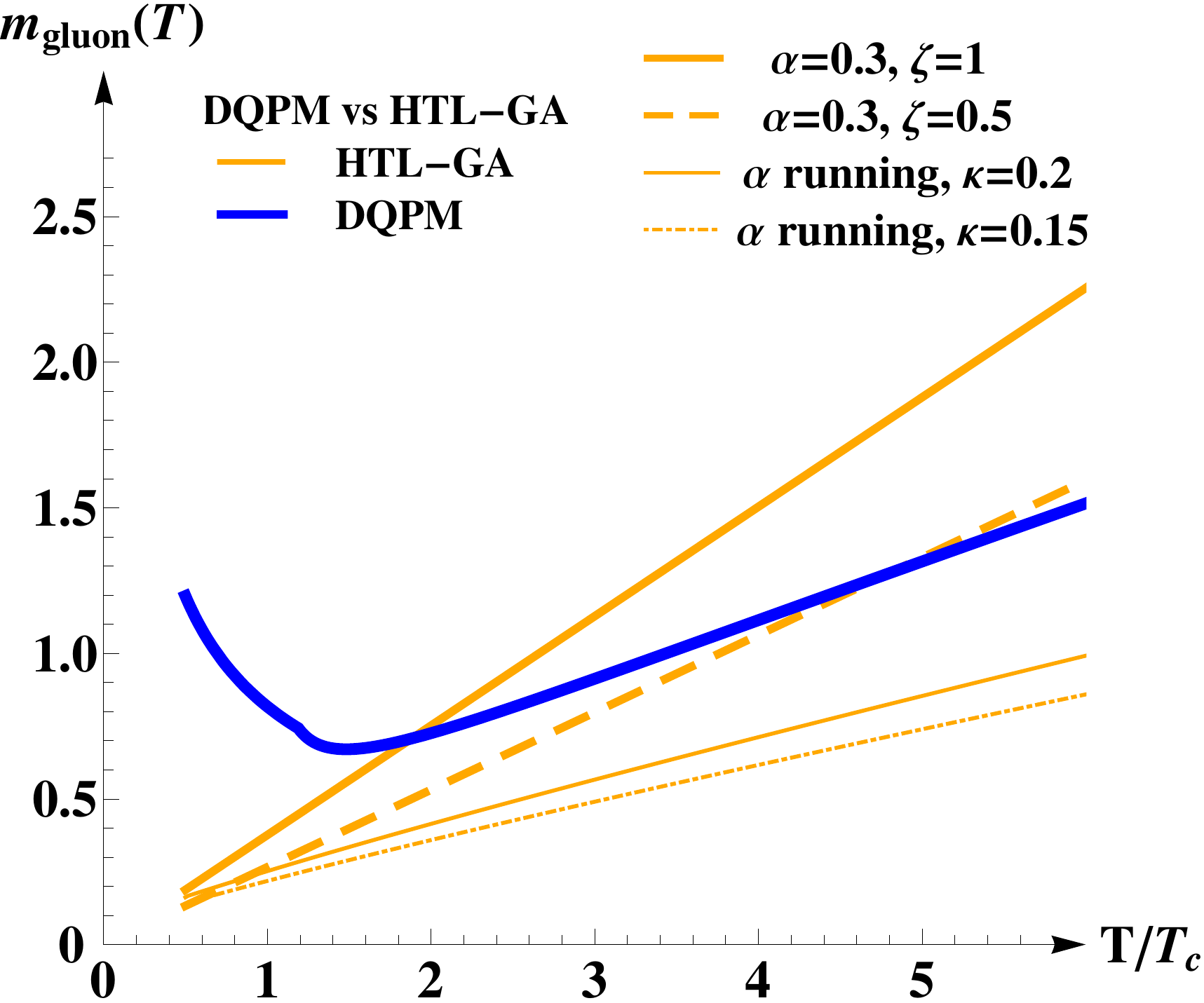}
     \end{minipage}\hfill
     \begin{minipage}[h!]{0.4\linewidth}
       \caption{\emph{(Color online) The gluon mass as a function of the medium scaled temperature $T/T_c$, with $T_c = 0.158$ GeV, according to the DQPM (\ref{equ:Sec2.5}) (blue) and the HTL-GA approaches (\ref{equ:Sec3.17}) (orange) for different settings of the coupling and infrared regulators (see legend). $m_{gluon} (T)$ is given in GeV.} \label{fig:mGluonDQPM-HTL}}
     \end{minipage}
   \end{center}
\end{figure}

  The total elastic cross section of a $c$ quark, which traverses a plasma at temperatures $T=2 T_c$ and $T=3 T_c$, calculated using the expression (\ref{equ:Sec3.6}), is shown in Fig. \ref{fig:SigmaqQDpQCD-HTLapproach} (a) as a function of $\sqrt{s}$. Apart from energies close to the threshold the cross sections show a rather smooth dependence on the invariant energy $\sqrt{s}$, however, differ substantially in magnitude. The HTL-GA with a running coupling yields a much larger cross sections than the DpQCD model, whereas for constant coupling ($\alpha_s=0.3$) lower cross sections are obtained owing to a sizeably larger infrared regulator. All cross sections decrease for the higher temperature of $3 T_c$ (dashed lines) in comparison to the $2 T_c$ case (solid lines). Because the values of the infrared regulators increase with increasing temperature, the cross sections decrease with $T$ for both the HTL-GA (in all variants) and the DpQCD approach, as shown in Fig. \ref{fig:SigmaqQDpQCD-HTLapproach} (b). Here we show explicitly the temperature dependence of the cross sections for $\sqrt{s}=4$ GeV and $\sqrt{s}= 7$ GeV. In the HTL-GA versions the running coupling increases the cross sections by a temperature independent factor with respect to the fixed $\alpha_s =0.3$. Whereas at high temperatures the results of the two models differ by roughly a constant factor, at low temperatures (close to $T_c$) the different parametrizations of the couplings yield a dramatically different $T$ dependence of the cross section which can be traced back to the infrared enhanced coupling in DpQCD. This is also reflected in the dependence of the gluon effective mass on temperature, as shown in Fig. \ref{fig:mGluonDQPM-HTL}. Here the effective gluon mass for both approaches is roughly proportional to $T$ for temperatures above $0.2$ GeV, however, with different proportionality constants. Note that the HTL-GA case with $\alpha_s=.3$ and $\zeta=1$ gives the largest masses for $T > 0.5 $ GeV and thus the largest infrared regulator, which in turn implies the lowest cross section.

%--------------------------------------------------------------------------------------------------------------------------------------------
\subsection[Massive light and heavy quarks of finite widths]{Massive light and heavy quarks of finite widths} \label{qQProcess.D} %----------
%--------------------------------------------------------------------------------------------------------------------------------------------

  So far we have worked with light and heavy quarks of fixed mass and the question arises if the spectral width of the degrees of freedom---owing to finite scattering rates---might have a sizable influence on the cross sections. In this respect we now calculate the $qQ$ elastic scattering by taking into account not only the finite masses of the partons, but also their spectral functions, i.e., their finite widths. For this purpose, we convolute the on-shell pQCD cross sections obtained before with the distribution of the quarks and gluons with different momenta and virtualities given by the DQPM spectral functions $A(m)$. Here, in principle, should appear a two-particle correlator, but because we work in a 2PI motivated scheme the partons in the sQGP can be characterized by (dressed) single-particle propagators.

  In leading order the dressed propagators and the strong coupling lead to substantial phase-space corrections. Furthermore, the relative contribution of the off-shell partons to the pQCD cross section is expected to change owing to different kinematical thresholds. Therefore, we first consider the general off-shell kinematics, when the participating quarks are massive, with the masses distributed according to the DQPM spectral functions. Denoting the masses of the incoming light and heavy quarks as $m_q^i$ and $M_Q^i$, the kinematical limits of the exchanged momentum change compared to the massless or massive on-shell case. The flux also changes to
%\begin{widetext}
{\setlength\arraycolsep{-10pt}
\begin{eqnarray}
\hspace*{-1cm} J = \frac{1}{2} \sqrt{(s - (m_q^i)^2 - (M_Q^i)^2) - 4 (m_q^i)^2 (M_Q^i)^2},
%\frac{1}{2} \sqrt{(k_i . p_i)^2 - (m_q^i)^2 (M_Q^i)^2}
%\\
%& & {} = \frac{1}{2} \sqrt{(s - (m_q^i)^2 - (M_Q^i)^2) - 4 (m_q^i)^2 (M_Q^i)^2},
%\nonumber
\label{equ:Sec3.19}
\end{eqnarray}}
%\end{widetext}
  compared to $J = s/2$ in the massless on-shell approximation. In addition to the kinematics, the matrix element---corresponding to the diagram in Fig. \ref{fig:qQFeynmanDiagram}---is modified in the general off-shell case compared to the matrix element for $qQ$ scattering of massless or massive on-shell quarks. Because the final light- and heavy-quark masses can be different from the initial ones, the $qQ$ elastic scattering process is considered as ``quasi-elastic''. The off-shell kinematical limits for the momentum transfer squared $t$ are
{\setlength\arraycolsep{0pt}
\begin{eqnarray}
\label{equ:Sec3.20}
& & t_{min}^{max} = - \frac{s}{2} (C_1 \pm C_2),
\nonumber\\
& & {} \textrm{where} \ C_1 = 1 - (\beta_1 + \beta_2 + \beta_3 + \beta_4) + (\beta_1 -  \beta_2)(\beta_3 - \beta_4), \hspace{0.5cm} C_2 = \sqrt{(1 - \beta_1 - \beta_2)^2 - 4 \beta_1 \beta_2} \sqrt{(1 - \beta_3 - \beta_4)^2 - 4 \beta_3 \beta_4},
\nonumber\\
& & {} \textrm{with} \ \beta_1 = (m_q^i)^2\!\!/\!s, \beta_2 = (M_Q^i)^2\!\!/\!s, \beta_3 = (m_q^f)^2\!\!/\!s, \beta_4 = (M_Q^f)^2\!\!/\!s.
\end{eqnarray}}
  Additionally, we note that there is a threshold in the invariant energy $\sqrt{s}$ for the $qQ$ interaction:
{\setlength\arraycolsep{0pt}
\begin{eqnarray}
& & s \geq max\{(m_q^i + M_Q^i)^2, (m_q^f + M_Q^f)\}.
%\nonumber\\
\label{equ:Sec3.21}
\end{eqnarray}}
  The off-shell cross section for the elastic $qQ$ interaction  has not been calculated before. Therefore, we provide here a short description of its evaluation. The light and heavy quarks are now described by the DQPM spectral functions with finite mass and width at fixed temperature $T$. The corresponding cross section for the process $q Q \rightarrow q Q$, $\sigma^{IEHTL}$ is deduced from $\sigma^{pQCD}$ according to the relations given in Sec. \ref{pPartons-DQP}. Considering off-shell initial and/or final particles the kinematics of the reaction $q Q \rightarrow q Q$, $\sigma^{IEHTL}$ changes compared to the on-shell case. The new expressions of the Mandelstam variables are given in Eq. (\ref{equ:SecA.1}).

  For the integration of the expression (iii) of Eq. (\ref{equ:Sec2.11}), one has to pay attention to the fact that the integration limits for the variables $m_q^i$, $m_q^f$, $M_Q^i$, and $M_Q^f$ are given according to Eq. (\ref{equ:Sec3.21}) by
{\setlength\arraycolsep{0pt}
\begin{eqnarray}
& & m_q^i \in [0, \sqrt{s}]; \hspace{0.5cm} M_Q^i \in [0, \sqrt{s}- m_q^i];  \hspace{0.5cm} m_q^f \in [0, \sqrt{s}];  \hspace{0.5cm} M_Q^f \in [0, \sqrt{s}- m_q^f].
\label{equ:Sec3.23}
\end{eqnarray}}

  Using the expression (\ref{equ:Sec3.4}) for $d \sigma/d t$ and integrating over $t$ we can determine the corresponding total cross section. We recall again that for the pQCD cross section we have considered $M_Q^i = M_Q^f$ and $m_q^i = m_q^f$. For $\sigma_{qQ}^{IEHTL}$, however, we have to consider the case where $m_q^i \neq m_q^f$ and $M_Q^i \neq M_Q^f$ for $qQ$ scattering. We use the DQPM spectral functions for the gluons and light and heavy quarks, and we consider the DQPM propagators (i.e., $t_{\pm}^{*} = t
- m_g^2 \pm 2 i \gamma_g q_0$, where $m_g$, $\gamma_g$ are, respectively, the effective gluon mass and total width at temperature $T$ and $q^0 = p_f^0 - p_i^0 = k_f^0 - k_i^0$). Thus the divergence in the gluon propagator in the $t$ channel is regularized. We note that the approximative HTL propagator \cite{PhysRevC.78.014904,PhysRevD.44.R2625,PhysRevD.44.1298} can be regained from the DQPM propagator in the case $m_q^i = m_q^f$ and $M_Q^i = M_Q^f$ and a zero width $\gamma_g$ of the thermal gluon.

  The energy averaged differential and total elastic cross section, which constitutes a first step towards the computation of transport coefficients, is obtained by integrating the differential (total) elastic cross sections $d\sigma/dt (T,s)$ ($\sigma(T,s)$) over $s$ taking into account all possible kinematic reactions for a given thermodynamical medium in equilibrium for fixed $T$. We give the expression for the energy-averaged total elastic cross section. The energy-averaged expression for the differential cross section is obtained straightforwardly by integration of $\sigma(T,s)$ over $s$, which is done differently for the case of on-shell and off-shell partons. For the case of on-shell partons we use \cite{Zhuang:1995uf,Sasaki201062,Marty:2013ita}:
\begin{widetext}
{\setlength\arraycolsep{0pt}
\begin{eqnarray}
& &   \bar{\sigma}_{qQ}^{\textrm{on}} (m_q^i, M_Q^i,m_q^f, M_Q^f, T) = \int \limits_{\text{Th}}^\infty d s \ \sigma_{qQ}^{\textrm{on}} (m_q^i, M_Q^i,m_q^f, M_Q^f,T,s) \ \mathcal{L}^{\textrm{on}} (m_q^i, M_Q^i,T,s),
%\nonumber\\
%& & {}
\label{equ:Sec3.24}
\end{eqnarray}}
  with the threshold $\text{Th} =$ max\{$(m_q^i + M_Q^i)^2$,$(m_q^f + M_Q^f)^2$\} [which depends on $(T)$] and $\mathcal{L}$ denoting the probability for a $q Q$ pair with the invariant energy $\sqrt{s}$ in the medium at finite temperature:
{\setlength\arraycolsep{0pt}
\begin{eqnarray}
\mathcal{L}^{\textrm{on}}(m_q^i, M_Q^i,T,s) & & \ = C'^{\textrm{on}} \int d^3p_1d^3p_2 \ f(E_1) f(E_2) \delta(\sqrt{s}-E_1^{cm}-E_2^{cm}) \delta^3(\bs{p}_1 + \bs{p}_2)
\nonumber\\
& & {}  = C^{\textrm{on}} \frac{E_1^{cm} (\sqrt{s}-E_1^{cm})}{\sqrt{s}} p^{cm} f(\sqrt{s}-E_1^{cm})f(E_1^{cm}),
\label{equ:Sec3.25}
\end{eqnarray}}
\end{widetext}
  with the c.m. momentum
{\setlength\arraycolsep{0pt}
\begin{eqnarray}
\hspace{-1.6cm} p_{\text{cm}}(m_q^i, M_Q^i, s) \! = \! \frac{\sqrt{\left(\!\! s-(m_q^i + M_Q^i)^2 \! \right) \!\! \left(\!\!s-(m_q^i - M_Q^i)^2 \!\right)}}{2 \sqrt{s}}\!,
\label{equ:Sec3.26}
\end{eqnarray}}
and
\vspace*{-0.5cm}
{\setlength\arraycolsep{0pt}
\begin{eqnarray}
   E_1^{\text{cm}}(m_q^i, M_Q^i, s) = \frac{\left(s-(m_q^i)^2 +(M_Q^i)^2 \right)}{2 \sqrt{s}},
\label{equ:Sec3.26}
\end{eqnarray}}
\vspace*{-0.1cm}
 while $C^{\textrm{on}}$ is a normalization factor fixed by
{\setlength\arraycolsep{0pt}
\begin{eqnarray}
\hspace{-1.3cm} [C^{\textrm{on}} (m_q^i, M_Q^i,T)]^{-1} = \int\limits_{\text{Th}}^\infty d s \ \mathcal{L}^{\textrm{on}} (m_q^i, M_Q^i,T,s).
\label{equ:Sec3.28}
\end{eqnarray}}
  In Eq. (\ref{equ:Sec3.25}) $f_{q,Q} (E) = 1/\exp[E/T + 1]$ is the Fermi-Dirac distribution for the light and heavy quark with the energy $E = \sqrt{p^2 + m^2}$. Similar calculations can be performed for cross sections with massive gluons by including the Bose-Einstein distribution in Eq. (\ref{equ:Sec3.25}).

\begin{figure}[h!]
       \begin{center}
       \includegraphics[width=8.5cm, height=7.85cm]{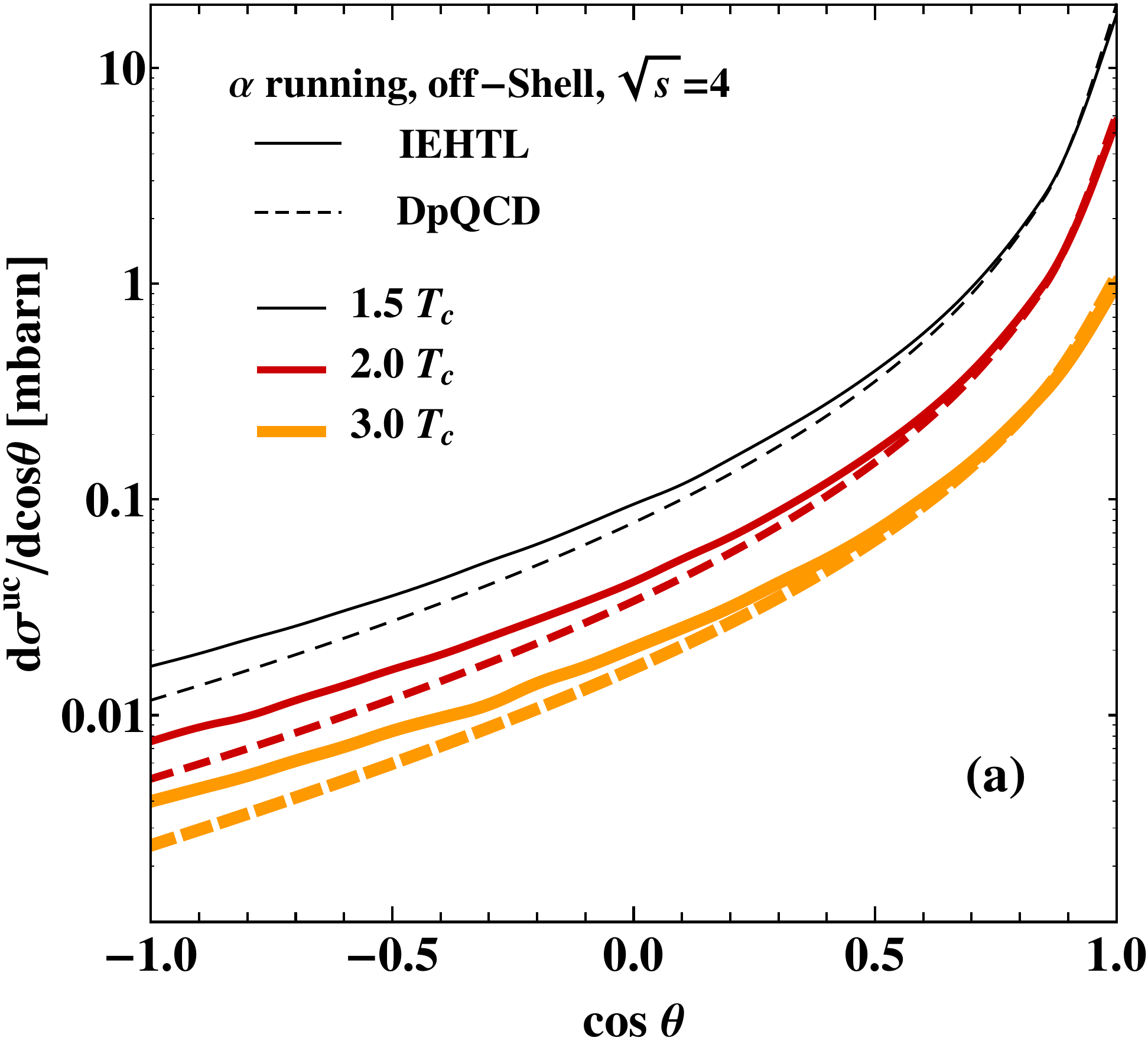}\hfill
%       \vspace*{0.3cm}
       \includegraphics[width=8.5cm, height=7.85cm]{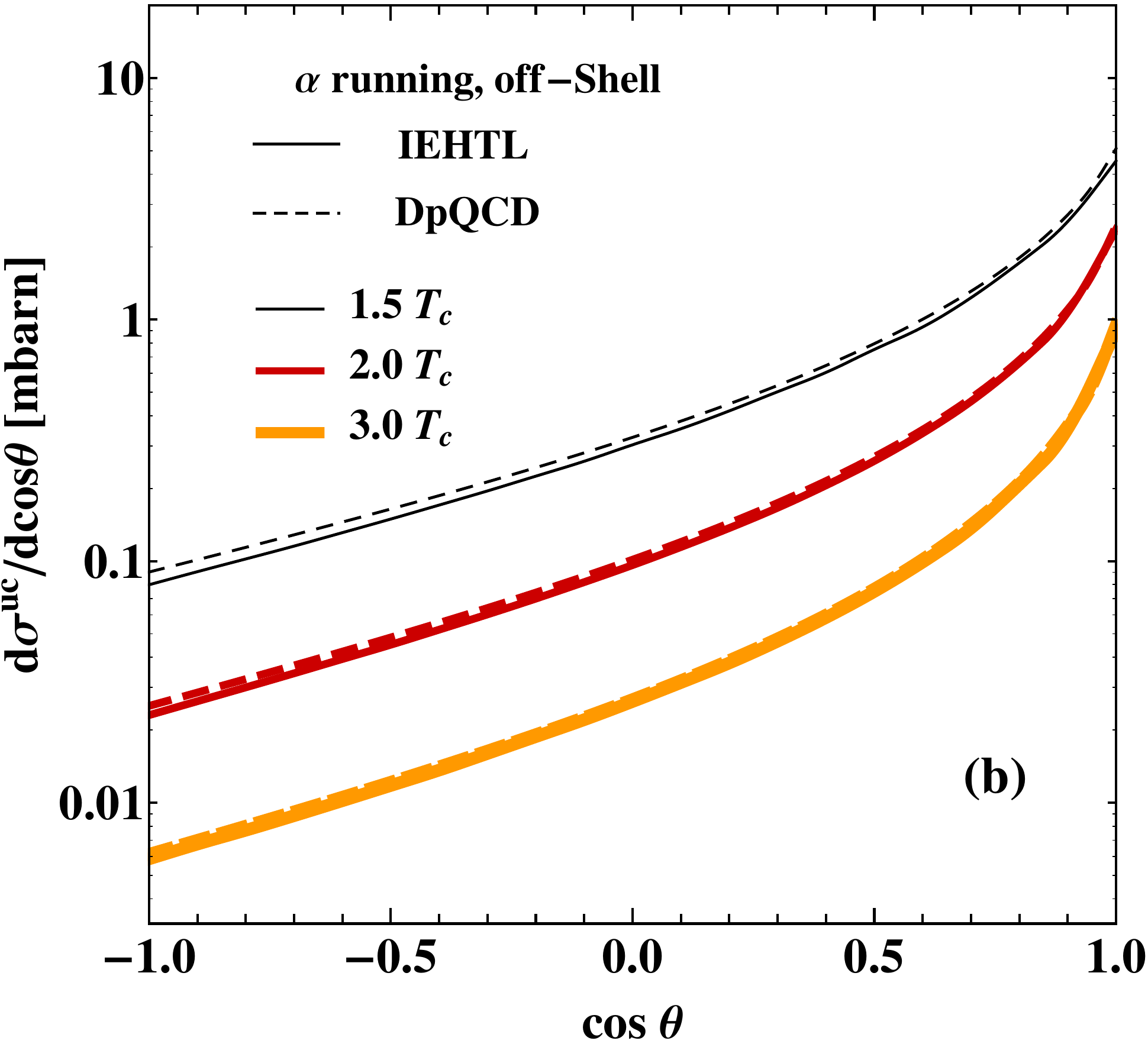}
       \end{center}
\vspace*{-0.3cm}
\caption{\emph{(color online) Differential elastic cross section of $u c \rightarrow u c$ for off-shell (solid lines) and on-shell partons (dashed lines) at three different temperatures (see legend). We consider the DQPM pole masses for the on-shell partons and the DQPM spectral functions for the off-shell degrees of freedom. \textbf{Left~ (a):} for $\sqrt{s} = 4$ GeV, \textbf{right~ (b):} averaged over energy.} \label{fig:dSigmadthetaqQDpQCD-IEHTL}}
\end{figure}

  The off-shell energy averaged differential $d\sigma_{qQ}^{\textrm{off}}/dt (T)$ and total [$\bar{\sigma}_{qQ}^{\textrm{off}}(T)$] elastic cross section are obtained in a similar way as the on-shell case, with new expressions for the probability function $\mathcal{L}$ and the normalization constant $C$. Therefore, for $\bar{\sigma}_{qQ}^{\textrm{off}}(T)$, one has:
{\setlength\arraycolsep{0pt}
\begin{eqnarray}
& &   \bar{\sigma}_{qQ}^{\textrm{off}} (T) = \int \limits_{\text{Th}}^\infty d s \ \sigma_{qQ}^{\textrm{off}} (T,s) \ \mathcal{L}^{\textrm{off}}(T,s),
%\nonumber\\
%& & {}
\label{equ:Sec3.29}
\end{eqnarray}}
 with
\begin{widetext}
\vspace*{-0.25cm}
{\setlength\arraycolsep{0pt}
\begin{eqnarray}
  \begin{aligned}
\hspace{-0.5cm} \mathcal{L}^{\textrm{off}}(T,s) = \int\limits_{0}^\infty d m_q^i \int\limits_{0}^\infty d M_Q^i \ \frac{E_1^{cm} (\sqrt{s}-E_1^{cm})}{\sqrt{s}} \ C^{\textrm{off}}(T) \ p_{\text{cm}}(m_q^i, M_Q^i, s) \ f_q\left(E_{1}^{cm} \right) f_{Q}\left(\sqrt{s} - E_1^{cm} \right) A_{q^i} (m_q^{i}) A_{Q^i} (M_Q^{i}),
  \end{aligned}
\label{equ:Sec3.30}
\end{eqnarray}}
\end{widetext}
and
\vspace*{-0.5cm}
{\setlength\arraycolsep{0pt}
\begin{eqnarray}
\hspace{-1.1cm}[C^{\textrm{off}} (T)]^{-1} \!\! =\!\!\! \int\limits_{\text{Th}}^\infty \!\!\! d s \!\!\! \int\limits_{0}^\infty \!\!\! d m_q^i \!\!\! \int\limits_{0}^\infty \!\!\! d M_Q^i  \mathcal{L}^{\textrm{off}} (T,s) A_{q^i} (m_q^{i}) A_{Q^i} (M_Q^{i}).
\label{equ:Sec3.31}
\end{eqnarray}}
  Figure \ref{fig:dSigmadthetaqQDpQCD-IEHTL} presents the off-shell differential cross section $d \sigma\!/\!d \cos \theta$ (solid lines) in comparison to the on-shell cross section (dashed lines) at $\sqrt{s}$ = 4 GeV (a) and averaged over energy (b) for temperatures of 1.5 $T_c$, 2.0 $T_c$, and 3.0 $T_c$. The importance of finite width corrections in the $qQ$ scattering process is illustrated by comparing the two differential cross sections. For the energy of $\sqrt{s} = 4$ GeV, one observes a deviation of the off-shell
results compared to on-shell ones only for large scattering angles.

\begin{figure}[h!]
       \begin{center}
       \includegraphics[width=8.6cm, height=8.3cm]{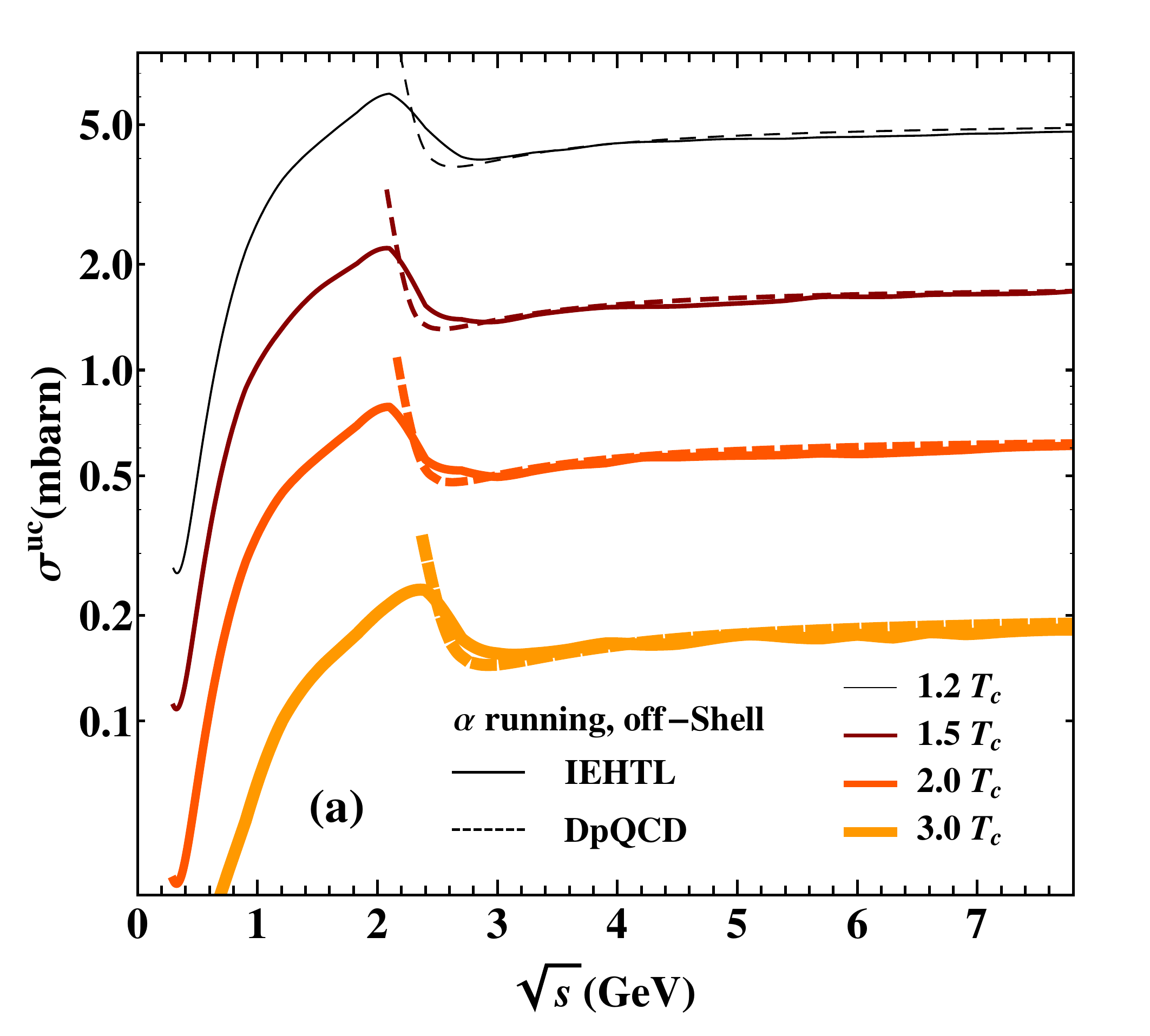}\hfill
%       \vspace*{0.3cm}
       \includegraphics[width=8.6cm, height=8.3cm]{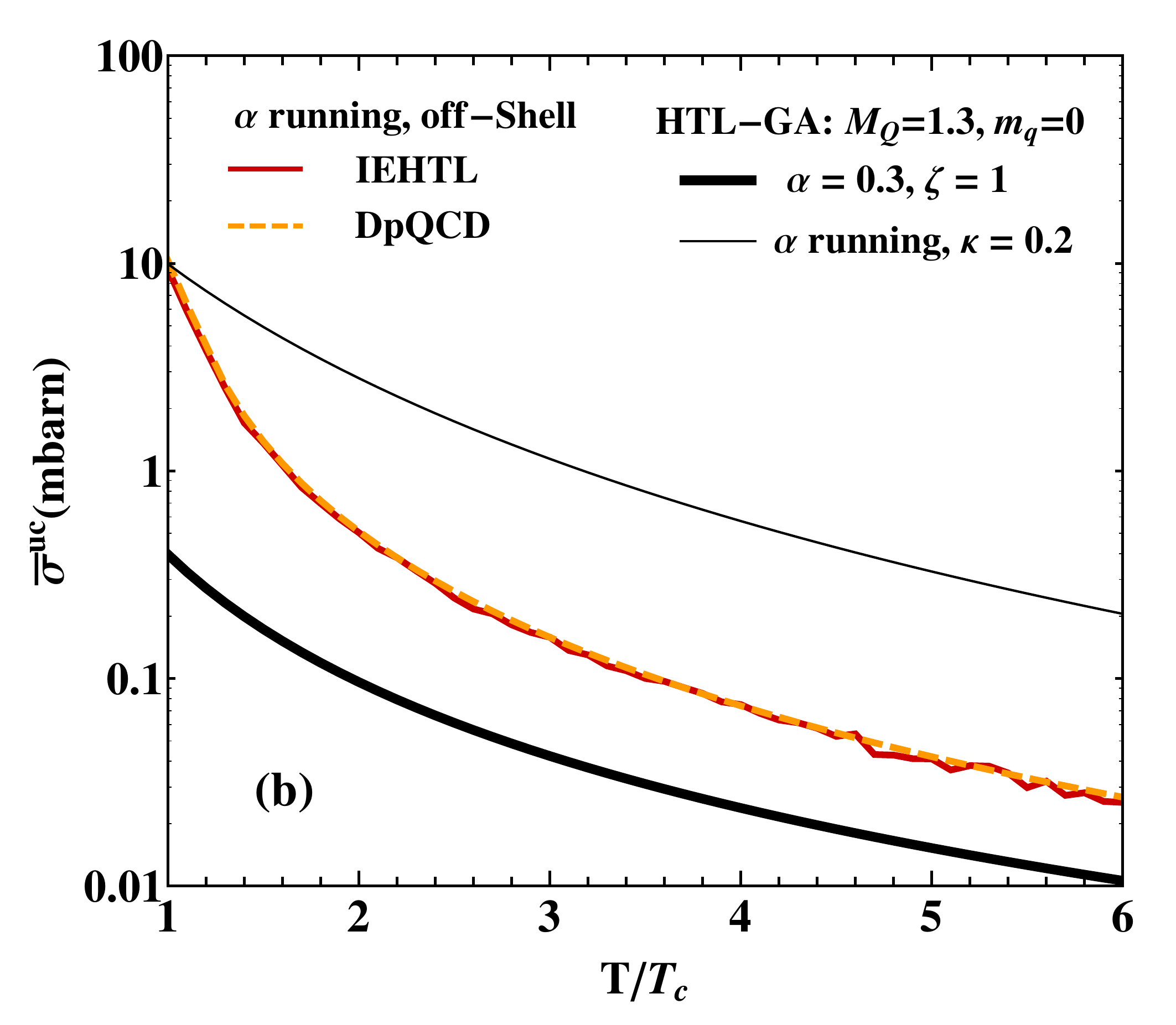}
       \end{center}
\vspace*{-0.3cm}
\caption{\emph{(Color online) Elastic cross section of $u c \rightarrow u c$ for off-shell (solid lines)  and on-shell partons (dashed lines) at different temperatures (see legend). We consider the DQPM pole masses for the on-shell partons and the DQPM spectral functions for the off-shell ones; \textbf{Left~ (a):} as a function of $\sqrt{s}$ for different temperatures, \textbf{right~ (b):} energy-averaged cross section as a function of the temperature $T/T_c$, where the results are also compared to those from the HTL-GA model with constant coupling (lower thick line) and running coupling (upper thin line) (see legend). $m_q$ and $M_Q$ are given in GeV.} \label{fig:SigmaqQDpQCD-IEHTL}}
\end{figure}

  When averaging over $s$ for a given temperature $T$, we observe the inverse situation at large angles; i.e., the off-shell differential cross section becomes slightly lower than the on-shell one. We may deduce that the off-shell differential cross section is larger than the on-shell one for small values of $\sqrt{s}$ and the situation is reversed for large $\sqrt{s}$. However, according to the small differences between the differential on-shell and off-shell cross sections one can conclude that the total on-shell cross sections do not change on a relevant scale when introducing off-shell masses. This is particularly important because the width of the heavy quark has been taken as an upper limit (cf. Sec. II).

  The off-shell cross sections, i.e. the numerical convolution of the on-shell cross sections with the spectral functions, are shown in Fig.  \ref{fig:SigmaqQDpQCD-IEHTL} (a) as a function of $\sqrt{s}$ for different temperatures. One observes by direct comparison of the solid lines (IEHTL) of Fig. \ref{fig:SigmaqQDpQCD-IEHTL} (a) with the dashed lines (DpQCD) that the effect of a parton spectral width on the $qQ$ elastic scattering is negligible for energies well above the on-shell threshold. This is attributable to the moderate parton widths considered in the DQPM model. At energies below the on-shell threshold the off-shell cross section increases with $\sqrt{s}$ because more and more masses can contribute. Whereas the on-shell cross section diverges at the threshold the off-shell cross section shows a maximum at the on-shell threshold and decreases then owing to the decrease of the on-shell cross section. After passing a slight minimum, for large $\sqrt{s}$ on- and off- shell cross section are increasing only very slowly. The shape of the cross section in $\sqrt{s}$ is approximately the same in this temperature range ($1.2 T_c < T < 3 T_c$) and allows for a simple parametrization with the absolute value driven by $g^2(T/T_c)$. From Fig. \ref{fig:SigmaqQDpQCD-IEHTL} (a) we deduce as well that an increasing medium temperature $T$ leads to an increase of the thermal gluon mass (infrared regulator) and hence to a decrease of the IEHTL and DpQCD $qQ$ elastic cross sections.

  Figure \ref{fig:SigmaqQDpQCD-IEHTL}(b) shows, furthermore, the energy averaged $qQ$ elastic cross section as a function of the medium temperature $T/T_c$ for the different approaches presented above (see legend in the figure). We notice different power laws in $T$, i.e., $\sim \displaystyle T^{-\beta}$ for the HTL-GA and DpQCD/IEHTL models. In fact, one can find that ($\displaystyle \beta^{T<1.2 T_c} \sim 2, \beta^{T>1.2 T_c} \sim 1.7$) for the HTL-GA versions, whereas ($\displaystyle \beta^{T<1.2 T_c} \sim 4, \beta^{T>1.2 T_c} \sim 2$) for the DpQCD/IEHTL approaches that are practically identical. The higher power coefficients in the DpQCD/IEHTL approaches can be traced back to the infrared enhancement of the effective coupling. These different  power laws in $T$ will have a sizable effect on the transport coefficients to be evaluated in the future. We stress again that the effect of the DQPM spectral function on $\displaystyle \bar{\sigma} (T)$ is negligible by comparing DpQCD/IEHTL energy-averaged cross sections in Fig. \ref{fig:SigmaqQDpQCD-IEHTL}(b).

\begin{figure}[h!]
    \begin{center}
    \begin{minipage}[h!]{0.6\linewidth}
       \includegraphics[width=8.3cm, height=7.5cm]{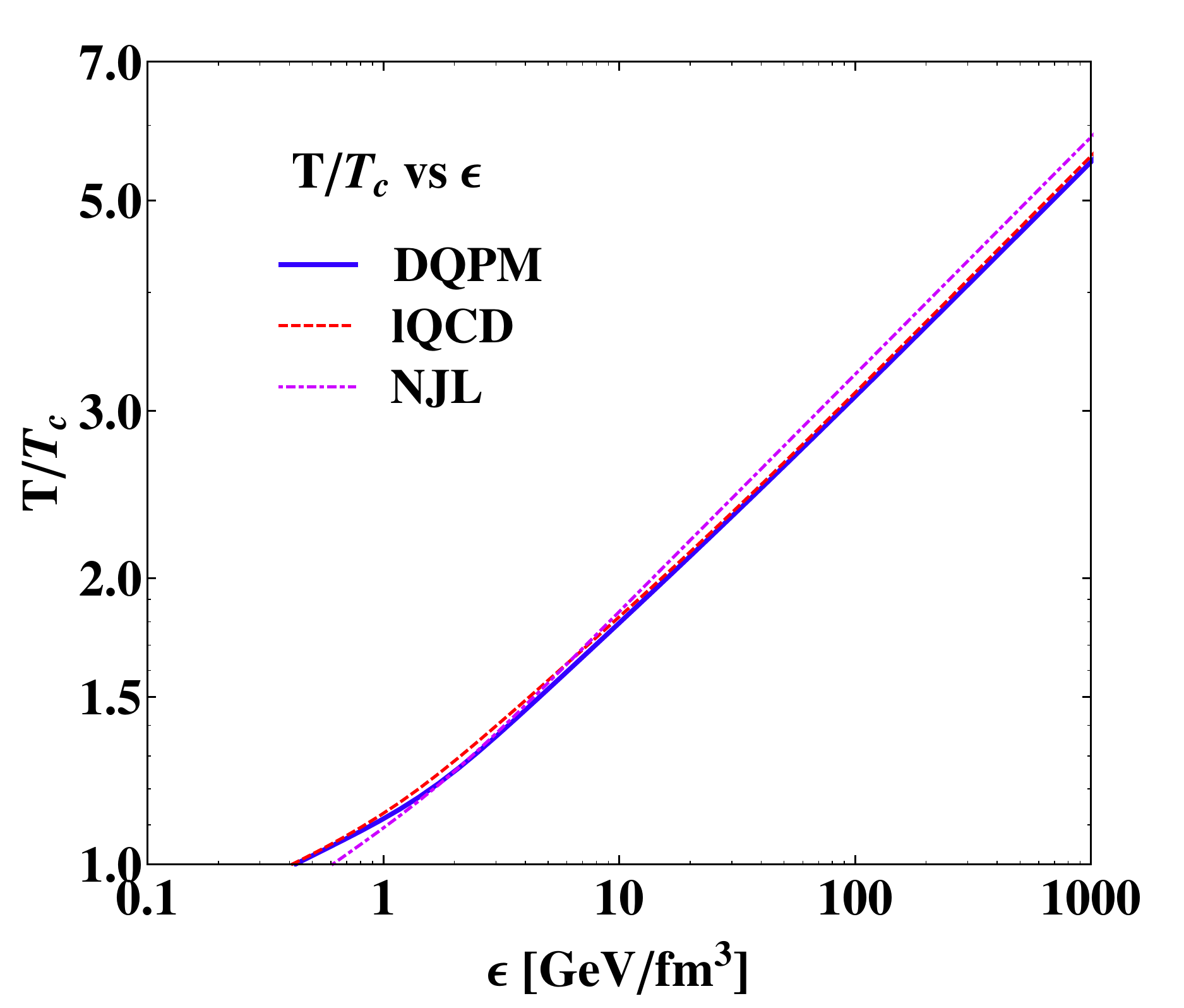}
%       \hspace*{0.4cm}
%       \includegraphics[width=8.5cm, height=7.47cm]{Images/qQProcess/SigmaqQIEHT-DpQCD-HTLvsEnergyDensity.png}
     \end{minipage}\hfill
     \begin{minipage}[h!]{0.4\linewidth}
\vspace*{-0.2cm}
\caption{\emph{(Color online) Temperature $T$ as a function of the energy density $\epsilon$ from lattice QCD (blue solid line) \cite{Borsanyi:2010bp,Borsanyi:2010cj} in the DQPM (red dashed line) \cite{Wcassing2009EPJS,Cassing:2009vt,Cassing2007365} and the
NJL model (pink dot-dashed line) \cite{Marty:2013ita}.} \label{fig:TvsEnergyDensity}}
   \end{minipage}
\end{center}
\end{figure}

\begin{figure}[h!]
       \begin{center}
    \begin{minipage}[h!]{0.6\linewidth}
       \includegraphics[width=8.6cm, height=7.8cm]{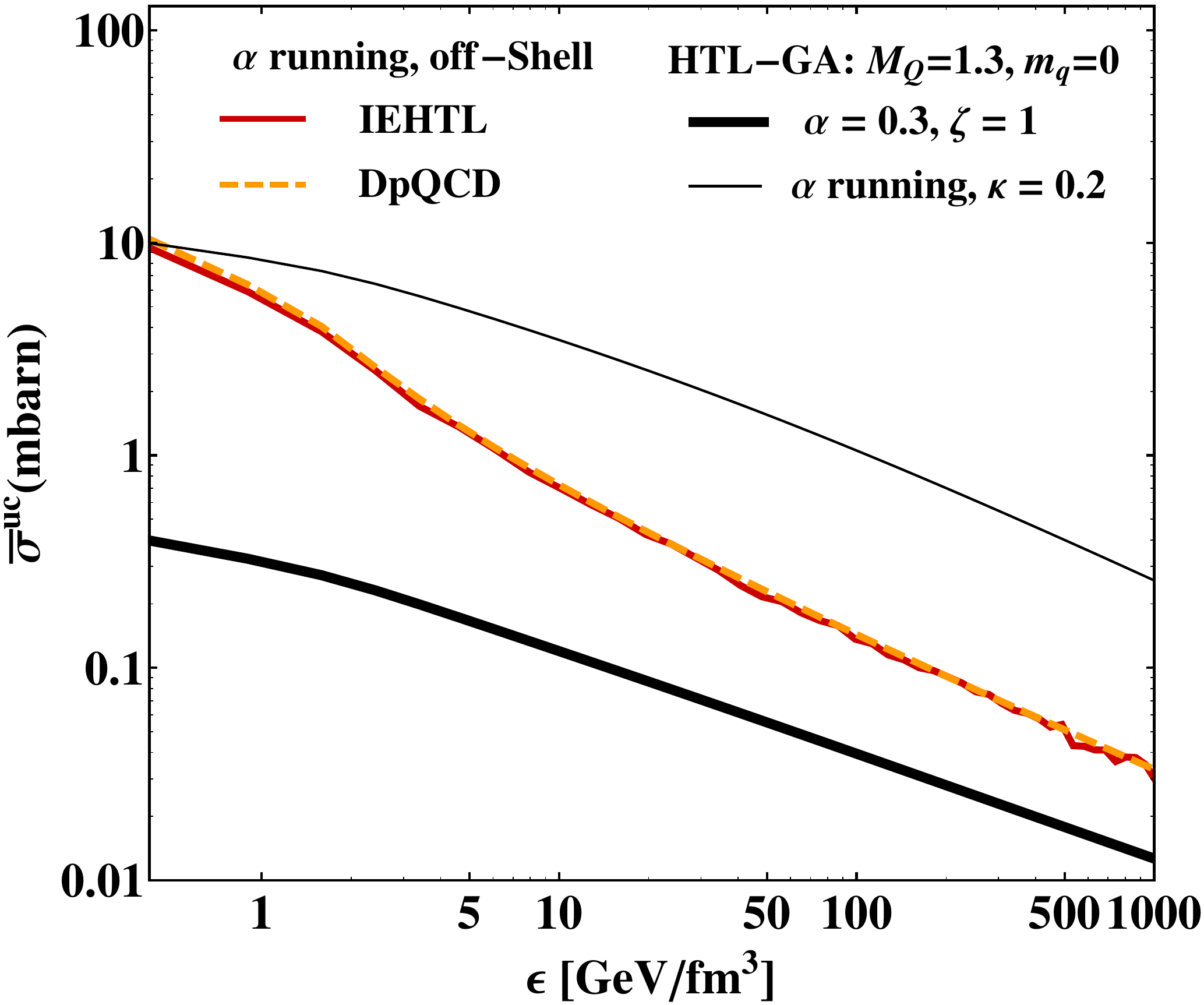}
     \end{minipage}\hfill
     \begin{minipage}[h!]{0.4\linewidth}
\vspace*{-0.2cm}
\caption{\emph{(Color online) Energy-averaged elastic cross section for $u c \rightarrow u c$ scattering for off-shell and on-shell partons as a function of the energy density $\epsilon$ (see legend). We consider the DQPM pole masses for the on-shell partons (DpQCD) and the DQPM spectral functions for the off-shell case (IEHTL). Also shown are the results for the HTL-GA versions with constant (thick solid lower line) and running coupling (upper thin line), where $m_q$ and $M_Q$ are given in GeV.} \label{fig:SigmaqQDpQCD-IEHTLvsEnergyDensity}}
   \end{minipage}
\end{center}
\end{figure}

  The last point we address for the $qQ$ scattering is the on-shell and off-shell total elastic cross section as a function of the energy density $\epsilon$. These quantities are displayed in Fig. \ref{fig:SigmaqQDpQCD-IEHTLvsEnergyDensity} using the inverted DQPM equation of state which gives the temperature as a function of the energy density $\epsilon$ as shown in Fig. \ref{fig:TvsEnergyDensity}. Note that the DQPM model describes the QCD energy density at temperatures even as low as $T \sim T_c$. Simulations in transport theory have shown that the local energy densities achieved in the course of HICs at SPS and RHIC energies reach at most 30 GeV fm$^{-3}$ and at LHC energies 300 GeV fm$^{-3}$. Therefore, one observes that the $qQ$ elastic cross section at the energy densities of interest is $\in$ [0.1--10] mb following the DpQCD/IEHTL approaches and $\in$ [0.08--0.8] mb ($\in$ [2--18] mb) following the HTL-GA approach with fixed coupling (running coupling constant). Actual PHSD transport calculations with the cross section computed so far in comparison to data at RHIC and LHC energies will clarify the validity or inadequacy of the models investigated here.


   In concluding this section we have presented the differential elastic scattering of charm quarks on light $u,d$ quarks of the QGP within different approaches. We have checked that the $s-c$ elastic scattering cross sections are almost identical to the $u,d-c$ cross sections in the DpQCD/IEHTL approaches owing to the large difference between the charm quark mass and those of the light quarks ($u$,$d$,$s$). The different masses of the light quarks have a minor effect on the elastic scattering with $c$ quarks except close to
threshold. The same holds for the effect of a finite spectral width of the degrees of freedom, which has been calculated here for the first time. The size of the elastic cross section is dominated by the infrared regulator, which in the finite temperature medium is determined by a dynamical gluon mass. Here the HTL-inspired models---fixed by elementary vacuum cross sections and decay amplitudes---provide quite different results especially with respect to the running coupling and the associated gluon (screening) mass as compared to the DQPM that has been fixed to lattice QCD calculations at finite temperature. Explicit transport calculations in comparison to experimental data are needed to figure out the appropriate scenario. However, one should point out that DpQCD/IEHTL has much larger $<\!t\!>$ than the HTL-GA model so that the transport coefficients like $\hat{q}$ might be not very different from one model to another.
  %---------------------------------------------------------------------------------------------------------------------------------------------
\section{$g Q \rightarrow g Q$ \ scattering}
\label{gQProcess}
%---------------------------------------------------------------------------------------------------------------------------------------------
\vskip -2mm

  In this section we study the $gQ$ elastic scattering in vacuum and in the QGP medium at finite temperature $T$. As in the previous sections, we consider the case of on-shell and off-shell heavy quarks and gluons. Here we present (for completeness) the well-known pQCD $gQ$ elastic cross section with massive heavy quarks and massless gluons, the cross section with heavy quarks and gluons dressed by effective masses (\emph{``DpQCD''}: dressed pQCD) and finally the case of off-shell heavy quarks and gluons using the DQPM spectral functions.

\begin{widetext}
\begin{figure}[h!]
\begin{center}
      \hspace*{1.3cm}\includegraphics[width=16cm]{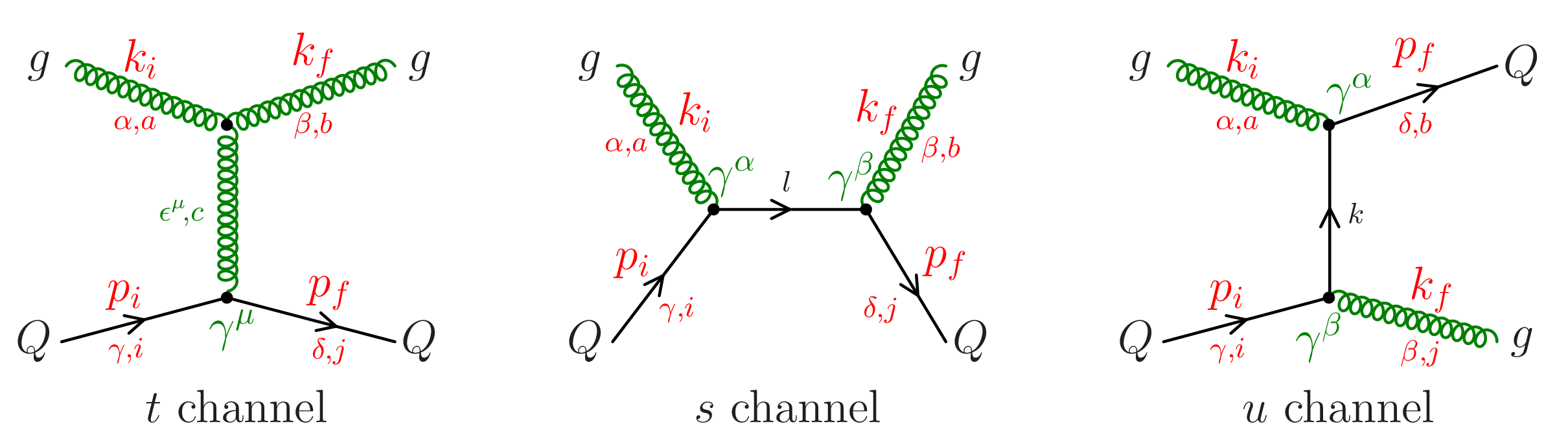}
\begin{minipage}{\textwidth}
\caption{\emph{
(Color online) Feynman diagrams for the $g Q \rightarrow g Q$ scattering process. The indices $i$,$j$,$k$,$l$ $= 1-3$
stand for quark color, while the indices $a$,$b$,$c$ $= 1-8$ represent gluon color. $\mu$,$\nu$,$\lambda$ are Lorentz indices while $k_i$ and $p_i$ ($k_f$ and $p_f$) denote the initial (final) momentum of the gluon and the heavy quark, respectively. The invariant energy squared is given by $s=(p_i + k_i)^2$, $t=(p_i - p_f)^2$, $u=(p_i - k_f)^2$.} \label{fig:gQFeynmanDiagram}}
\end{minipage}
\end{center}
\end{figure}
\end{widetext}

  The Feynman diagrams for the process $g Q \rightarrow g Q$ are given in Fig. \ref{fig:gQFeynmanDiagram} and represent $t$, $s$, and $u$ channels.

%-----------------------------------------------------------------------------------------------------------------------------------------------------------------------------
\subsection[Massive heavy quarks and massless gluons of zero widths]{Massive heavy quarks and massless gluons of zero widths} \label{gQProcess.B} %--------------------------- %-----------------------------------------------------------------------------------------------------------------------------------------------------------------------------

  The invariant amplitudes for the three graphs (shown in Fig.\ref{fig:gQFeynmanDiagram}) for the case of massive heavy quarks and massless  gluons is given according to Combridge\cite{Combridge1979429} and Cutler and Sivers \cite{PhysRevD.17.196} and expressed in the appendix by Eq. (\ref{equ:SecA.2}). The lowest-order amplitude for the process $g Q \rightarrow g Q$ is obtained from the Feynman rules of the gauge theory by the sum of the amplitudes (\ref{equ:SecA.2}). To obtain the correct result for the squared matrix element
{\setlength\arraycolsep{0pt}
\begin{eqnarray}
<|\mathcal{M}|^2> = \frac{1}{4} \sum_{spins} T_{\alpha \beta} T_{\alpha' \beta'}^{\star} \epsilon_i^{\alpha} \epsilon_i^{\star \alpha'} \epsilon_f^{\beta} \epsilon_f^{\star \beta'},
%\nonumber\\
%& & {}
\label{equ:Sec4.3}
\end{eqnarray}}
  we have either to use appropriate projection operators for the transverse polarization states,
{\setlength\arraycolsep{-9pt}
\begin{eqnarray}
& & \sum_{spins} \epsilon_i^{\alpha} \epsilon_i^{\star \alpha'} = - g^{\alpha \alpha'} + \frac{2}{s} (p_i^{\alpha} k_i^{\alpha'} + p_i^{\alpha'} k_i^{\alpha}), \hspace{0.5cm} \sum_{spins} \epsilon_f^{\beta} \epsilon_f^{\star \beta'} = - g^{\beta \beta'} + \frac{2}{s} (p_f^{\beta} k_f^{\beta'} + p_f^{\beta'} k_f^{\beta}).
\label{equ:Sec4.7}
\end{eqnarray}}
 or to introduce Fadeev-Popov ghosts. For a detailed discussion we refer the reader to Ref. \cite{PhysRevD.17.196}. The result for the squared amplitude is
\begin{widetext}
{\setlength\arraycolsep{0pt}
\begin{eqnarray}
\label{equ:Sec4.10}
< |\mathcal{M}|^2 > = \pi^2 \alpha_s^2 (Q^2) & & \Biggl[\frac{32(s - M_Q^2)(M_Q^2 - u)}{t^2} + \frac{64}{9} \frac{(s - M_Q^2)(M_Q^2 - u) + 2 M_Q^2 (s + M_Q^2)}{(s - M_Q^2)^2}
\nonumber\\
& & {} + \frac{64}{9} \frac{(s - M_Q^2)(M_Q^2 - u) + 2 M_Q^2 (M_Q^2 + u)}{(M_Q^2 - u)^2} + \frac{16}{9} \frac{M_Q^2 (4 M_Q^2 - t)}{(s - M_Q^2)(M_Q^2 - u)}
\\
& & {} + 16 \frac{(s - M_Q^2)(M_Q^2 - u) + M_Q^2 (s - u)}{t (s - M_Q^2)} - 16 \frac{(s - M_Q^2)(M_Q^2 - u) - M_Q^2 (s - u)}{t (M_Q^2 - u)} \Biggr].
\nonumber
\end{eqnarray}}
\end{widetext}
 which becomes for $M_Q = 0$ identical to the expression given by Cutler and Sivers \cite{PhysRevD.17.196} for the scattering of massless light quarks with massless gluons,
{\setlength\arraycolsep{0pt}
\begin{eqnarray}
& & <|\mathcal{M}|^2> = 1 - \frac{4}{9} \frac{u}{s} - \frac{4}{9} \frac{u}{s} - \frac{2 u s}{t^2}.
%\nonumber\\
%& & {}
\label{equ:Sec4.8}
\end{eqnarray}}

 For the process $g Q \rightarrow g Q$, we choose for the running coupling $Q^2 = \frac{1}{2} [- t + (M_Q^2 - u)] = \frac{1}{2}(s - M_Q^2)$. We also examine the effect on our results by taking $Q^2 = s$.

 As in the process $q Q \rightarrow q Q$, the resulting cross section, $\sigma (s)$, given by Eq. (\ref{equ:Sec4.11}), diverges owing to the pole at $t = 0$ which corresponds to soft $gQ$ interactions in the graphs of Fig.  \ref{fig:gQFeynmanDiagram}. For the same physical argumentation as in case of  $q Q \rightarrow q Q$ scattering, we reduce the upper limit of integration in the evaluation of $d \sigma/dt$ to $t = - Q_0^2$ but require in addition for $g Q \rightarrow g Q$ process that also $u - m^2 < - Q_0^2$ to avoid the $u$-channel pole. Performing this integration we obtain
\begin{widetext}
{\setlength\arraycolsep{0pt}
\begin{eqnarray}
\sigma_{gQ \rightarrow gQ} (s) = \frac{\pi \alpha_s^2 (Q^2)}{(s - M_Q^2)^2} & &  \biggl[\left(1 + \frac{4}{9} \bigg\{\frac{s + M_Q^2}{s - M_Q^2} \biggr\}^2 \right) (L - Q_0^2) + \frac{2}{9} \frac{(Q_0^4 - L^2)}{(s - M_Q^2)}
\nonumber\\
& & {} + 2 (s + M_Q^2) \ln \frac{Q_0^2}{L} + \frac{4}{9} \frac{s^2 - 6 M_Q^2 s + 6 M_Q^4}{s - M_Q^2} \ln \frac{s - M_Q^2 - Q_0^2}{s - M_Q^2 - L}
\nonumber\\
& & {} + 2 (s - M_Q^2)^2 \left(\frac{1}{Q_0^2} - \frac{1}{L} \right) + \frac{16}{9} M_Q^4 \left(\frac{1}{s - M_Q^2 - L} - \frac{1}{s - M_Q^2 - Q_0^2} \right) \Biggr],
\label{equ:Sec4.11}
\end{eqnarray}}
\end{widetext}
 with $L = min \left(s - M_Q^2 - Q_0^2, (s - M_Q^2)^2/s \right)$. As in the $q Q \rightarrow q Q$ process, the cuts on the integration range have also the effect of raising the threshold from $s > M_Q^2$ to $s > M_Q^2 + \frac{1}{2} Q_0^2 + \sqrt{M_Q^2 Q_0^2 + \frac{1}{4} Q_0^4}$ if $Q_0^2 < \frac{1}{2} M_Q^2$ and to $s > m^2 + 2 Q_0^2$ if $Q_0^2 < \frac{1}{2} M_Q^2$. The coupling constant $\alpha_s$ can be taken as a fixed value or as the running $\alpha_s (Q^2)$ given by Eq. (\ref{equ:Sec3.12}).

\begin{figure}[h!] %tbh!
\begin{center}
%   \begin{minipage}[c]{.49\linewidth}
     \includegraphics[width=8.6cm, height=6.41cm]{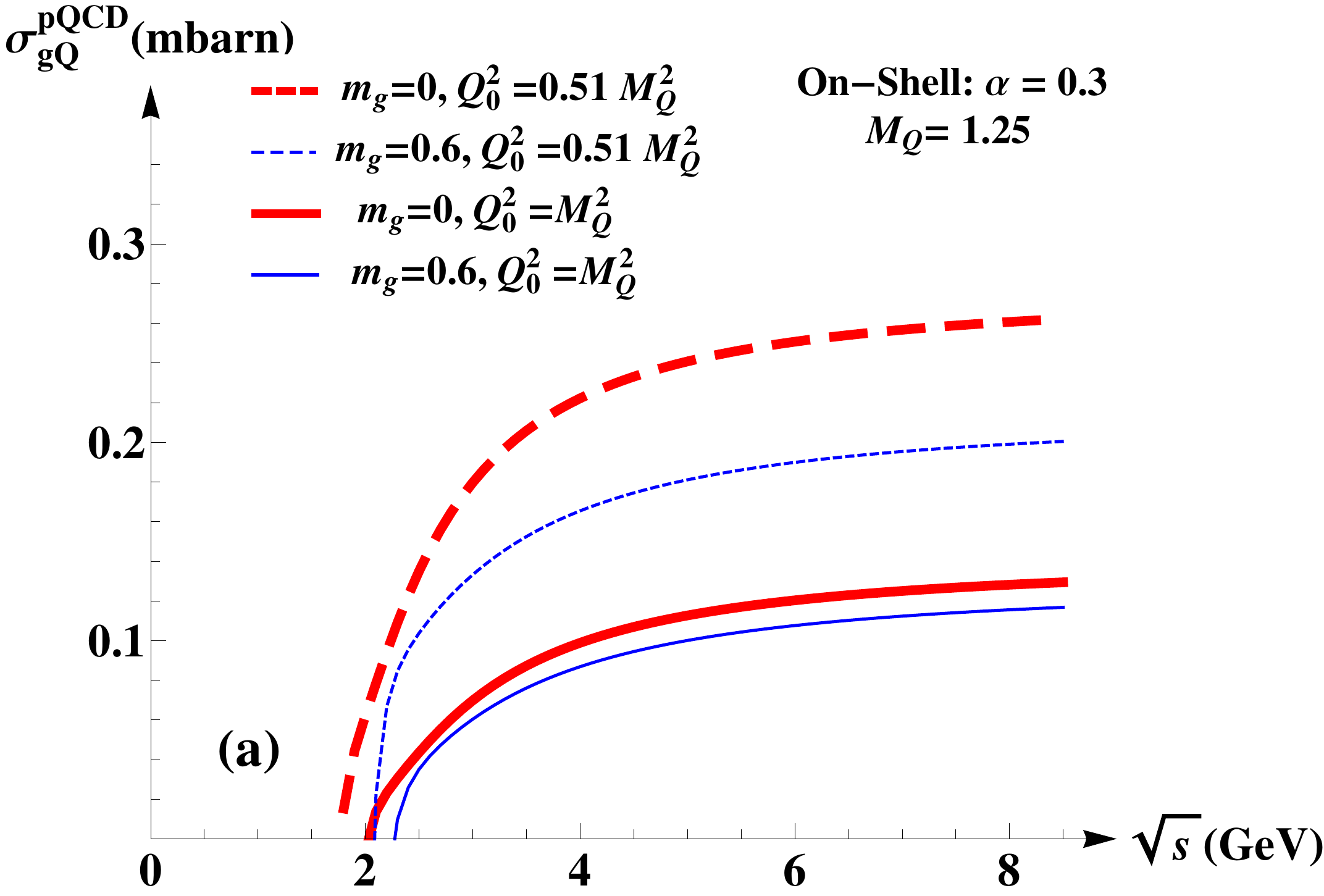}
%   \end{minipage} \hfill
%   \begin{minipage}[c]{.49\linewidth}
     \includegraphics[width=8.6cm, height=6.41cm]{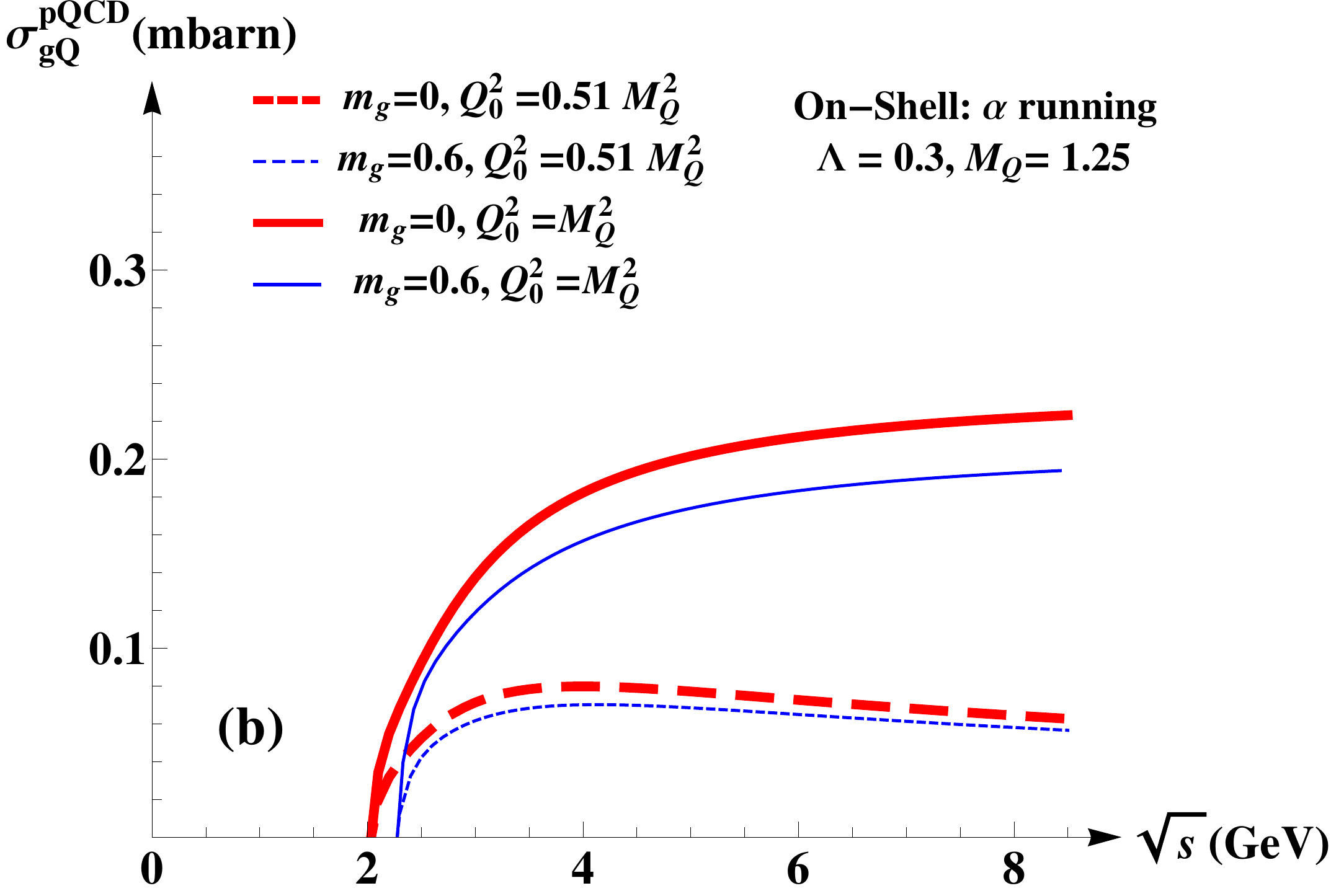}
%\end{minipage}
\hspace{-0.1cm}
\caption{\emph{(Color online) Elastic cross section for the process $g Q \rightarrow g Q$ as a function of $\sqrt{s}$ for different
gluon masses (red, $m_g$ = 0; blue, $m_g$ = 0.6 GeV), $M_Q = 1.25$ GeV and two choices of $Q_0^2$ (see legend). (a) For constant $\alpha = \alpha_s=0.3$; (b) for the running coupling $\alpha_s(Q^2)$ (\ref{equ:Sec3.12}).}}
\label{fig:qgSigmapQCD_Combrdige}
\end{center}
\end{figure}

  Figure \ref{fig:qgSigmapQCD_Combrdige}(a) presents the results of Eq. (\ref{equ:Sec4.11}) for the elastic scattering cross section $\sigma_{g Q \rightarrow g Q} (s)$ for the case of massive heavy quarks and massless gluons. We see how the increasing infrared regulator $Q^2_0$ suppresses the cross section. Also a finite gluon mass (of 0.6 GeV) reduces the cross section as compared to a zero mass. For a running coupling  the form of the cross section
is similar but the absolute value increases, as can be seen from  Fig. \ref{fig:qgSigmapQCD_Combrdige}(b).

  Owing to time-reversal invariance one can deduce that the Eqs. (\ref{equ:Sec4.3})--(\ref{equ:Sec4.11}) describe also the elastic cross section for the process $g \bar{Q} \rightarrow g \bar{Q}$.

%---------------------------------------------------------------------------------------------------------------------------------------------
\subsection[Massive heavy quarks and gluons of zero widths]{Massive heavy quarks and massive gluons of zero widths} \label{gQProcess.C} %-----
%---------------------------------------------------------------------------------------------------------------------------------------------

\begin{figure}[h!] %tbh!
\begin{center}
%   \begin{minipage}[c]{.49\linewidth}
\includegraphics[width=8.9cm,height=7cm]{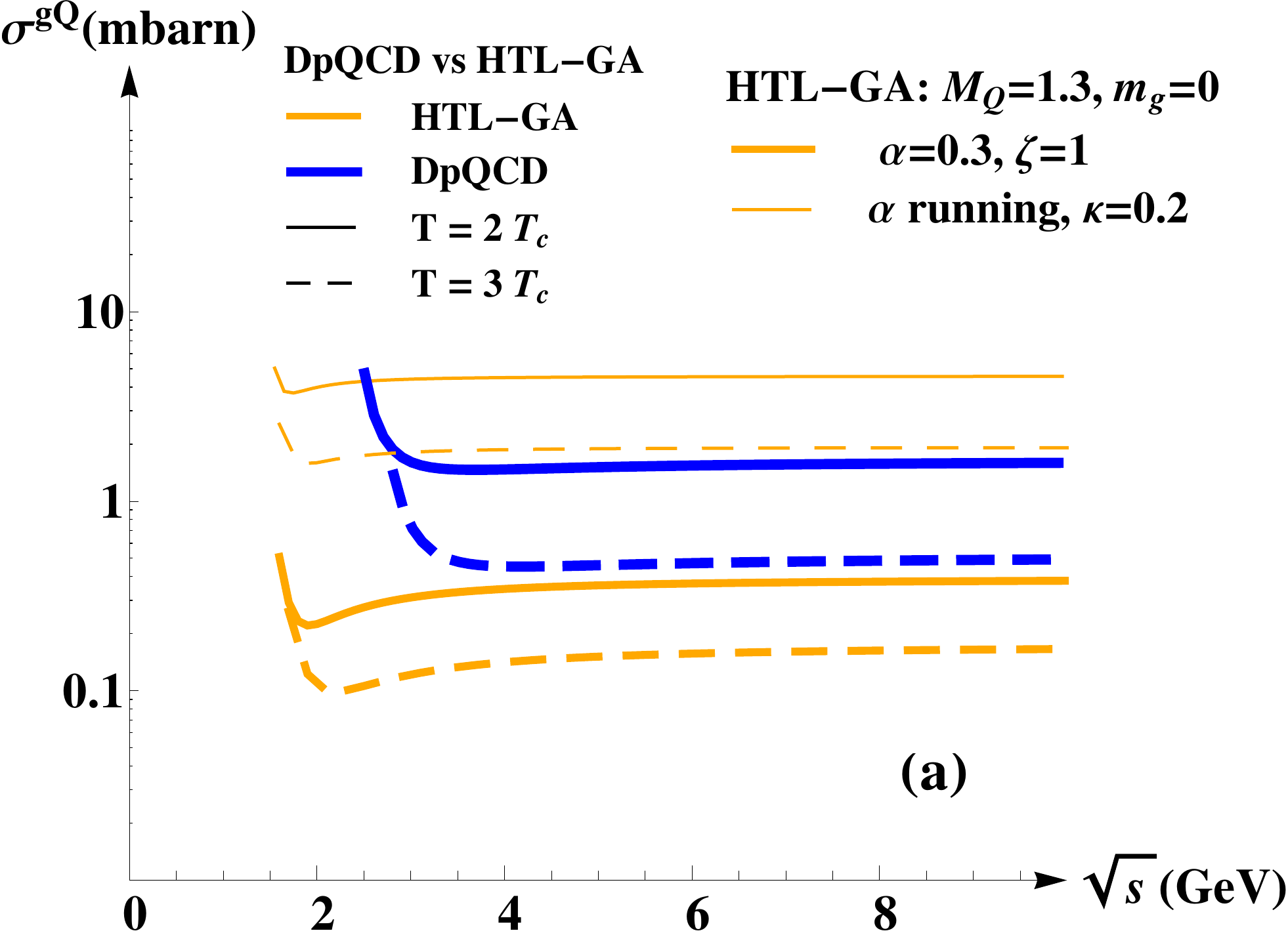} \hspace*{0.3cm}
%   \end{minipage}
%   \begin{minipage}[c]{.49\linewidth}
\includegraphics[width=8.67cm]{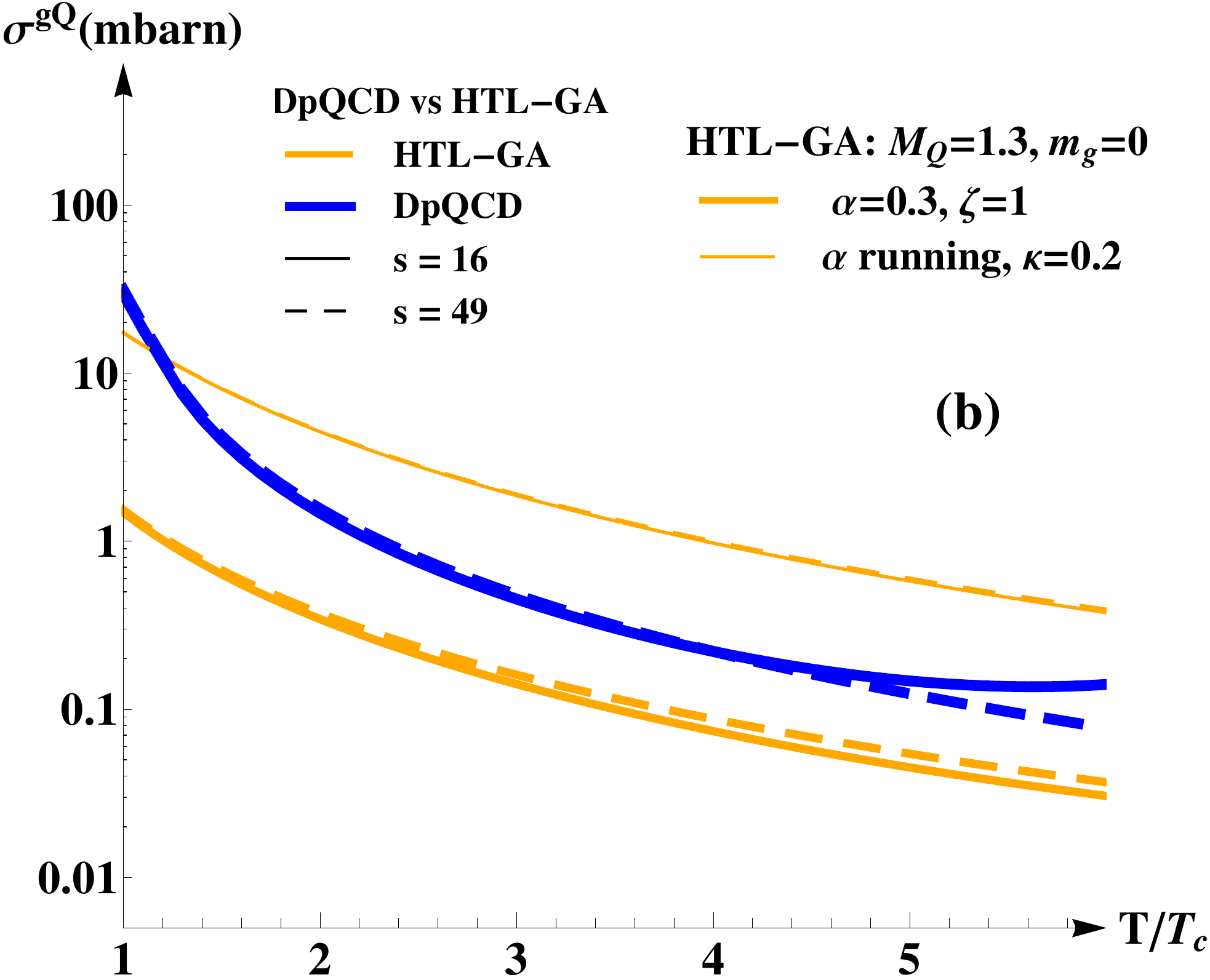}
%\end{minipage}
\caption{\emph{(Color online) Comparison of $\sigma^{gQ}$ calculated within the HTL-GA (orange lines) and DpQCD (blue lines) approaches. $m_g$ and $M_Q$ are given in GeV. (a) As a function of $\sqrt{s}$ for different temperatures $T$ (see legend); (b) as a function of the scaled temperature $T/T_c$, with $T_c = 0.158$ GeV for $\sqrt{s}$= 4 and 7 GeV.}}
\label{fig:SigmagQDpQCD-HTLapproach}
\end{center}
\end{figure}

  If not only the quarks but also the gluons are massive the invariant amplitudes for the three graphs in Fig. \ref{fig:gQFeynmanDiagram} are given by (\ref{equ:SecA.4}) in the Appendix. We recall that for vector fields with nonzero Lagrangian mass there is no gauge freedom anymore. The massive vector field  $A_{\mu}$ only has to fulfill  the condition $\partial^{\mu} A_{\mu} = 0$. Therefore, the propagator for a massive vector gluon is given in the Appendix by Eq. (\ref{equ:SecA.9}) and the sum over the initial and final gluon polarizations is fixed by the expressions in Eq. (\ref{equ:Sec4.20}):
{\setlength\arraycolsep{-5pt}
\begin{eqnarray}
& & \sum_{pol, i} \epsilon_{i,\alpha} \epsilon_{i, \alpha' } =  g_{\alpha \alpha'} - \frac{k_{i, \alpha} \ k_{i, \alpha'}}{(m_g^i)^2}, \hspace{0.5cm} \sum_{pol, f} \epsilon_{f,\beta} \epsilon_{f, \beta'} = g_{\beta \beta'} - \frac{k_{f, \beta} \ k_{f, \beta'}}{(m_g^f)^2}.
\label{equ:Sec4.20}
\end{eqnarray}}
 As in the case of $qQ$ elastic scattering, we examine different assumptions on the value of initial and final heavy quarks and gluons and the exchanged gluon:

\begin{itemize}

\item[\emph{i)}] $\sigma^{pQCD}$ with $\alpha_s (Q^2)$ running in HTL-GA approximation. As explained in Sec. \ref{qQProcess.D}, we determine the pQCD $gQ$ scattering cross section following the Peshier-Gossiaux-Aichelin approach. The gluons in the external legs are massless whereas the exchanged gluon is given by the modified Debye mass $\mu = \sqrt{\kappa} \tilde{m}_D$ as in Refs. \cite{Gossiaux:2004qw,PhysRevC.78.014904,Gossiaux2009203c,Gossiaux:2009hr,Gossiaux:2009mk,Gossiaux:2010yx,Gossiaux:2006yu} and arbitrary mass is attributed to the heavy quark ($M_Q = 1.3$ GeV). Figures \ref{fig:dSigmadtgQDpQCD-HTLapproach}(a) [respectively (b)] represent ${d \sigma}/{d t}$ [respectively ${d \sigma}/{d \cos \theta}$] (orange lines) for the $gQ$ elastic scattering at $\sqrt{s}= \sqrt{40} GeV$. Figure \ref{fig:SigmagQDpQCD-HTLapproach} shows the corresponding total cross section as a function of $\sqrt{s}$ (a) and as a function of temperature $T$ (b).

\item[\emph{ii)}] $\sigma^{pQCD}$ with $\alpha_s (T)$ and the DQPM gluon propagator [finite mass, zero width (DpQCD approach)].
$\sigma^{DpQCD}$ is obtained by using the DQPM propagator and keeping particles on-shell. We take the DQPM pole masses for the exchanged gluon
and the external gluons and heavy quarks.
\begin{figure}[h!]
       \begin{center}
       \includegraphics[width=8.6cm,height=7.3cm]{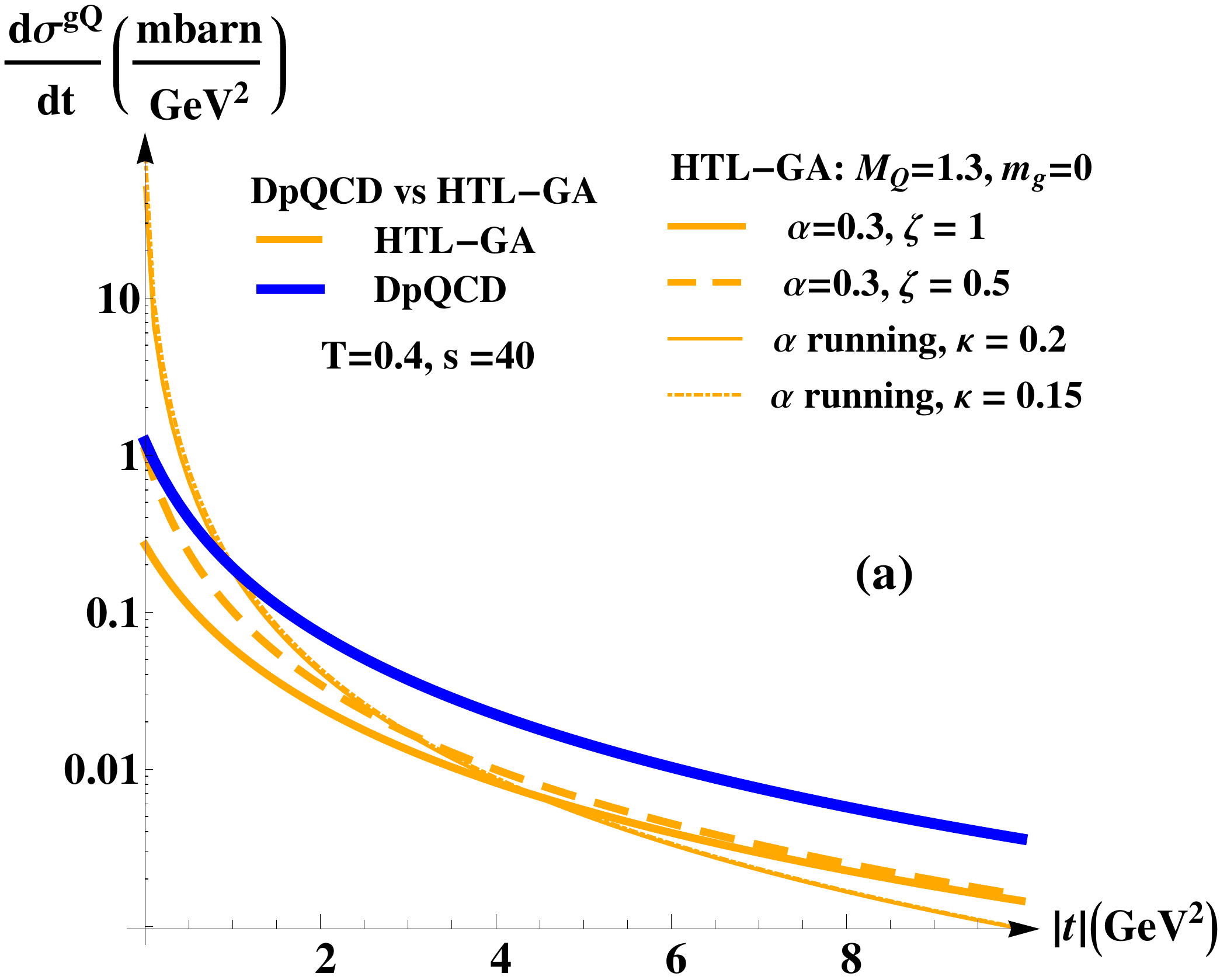}
       %\vspace*{0.8cm}\\
       \hspace*{0.5cm}
       \includegraphics[width=8.6cm,height=7.3cm]{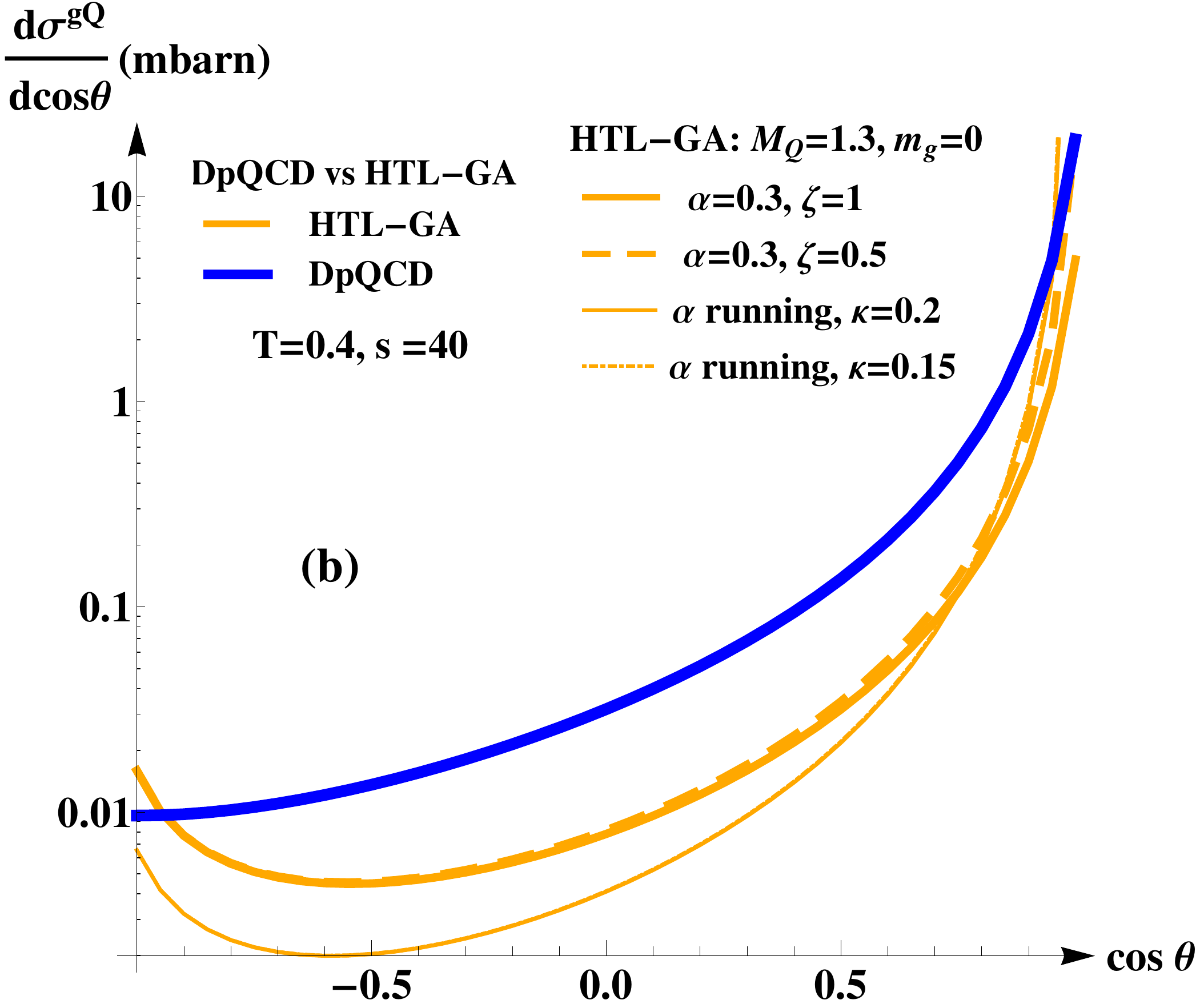}
       \end{center}
       %\vspace{-0.4cm}
\caption{\emph{(Color online) Differential elastic cross section of $g c \rightarrow g c$ $\frac{d \sigma}{d t}$ at $\sqrt{s}= \sqrt{40}$ GeV (a) and $\frac{d \sigma}{d cos \theta}$ (b) for different choices of the strong coupling constant and of the infrared regulator (see legend).
The gluons are part of a heat bath of  temperature $T = 0.4$ GeV. $m_g$ and $M_Q$ are given in GeV.} \label{fig:dSigmadtgQDpQCD-HTLapproach}}
\end{figure}

 Figures \ref{fig:dSigmadtgQDpQCD-HTLapproach}(a) [\ref{fig:dSigmadtgQDpQCD-HTLapproach}(b)] show ${d \sigma}/{d t}$ [${d \sigma}/{d cos \theta}$] for $gQ$ elastic scattering at $\sqrt{s} = \sqrt{40}$ GeV as a function of $|t|$ (or the angle $\theta$) for the two approaches. The same conclusions as in the study of $qQ$ elastic scattering can be drawn, however, with cross sections for the $gQ$ elastic scattering that are larger than the cross sections for $qQ$ scattering by roughly a factor of $9/4$, which is ratio of the different color Casimir operators (squared).

\end{itemize}

 We see again that the cross section in both approaches is of the same order of magnitude and also the angular distribution is similar for a given choice of couplings. The running coupling in the HTL-GA model essentially enhances the cross section at small $t$ (or forward angles).

 Figures \ref{fig:SigmagQDpQCD-HTLapproach}(a) and \ref{fig:SigmagQDpQCD-HTLapproach}(b) present the total elastic scattering of $gQ$ calculated within the two different approaches presented previously, (a) as a function of $\sqrt{s}$ for different temperatures ($T=2 T_c, T= 3 T_c$), (b) as a function of temperature  for $\sqrt{s}$= 4 and 7 GeV. Apart from threshold effects, the cross sections are rather independent on $\sqrt{s}$. These figures show clearly the different values for $\sigma^{gQ}$ for the different choices of couplings and infrared regulators with the largest cross section given by the running coupling in the HTL-GA approach (thin orange lines) and the smallest cross section in HTL-GA model with fixed $\alpha_s=0.3$ (thick orange lines). The variation in $\sqrt{s}$ and temperature is qualitatively very similar to the case of $qQ$ scattering. Note again the large enhancement of the DpQCD cross section for temperatures close to $T_c$, which is attributable to the infrared enhancement of the effective coupling.

%--------------------------------------------------------------------------------------------------------------------------------------------
\subsection[Massive heavy quarks and gluons of finite widths]{Massive heavy quarks and gluons of finite widths} \label{gQProcess.D} %--------
%--------------------------------------------------------------------------------------------------------------------------------------------

 For the case of finite masses and widths of the scattering quarks and gluons, the expressions (\ref{equ:SecA.4})--(\ref{equ:SecA.10}) are still describing the $gQ$ elastic scattering amplitude but, in addition of taking the spectral functions for the heavy quark and gluon masses into account, one has to change the denominator of the quark and gluon propagators. Indeed, instead of using the usual expression for the quark and gluon propagator given in the appendix by Eq. (\ref{equ:SecA.9}), we consider the following expressions for the case of massive vector gluons with finite lifetime $G_F^{\mu \nu} (q, m_g)$ and for the case of massive fermions with finite lifetime $S_F (p, m_q)$:
{\setlength\arraycolsep{-5pt}
\begin{eqnarray}
\label{equ:Sec4.22}
& & G_F^{\mu \nu} (q, m_g) = - i \ \frac{g^{\mu \nu} - q^{\mu} q^{\nu}/m_g^2}{q_0^2 - \bs{q}^2 - m_g^2 + i 2 \gamma_g q_0}, \hspace{0.5cm} S_F (p, M_Q) = \frac{\slashed{p} + M_Q}{p_0^2 - \bs{p}^2 - M_Q^2 + i 2 \gamma_Q p_0},
\nonumber\\
& & {} {} G_F^{t} = - i \frac{g^{\mu \nu} - q^{\mu} q^{\nu}/m_g^2}{t - m_g^2 + i 2 \gamma_g (p_0^i - p_0^f)}, \hspace{0.5cm} S_F^{u} = \! \frac{\slashed{p} + M_Q}{u - M_Q^2 + i 2 \gamma_Q (p_0^i - k_0^f)}, \hspace{0.5cm} S_F^{s} = \! \frac{\slashed{p} + M_Q}{s - M_Q^2 + i 2 \gamma_Q (p_0^i + k_0^i)},
\end{eqnarray}}
 where $m_g$, $\gamma_g$ ($M_Q$, $\gamma_Q$) are the mass and width of the gluon or the heavy quark. In expressions (\ref{equ:Sec4.22}), $q_0$ $(p_0)$ present in $G_F^{\mu \nu}$ ($S_F$) is the energy of the gluon in the $t$ channel (the heavy quark in the $u$ or $s$ channel).

\begin{figure}[h!]
       \begin{center}
       \includegraphics[width=8.5cm, height=7.83cm]{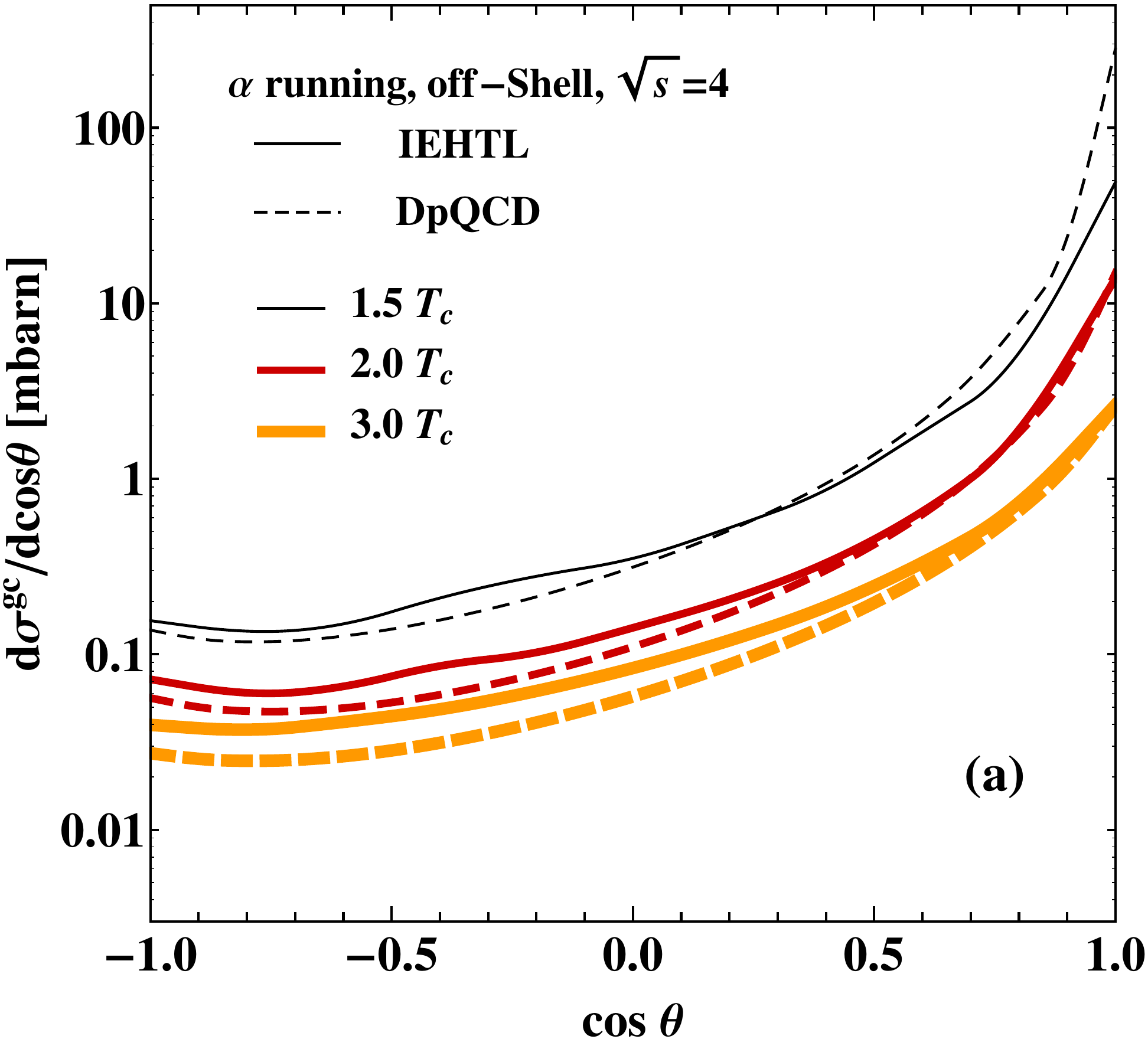}\hfill
%       \vspace*{0.3cm}
       \includegraphics[width=8.5cm, height=7.8cm]{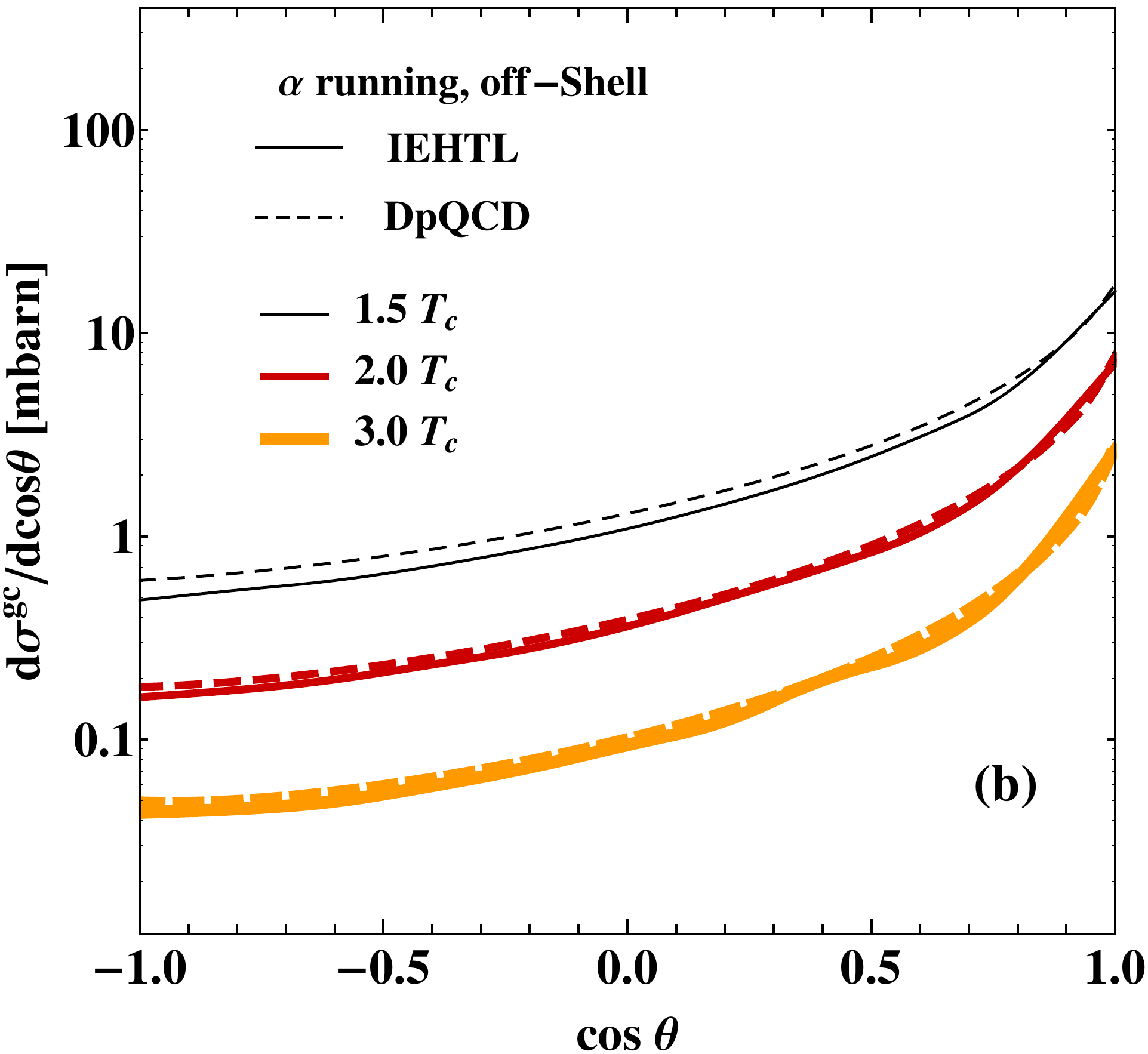}
       \end{center}
       \vspace{-0.5cm}
\caption{\emph{(Color online) Differential elastic cross section for $g c  \rightarrow g c$ scattering for off-shell (IEHTL, solid lines)
and on-shell (DpQCD, dashed lines) partons at three different temperatures (see legend). We consider the DQPM pole masses for the on-shell partons and the DQPM spectral functions for the off-shell ones: (a) for $\sqrt{s} = 4$ GeV, (b) energy averaged.}
\label{fig:dSigmadthetagQDpQCD-IEHTL}}
\end{figure}

\begin{figure}[h!]
  \begin{center}
    \begin{minipage}[h!]{0.6\linewidth}
      \includegraphics[width=8.7cm, height=8cm]{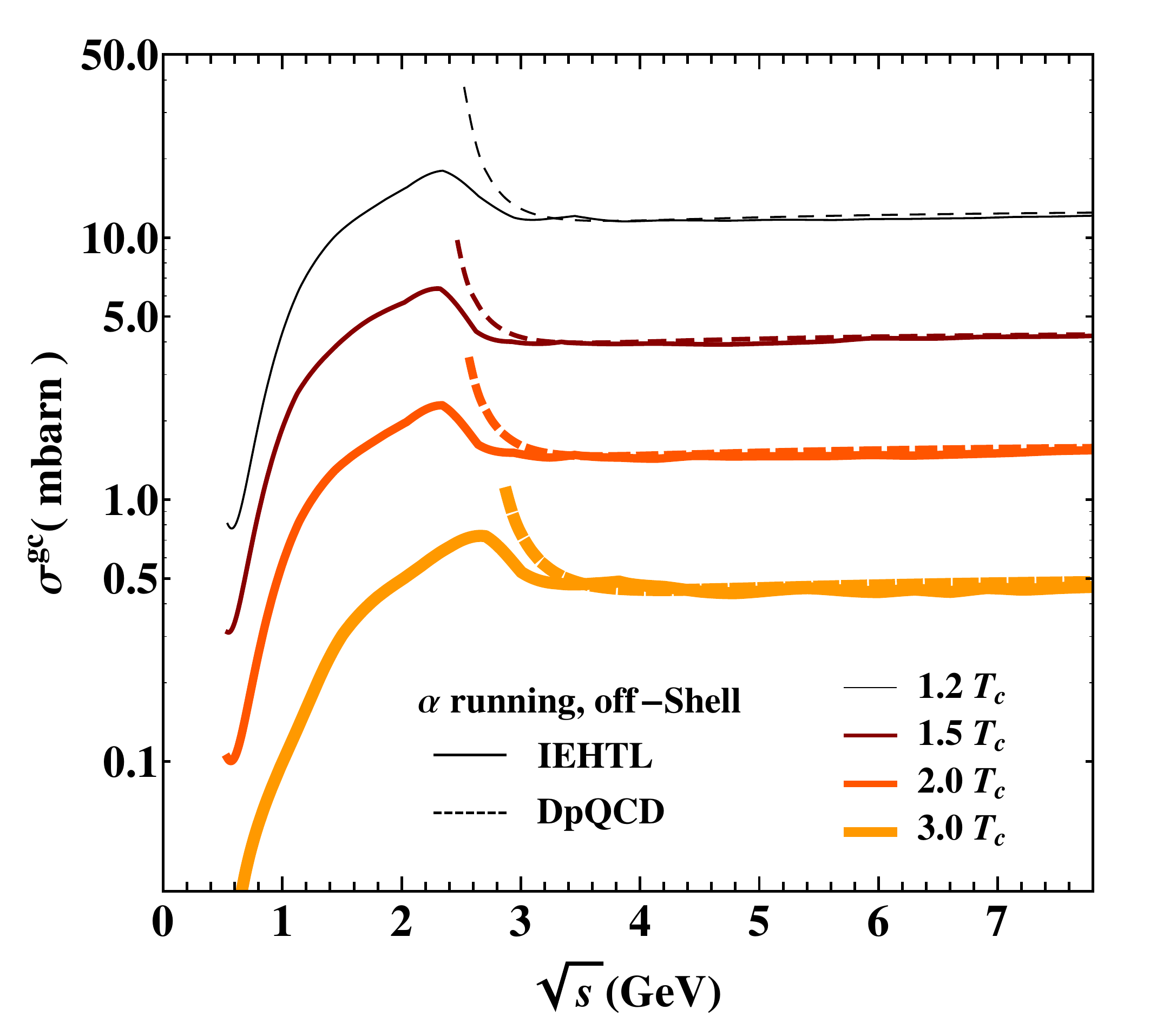}\hspace{5mm}
    \end{minipage}\hfill
    \begin{minipage}[h!]{0.4\linewidth}
\caption{\emph{(Color online) Elastic cross section of $g c \rightarrow g c$ scattering for off-shell (IEHTL, solid lines) and on-shell (DpQCD, dashed lines) partons as a function of $\sqrt{s}$ for different temperatures (see legend). We consider the DQPM pole masses for the on-shell partons and the DQPM spectral functions for the off-shell ones.} \label{fig:SigmagQDpQCD-IEHTLs}}
    \end{minipage}
\end{center}
\end{figure}

\begin{figure}[h!]
       \begin{center}
       \includegraphics[width=8.5cm, height=7.8cm]{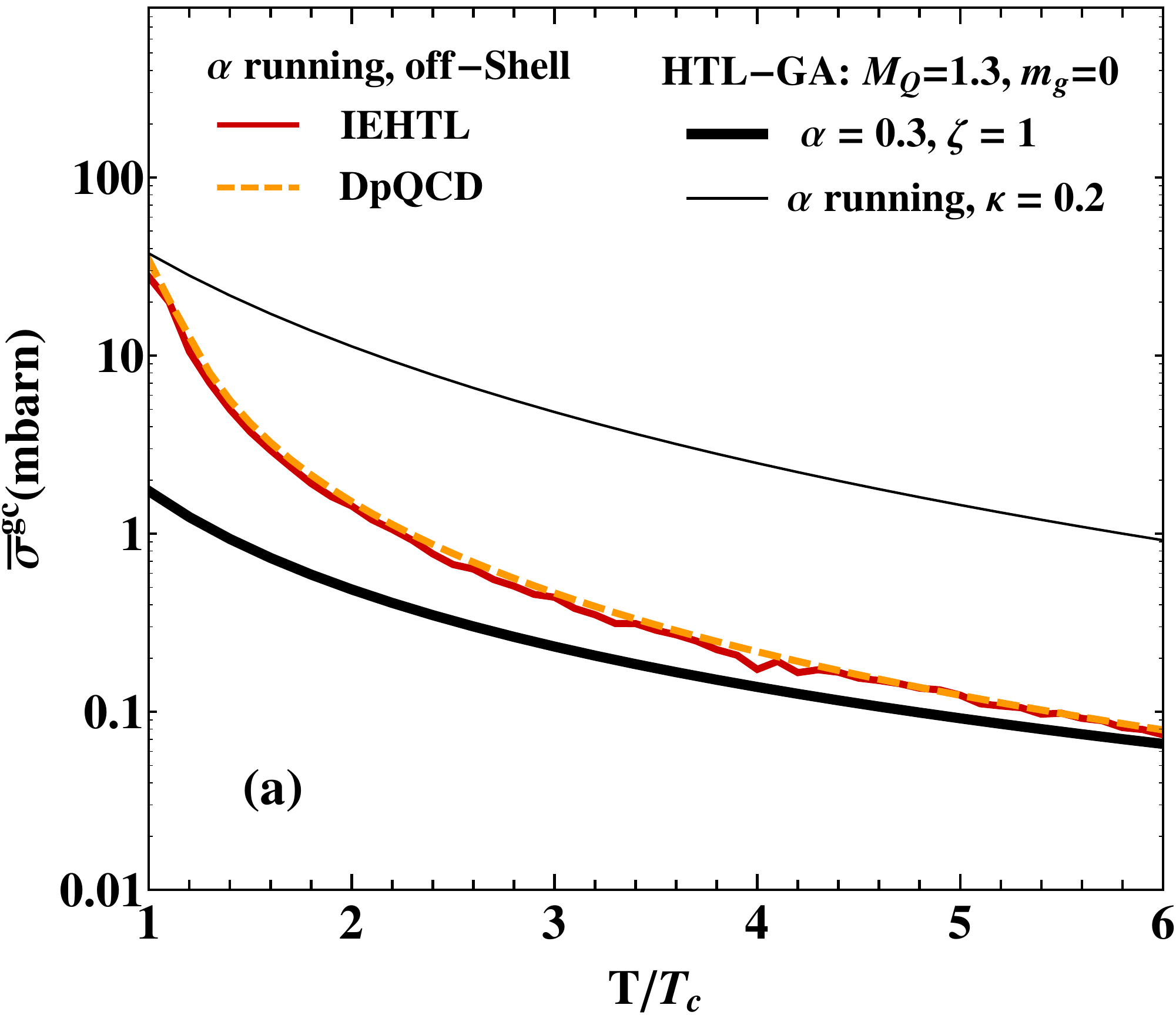}\hfill
       \vspace*{0.5cm}
       \includegraphics[width=8.5cm, height=7.8cm]{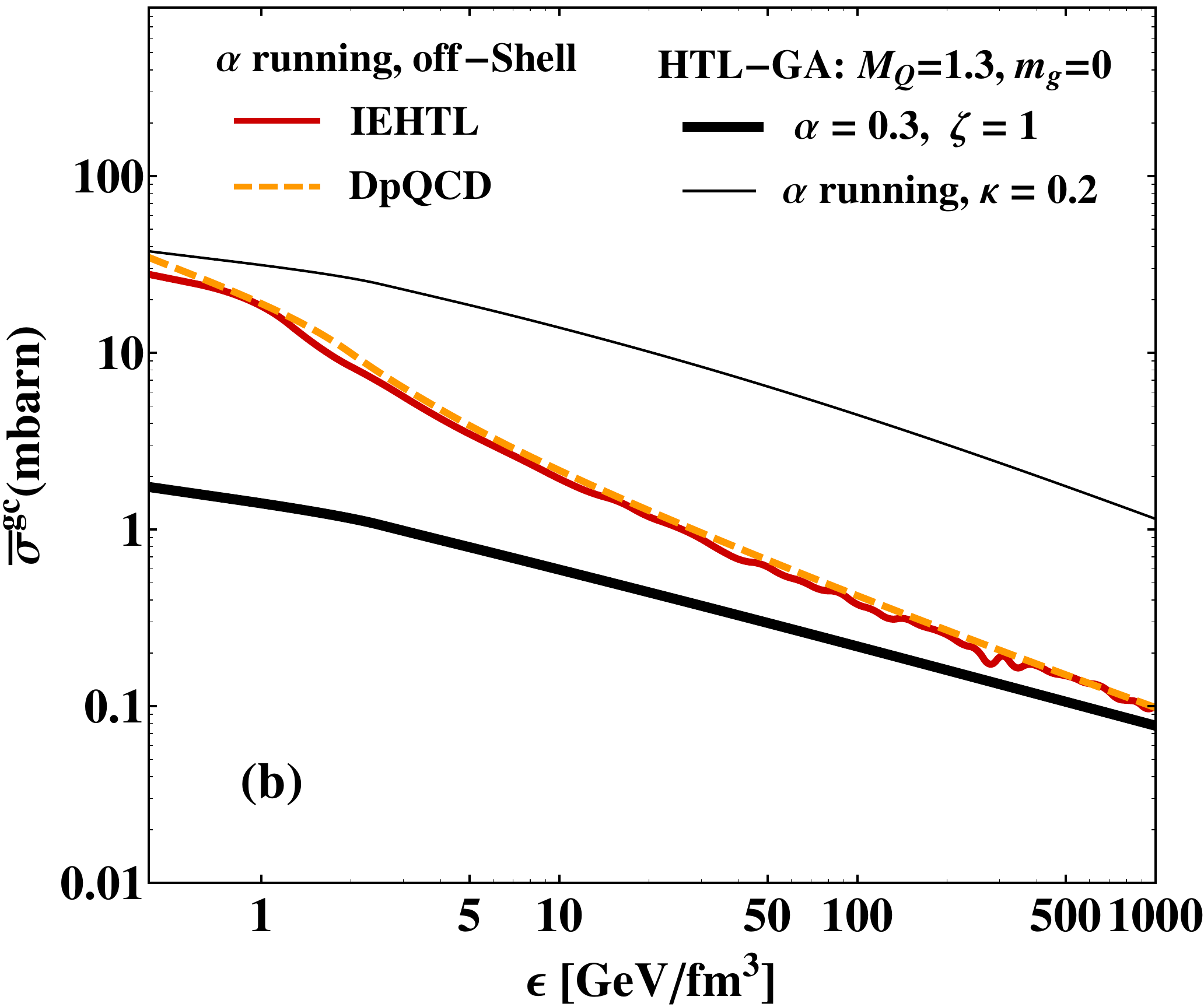}
       \end{center}
\vspace{-0.5cm}
\caption{\emph{(Color online) Elastic cross section for $g c \rightarrow g c$ scattering for off-shell (IEHTL, red line) and on-shell (DpQCD, yellow line) partons. We consider the DQPM pole masses for the on-shell partons and the DQPM spectral functions for the off-shell ones: (a) energy-averaged cross section as a function of the scaled temperature in comparison to results in the HTL-GA approaches where $m_g$ and $M_Q$ are given in GeV (see legend) and (b) energy-averaged cross section as a function of the energy density $\epsilon$}. \label{fig:SigmagQDpQCD-IEHTLeps}}
\end{figure}

 The kinematic limits for the $s$, $t$, and $u$ channels in the off-shell $gQ$ process are analogous to those in the $qQ$ case (cf. Sec. \ref{qQProcess}). In the off-shell case, the kinematical limits on the momentum transfer $t$ are given by Eqs. (\ref{equ:Sec3.20})--(\ref{equ:Sec3.23}), with
{\setlength\arraycolsep{0pt}
\begin{eqnarray}
\label{equ:Sec4.23}
& & \beta_1 = (m_g^i)^2/s, \ \beta_2 = (M_Q^i)^2/s, \ \beta_3 = (m_g^f)^2/s, \ \beta_4 = (M_Q^f)^2/s,
\end{eqnarray}}
 while
\vspace{-0.5cm}
{\setlength\arraycolsep{0pt}
\begin{eqnarray}
& & s \geq max\{(m_g^i + M_Q^i)^2, (m_g^f + M_Q^f)\}.
%\nonumber\\
\label{equ:Sec4.24}
\end{eqnarray}}
  Again, to obtain the off-shell $gQ$ elastic cross section, the elementary modified pQCD cross section has to be convoluted with the effective spectral functions for the heavy quarks and gluons. We show a comparison of the off-shell (IEHTL, solid lines) and on-shell (DpQCD) versions in Figs. \ref{fig:dSigmadthetagQDpQCD-IEHTL}(a) and \ref{fig:dSigmadthetagQDpQCD-IEHTL}(b) for the differential cross sections and in Figs. \ref{fig:SigmagQDpQCD-IEHTLs} and \ref{fig:SigmagQDpQCD-IEHTLeps} for the total cross sections as a function of $\sqrt{s}$, temperature $T/T_c$, and energy density $\epsilon$. From Figs. \ref{fig:dSigmadthetagQDpQCD-IEHTL}(a) and \ref{fig:dSigmadthetagQDpQCD-IEHTL}(b), we see that effects from the off-shell masses appear only for large scattering angles as in case of $qQ$ elastic scattering for all temperatures of the thermal bath considered.

  The total off-shell $gQ$ elastic cross section is presented as a function of $\sqrt{s}$ in Fig. \ref{fig:SigmagQDpQCD-IEHTLs} for different temperatures from 1.2 $T_c$ to 3 $T_c$ and demonstrates that the off-shell mass distributions only have a sizable impact at the threshold given by the pole masses as in case of $qQ$ scattering. The energy-averaged $gQ$ cross sections are displayed as a function of $T/T_c$ in Fig.  \ref{fig:SigmagQDpQCD-IEHTLeps}(a) in comparison to the HTL-GA model for fixed coupling (lower solid thick line) and running coupling (upper solid thin line) and finally as a function of the energy density $\epsilon$ in Fig. \ref{fig:SigmagQDpQCD-IEHTLeps}(b). Again we find---as in the case of $qQ$ scattering---that the off-shellness in mass has practically no effect on the energy-averaged cross sections in the whole temperature range considered. Here the energy averaged cross sections are computed in line with Sec. \ref{qQProcess} using the Bose-Einstein distribution for the gluons. The different power laws in temperature for the DpQCD and HTL-GA approaches are the same as for $qQ$ scattering and owing to the different regularization schemes. However, one has to point out that the amplitude for $gQ$ elastic scattering is larger than that for $qQ$ elastic scattering. In fact, the scattering of heavy quarks with gluons proceeds according to $t$, $s$, and $u$ channels, whereas one has only the $t$ channel for $qQ$ elastic scattering.
  %--------------------------------------------------------------------------------------------------------------------------------------------
\section{Summary}
\label{Summary}
%--------------------------------------------------------------------------------------------------------------------------------------------
\vskip -3mm

  In this study we have presented a detailed calculation of elastic scattering of heavy quarks with light quarks and gluons in a QGP medium using pQCD first-order Born diagrams. We have compared two approaches based on different regularization schemes. (i) The DQPM \cite{Cassing2007365} in which quarks and gluons have a finite mass and width that vary with temperature $T$. The functional form of both, as well of the running coupling $g(T/T_c)$, is given by the DQPM model \cite{Wcassing2009EPJS} and the very few parameters for the infrared enhanced coupling $g(T/T_c)$ are fitted to the equation of state obtained from lattice gauge calculations with 2+1 light flavors. We note that the DQPM is an ``effective model'' that incorporates  broad features of finite temperature QCD and should not be mixed up with first-principle pQCD calculations. However, it properly describes the equation of state from lattice QCD as well as correlators such as shear and bulk viscosities, the electric
conductivity and heat conductivity, and the electromagnetic correlator in comparison to lQCD results \cite{Marty:2013ita}. In case of a finite width the cross sections of these ``particles'' have to be calculated using spectral functions. In the DQPM model the regularization proceeds via the dynamical gluon mass in the thermal QGP medium. (ii) The Peshier-Gossiaux-Aichelin approach \cite{PhysRevC.78.014904,Gossiaux:2009hr,Gossiaux:2009mk}, which uses massless partons, a running coupling $\alpha_s(Q^2)$, and an infrared regulator, which have been adjusted to reproduce the heavy-quark energy loss in HICs at RHIC energies. The cross sections are calculated in a HTL ``inspired'' approach. This model allows as well for calculations with a fixed coupling $\alpha_s$.  They have shown  that by using a fixed coupling and a Debye mass ($m_D \approx g T$) as an infrared regulator, pQCD calculations are not able to reproduce the data at RHIC, neither the energy loss nor the azimuthal ($v_2$) distribution. These authors have proposed, furthermore,  that by employing a running coupling and by replacing the Debye mass $m_D$ by a more realistic HTL calculation, a substantial increase in the collisional energy loss can be achieved, which brings the elliptic flow $v_2 (p_T)$ as well as ratio $R_{AA} (p_T)$ closer to the experimental data.

  Our detailed studies have demonstrated that the finite width of the DQPM model---which encodes the multiple partonic scattering---has little influence on the cross section for $q Q \rightarrow q Q$ as well as $g Q \rightarrow g Q$ scattering except close to thresholds. Only at very large scattering angles do we find a difference between cross sections calculated with the spectral function and those incorporating only the pole mass. Thus, when studying the dynamics of energetic heavy quarks in a QGP medium the spectral width of the degrees of freedom may be discarded in actual transport simulations.

   As shown in Secs. \ref{qQProcess.B} and \ref{gQProcess.B} the finite  gluon mass in the DQPM screens the infrared singularity and shifts the kinematical thresholds accordingly. Owing to different kinematical boundaries the differential cross sections for heavy-quark scattering on quarks and gluons change substantially as compared to the case of massless partons, i.e., in the magnitude of the cross section as well as in the angular distributions. The angle-integrated cross sections show a very smooth dependence on the invariant energy $\sqrt{s}$ at all temperatures of the thermal bath considered and are lower than 1 mb for temperatures $T > 2 T_c$, however, increase up to about 10 mb when going down in temperature close to the critical temperature $T_c$ (cf. Secs. \ref{qQProcess.C}, \ref{qQProcess.D} and \ref{gQProcess.C}, \ref{gQProcess.D}). This dependence on temperature $T$ can be traced back to the infrared enhancement of the effective coupling which has proven to be vital for a proper description of transport coefficients. Nevertheless, using these cross sections and the DQPM densities for quarks and gluons the relaxation times for $c$ quarks are above 1 fm/$c$ up to temperatures of 3 $T_c$.

  In the Peshier-Gossiaux-Aichelin model the size of the elastic cross section and its angular distribution is dominated by the choice of the infrared regulator and the strong coupling $\alpha_s$ (running or fixed). The HTL-inspired models---essentially fixing the regulators by elementary vacuum cross sections and decay amplitudes---provide quite different results especially with respect to the temperature dependence of the $qQ$ and $gQ$ cross sections (in all settings). Accordingly, the transport properties of heavy quarks should be different as a function of temperature when compared to DQPM results.

  The differential cross sections obtained in this study will form the basis for the calculation of heavy-quark production and propagation in HICs at GSI Facility for Antiproton and Ion Research (FAIR), SPS, RHIC, and LHC energies. They will be implemented into the PHSD transport approach \cite{Cassing:2009vt}, which is based on the DQPM propagators in the partonic phase. A comparison of the DQPM cross sections with those of the Peshier-Gossiaux-Aichelin model---which have been used successfully to describe heavy-quark dynamics at RHIC energies---in a heavy-ion environment at different bombarding energies  will be mandatory to find out which approach is favored by experimental data.

  We finally point out that our calculations have employed a couple of theoretical approximations and model assumptions. Main uncertainties we encounter in the description of the $qqg$ and $ggg$ vertices, quark and gluon propagators, and parton spectral functions at finite temperature. Although in the weak coupling limit all HTL-dressed $n$-point functions are known analytically \cite{Braaten:1991gm,Braaten:1989mz} the case of the three-gluon vertex for strong coupling is even more complicated and its structure at finite temperature is not well determined yet. For this reason, we naively have used the perturbative three-gluon vertex at zero temperature. The perspectives of our study thus are to go beyond the actual approximations in future calculations and to test the resulting approaches in comparison to data from relativistic nucleus-nucleus reactions.

  \vspace{3mm}
\section*{Acknowledgment}
\vskip -3mm

 H. Berrehrah thanks R. Marty and O. Linnyk for fruitful discussions and their continuous interest. He also appreciates the ``HIC for FAIR'' framework of the ``LOEWE'' program for support. The computational resources have been provided by the LOEWE-CSC.

  \section*{Appendix}\appendix
\label{appendix}

\section{$q Q \rightarrow q Q$ \ scattering: massive light and heavy quarks of finite widths} %-----------------------------------------------------------------------------

 The expressions of the Mandelstam variables in the case of off-shell initial and/or final particles in the reaction $q Q \rightarrow q Q$ are given by:
\begin{widetext}
{\setlength\arraycolsep{0pt}
\begin{eqnarray}
& & s \ = (M_Q^i)^2 + (m_q^i)^2 + 2 \biggl(\!\! \sqrt{((M_Q^i)^2 + p^2)((M_q^i)^2 + p^2)} + p^2 \!\!\biggr) = (M_Q^f)^2 + (m_q^f)^2 + 2 \biggl(\!\! \sqrt{((M_Q^f)^2 + p^2)((M_q^f)^2 + p^2)} + p^2 \!\! \biggr)
\\
& & {} u \ = (M_Q^i)^2 + (m_q^f)^2 - 2 \biggl(\!\! \sqrt{((M_Q^i)^2 + p^2)((m_q^f)^2 + p^2)} + p^2 \cos \theta \!\! \biggr) = (M_Q^f)^2 + (m_q^i)^2 - 2 \biggl(\!\! \sqrt{((M_Q^f)^2 + p^2)((M_q^i)^2 + p^2)} + p^2 \cos \theta \!\!\biggr)
\nonumber\\
& & {} t \ = (m_q^i)^2 + (m_q^f)^2 - 2 \biggl(\!\!\sqrt{((m_q^i)^2 + p^2)((m_q^f)^2 + p^2)} - p^2 \cos \theta \!\! \biggr) = (M_Q^i)^2 + (M_Q^f)^2 - 2 \biggl(\!\! \sqrt{((M_Q^f)^2 + p^2)((M_Q^f)^2 + p^2)} - p^2 \cos \theta \!\!\biggr).
\nonumber
\label{equ:SecA.1}
\end{eqnarray}}
%\end{widetext}

\section{$g Q \rightarrow g Q$ \ scattering: massive heavy quarks and massless gluons of zero widths} %---------------------------------------------------------------------

   The invariant amplitudes for the three graphs (shown in Fig.\ref{fig:gQFeynmanDiagram}) for the case of massive heavy quarks and massless  gluons is given according to Combridge\cite{Combridge1979429} and Cutler and Sivers \cite{PhysRevD.17.196} by:
%\begin{widetext}
{\setlength\arraycolsep{-9pt}
\begin{eqnarray}
& & \mathcal{M}_t (g_a Q_i  \rightarrow g_b Q_j) = \frac{g^2}{t} f^{c a b}  T_{i j}^{c} \epsilon_i^{\alpha} \epsilon_f^{\beta} C_{\mu \alpha \beta} (k_i - k_f, - k_i, k_f) \bar{u}_j (p_f) \gamma^{\lambda} u_i (p_i),
\nonumber\\
& & {} \mathcal{M}_u (g_a Q_i  \rightarrow g_b Q_j) = - \frac{i g^2}{u - M_Q^2}  T_{i k}^{b} T_{k j}^{a} \bar{u}_j (p_f) \slashed{\epsilon}_i (\slashed{p}_i - \slashed{k}_f + M_Q) \slashed{\epsilon}_f u_i (p_i),
\nonumber\\
& & {} \mathcal{M}_s (g_a Q_i  \rightarrow g_b Q_j) = - \frac{i g^2}{s - M_Q^2}  T_{i l}^{a} T_{l j}^{b} \bar{u}_j (p_f) \slashed{\epsilon}_f (\slashed{p}_i + \slashed{k}_i + M_Q) \slashed{\epsilon}_i u_i (p_i),
\label{equ:SecA.2}
\end{eqnarray}}
 where we have suppressed the flavor indices and defined the Lorentz tensor $C$---appearing in the three-gluon vertex---by
{\setlength\arraycolsep{-9pt}
\begin{eqnarray}
& & C^{\mu \lambda \nu} (q_1,q_2, q_3) \equiv [(q_1 - q_2)^{\nu} g^{\mu \nu} + (q_2 - q_3)^{\mu} g^{\lambda \nu} + (q_3 - q_1)^{\lambda} g^{\mu \nu}].
%\nonumber\\
%& & {}
\label{equ:SecA.3}
\end{eqnarray}}
\end{widetext}

  In Eq. (\ref{equ:SecA.2}), the Latin (Greek) subscripts denote color (spin) indices assigned as in Fig. \ref{fig:gQFeynmanDiagram}, $M_Q$ is the heavy-quark mass, the $T$'s are the well-known Gell-Mann $SU(3)$ matrices, and  $\epsilon_i^{\alpha}$ ($\epsilon_f^{\beta}$) are the initial (final) gluon polarization vectors.
  
\section{$g Q \rightarrow g Q$ \ scattering: massive heavy quarks and massive gluons of zero widths} %---------------------------------------------------------------------
  
   If not only the quarks but also the gluons are massive, the invariant amplitudes for the three graphs in Fig. \ref{fig:gQFeynmanDiagram} are given by:
\begin{widetext}
{\setlength\arraycolsep{-5pt}
\begin{eqnarray}
& & \mathcal{M}_t (g_a Q_i \rightarrow g_b Q_j) = \frac{g^2}{t - m_g^2} f^{c a b}  T_{i j}^{c} \epsilon_{i, \alpha} \epsilon_{f, \beta} C^{\alpha\mu' \beta} (k_i - k_f, - k_i, k_f) \Biggl[ g_{\mu \mu'}  - \frac{q_{\mu} q_{\mu'}}{m_g^2} \Biggr ] \bar{u}_j (p_f) \gamma^{\mu} u_i (p_i),
\nonumber\\
& & {} \mathcal{M}_u (g_a Q_i \rightarrow g_b Q_j) = - \frac{i g^2}{u - (M_Q^i)^2}  T_{i k}^{b} T_{k j}^{a} \bar{u}_j (p_f) \slashed{\epsilon}_i (\slashed{p}_i - \slashed{k}_f + M_Q^i) \slashed{\epsilon}_f u_i (p_i),
\nonumber\\
& & {} \mathcal{M}_s (g_a  Q_i\rightarrow  g_b Q_j) = - \frac{i g^2}{s - (M_Q^i)^2}  T_{i l}^{a} T_{l j}^{b} \bar{u}_j (p_f) \slashed{\epsilon}_f (\slashed{p}_i + \slashed{k}_i + M_Q^i) \slashed{\epsilon}_i u_i (p_i),
\label{equ:SecA.4}
\end{eqnarray}}
%\end{widetext}
 where $u_i(p_i,M_Q^i)$ ($\bar{u}_j(p_f,M_Q^f)$) is a Dirac spinor for the incoming (outgoing) heavy quark with momentum $p^i$ ($p^f$), mass $M_Q^i$ ($M_Q^f$) and color $i$ ($j$). The different amplitudes squared, summed (averaged) over the final (initial) degrees of freedom, are given by
%\begin{widetext}
{\setlength\arraycolsep{-5pt}
\begin{eqnarray}
\label{equ:SecA.5}
& & <\!\! |\mathcal{M}_t |^2 \!\!>  = \frac{g^4}{8 (t - m_g^2)^2} \ tr \Biggl[(\slashed{p}_f + M_Q^f) \gamma^{\mu} (\slashed{p}_i + M_Q^i) \gamma^{\nu} C^{\alpha \mu' \beta} (k_i - k_f, - k_i, k_f) \biggl[g_{\mu \mu'}  - \frac{q_{\mu} q_{\mu'}}{m_g^2} \biggr]
\nonumber\\
& & {} \hspace{4.5cm} \times \ ^\star C^{\lambda \nu' \rho} (k_i - k_f, - k_i, k_f)  \biggl[g_{\nu \nu'}  - \frac{q_{\nu} q_{\nu'}}{m_g^2} \biggr ] \Biggr ] \sum_{pol, i} \epsilon_{i,\alpha} \epsilon_{i, \lambda } \ \sum_{pol, f} \epsilon_{f,\beta} \epsilon_{f, \rho},
\\
& & {} <\!\! |\mathcal{M}_s|^2 \!\!>  = \! \frac{2 g^4/36}{(s - (M_Q^i)^2)^2} \ tr \Biggl[(\slashed{p}_f + M_Q^f) \gamma^{\beta} (\slashed{p}_i + \slashed{k}_i + M_Q^i) \gamma^{\alpha} (\slashed{p}_i + M_Q^i) \gamma^{\lambda} (\slashed{p}_i + \slashed{k}_i + M_Q^i) \gamma^{\rho} \Biggr] \! \sum_{pol, i} \epsilon_{i,\alpha} \epsilon_{i, \lambda } \! \sum_{pol, f} \epsilon_{f,\beta} \epsilon_{f, \rho},
\nonumber\\
& & {} <\!\! |\mathcal{M}_u|^2 \!\!>  = \! \frac{2 g^4/36}{(u - (M_Q^i)^2)^2} \ tr \Biggl[(\slashed{p}_f + M_Q^f) \gamma^{\alpha} (\slashed{p}_i - \slashed{k}_f + M_Q^i) \gamma^{\beta} (\slashed{p}_i + M_Q^i) \gamma^{\rho} (\slashed{p}_i - \slashed{k}_f + M_Q^i) \gamma^{\lambda} \Biggr] \! \sum_{pol, i} \epsilon_{i,\alpha} \epsilon_{i, \lambda } \! \sum_{pol, f} \epsilon_{f,\beta} \epsilon_{f, \rho},
\nonumber
%& & {}
\end{eqnarray}}

 and for the interference terms, we have for the $u$-$s$ interference (the star denotes complex conjugation)
{\setlength\arraycolsep{-5pt}
\begin{eqnarray}
& & <\!\! \mathcal{M}_s \mathcal{M}_u^{\star} \!\!> = \! \frac{- g^4}{144 (s - (M_Q^i)^2) (u - (M_Q^i)^2)} tr \Biggl[\!(\slashed{p}_f + M_Q^f) \gamma^{\beta} (\slashed{p}_i + \slashed{k}_i + M_Q^i) \gamma^{\alpha} (\slashed{p}_i + M_Q^i) \gamma^{\rho} (\slashed{p}_i - \slashed{k}_f + M_Q^i) \gamma^{\lambda} \! \Biggr]
\nonumber\\
& & {} \hspace{7cm} \times \sum_{pol, i} \epsilon_{i,\alpha} \epsilon_{i, \lambda } \ \sum_{pol, f} \epsilon_{f,\beta} \epsilon_{f, \rho},
\nonumber\\
& & {} <\!\! \mathcal{M}_u \mathcal{M}_s^{\star} \!\!> = \! \frac{- g^4}{144(s - (M_Q^i)^2) (u - (M_Q^i)^2)} tr \Biggl[\!(\slashed{p}_f + M_Q^f) \gamma^{\alpha} (\slashed{p}_i - \slashed{k}_f + M_Q^i) \gamma^{\beta} (\slashed{p}_i + M_Q^i) \gamma^{\lambda} (\slashed{p}_i + \slashed{k}_i + M_Q^i) \gamma^{\rho} \! \Biggr]
\nonumber\\
& & {} \hspace{7cm} \times \sum_{pol, i} \epsilon_{i,\alpha} \epsilon_{i, \lambda } \ \sum_{pol, f} \epsilon_{f,\beta} \epsilon_{f, \rho},
\nonumber\\
\label{equ:SecA.6}
\end{eqnarray}}

 for the $t$-$s$ interference,

{\setlength\arraycolsep{-5pt}
\begin{eqnarray}
& &  <\!\! \mathcal{M}_t \mathcal{M}_s^{\star} \!\!> = \! \frac{- g^4}{16 (t - M_Q^2) (s - (M_Q^i)^2)} tr \Biggl[\!(\slashed{p}_f + M_Q^f) \gamma^{\mu} (\slashed{p}_i + M_Q^i) \gamma^{\lambda} (\slashed{p}_i + \slashed{k}_i + M_Q^i) \gamma^{\rho} C^{\alpha\mu' \beta} (k_i - k_f, - k_i, k_f)
\nonumber\\
& & {} \hspace{6.5cm} \times \biggl[ g_{\mu \mu'}  - \frac{q_{\mu} q_{\mu'}}{m_g^2} \biggr] \Biggr ] \sum_{pol, i} \epsilon_{i,\alpha} \epsilon_{i, \lambda } \ \sum_{pol, f}, \epsilon_{f,\beta} \epsilon_{f, \rho},
\nonumber\\
& & {} <\!\! \mathcal{M}_s \mathcal{M}_t^{\star} \!\!>  = \frac{- g^4}{16 (t - M_Q^2) (s - (M_Q^i)^2)} tr \Biggl[(\slashed{p}_f + M_Q^f) \gamma^{\beta} (\slashed{p}_i + \slashed{k}_i + M_Q^i) \gamma^{\alpha} (\slashed{p}_i + M_Q^i) \gamma^{\nu} \ ^\star C^{\lambda \nu' \rho} (k_i - k_f, - k_i, k_f)
\nonumber\\
& & {} \hspace{6.5cm} \times \biggl[ g_{\nu \nu'}  - \frac{q_{\nu} q_{\nu'}}{m_g^2} \biggr] \Biggr ] \sum_{pol, i} \epsilon_{i,\alpha} \epsilon_{i, \lambda } \ \sum_{pol, f} \epsilon_{f,\beta} \epsilon_{f, \rho},
%\nonumber\\
\label{equ:SecA.7}
\end{eqnarray}}

 and finally for the $t$-$u$ interference,

{\setlength\arraycolsep{-5pt}
\begin{eqnarray}
& &  <\!\! \mathcal{M}_t \mathcal{M}_u^{\star} \!\!>  = \frac{g^4}{16 (t - m_g^2) (u - (M_Q^i)^2)} tr \Biggl[(\slashed{p}_f + M_Q^f) \gamma^{\mu} (\slashed{p}_i + M_Q^i) \gamma^{\rho} (\slashed{p}_i - \slashed{k}_i + M_Q^f) \gamma^{\lambda} C^{\alpha \mu' \beta} (k_i - k_f, - k_i, k_f)
\nonumber\\
& & {} \hspace{6.5cm} \times \biggl[ g_{\mu \mu'}  - \frac{q_{\mu} q_{\mu'}}{m_g^2} \biggr] \Biggr ] \sum_{pol, i} \epsilon_{i,\alpha} \epsilon_{i, \lambda } \ \sum_{pol, f} \epsilon_{f,\beta} \epsilon_{f, \rho},
\nonumber\\
& & {} <\!\! \mathcal{M}_u \mathcal{M}_t^{\star} \!\!>  = \frac{g^4}{16 (t - M_Q^2) (s - (M_Q^i)^2)} tr \Biggl[(\slashed{p}_f + M_Q^f) \gamma^{\alpha} (\slashed{p}_f - \slashed{k}_i + M_Q^f) \gamma^{\beta} (\slashed{p}_i + M_Q^i) \gamma^{\nu} \ ^\star C^{\lambda \nu' \rho} (k_i - k_f, - k_i, k_f)
\nonumber\\
& & {} \hspace{6.5cm} \times \biggl[ g_{\nu \nu'}  - \frac{q_{\nu} q_{\nu'}}{m_g^2} \biggr] \Biggr ] \sum_{pol, i} \epsilon_{i,\alpha} \epsilon_{i, \lambda } \ \sum_{pol, f} \epsilon_{f,\beta} \epsilon_{f, \rho}.
\label{equ:SecA.8}
\end{eqnarray}}
\end{widetext}

 In the expressions (\ref{equ:SecA.4})--(\ref{equ:SecA.8}) we have defined $q = k_f - k_i$ as the 4-momentum of the exchanged gluon and have used the quark and gluon propagators

{\setlength\arraycolsep{-5pt}
\begin{eqnarray}
& & \hspace{-1.4cm} G_F^{\mu \nu} (\!q, m_g\!) = \! - i \frac{g_{\mu \nu} - q_{\mu} q_{\nu}/m_g^2}{q^2 - m_g^2}, \hspace{0.5cm} S_F (\!p, M_Q\!) = \! \frac{\slashed{p} + M_Q}{p^2 - \! M_Q^2}\!,
\label{equ:SecA.9}
\end{eqnarray}}
 and also for $C^{\alpha \mu' \beta} (k_i - k_f, - k_i, k_f)$ the expression
{\setlength\arraycolsep{-5pt}
\begin{eqnarray}
C^{\alpha \mu' \beta} (k_i - k_f, - k_i, k_f) = & & \biggl[(2 k_i - k_f)_{\beta} \ g_{\mu' \alpha} + (- k_i - k_f)_{\mu} \ g_{\alpha \beta} \ + (2 k_f - k_i)_{\alpha} \ g_{\mu \beta} \biggr].
\label{equ:SecA.10}
\end{eqnarray}}
  The total amplitude for the process $g Q \rightarrow g Q$ is then evaluated according to:
{\setlength\arraycolsep{0pt}
\begin{eqnarray}
\label{equ:SecA.11}
<\!\! |\mathcal{M}|^2 \!\!> \ = \ & &  <\!\! |\mathcal{M}_t |^2 \!\!> + <\!\! |\mathcal{M}_s|^2 \!\!> + <\!\! |\mathcal{M}_u|^2 \!\!>  + <\!\! \mathcal{M}_s \mathcal{M}_u^{\star} \!\!> + <\!\! \mathcal{M}_u \mathcal{M}_s^{\star} \!\!> + <\!\! \mathcal{M}_t \mathcal{M}_s^{\star} \!\!> 
\nonumber\\
& & {} + <\!\! \mathcal{M}_s \mathcal{M}_t^{\star} \!\!>  +  <\!\! \mathcal{M}_t \mathcal{M}_u^{\star} \!\!> + <\!\! \mathcal{M}_u \mathcal{M}_t^{\star} \!\!>.
%\nonumber
\end{eqnarray}}

  \bibliography{References}

\end{document}